%% file: AstroBH_notesM2.tex
\documentclass[12pt, a4paper, twoside]{article}

\usepackage[
   paper=a4paper, 
	inner=2.5cm, 
	outer=3.8cm, 
	bindingoffset=.5cm, 
	top=3cm, 
	bottom=2.7cm, 
	]{geometry}

\usepackage[hidelinks]{hyperref}
\usepackage{amsfonts}
\usepackage{amsmath}
\usepackage{empheq}
\usepackage{setspace}
\usepackage{color}
\usepackage{esint}
\usepackage{slashed}
\usepackage{enumitem}
\usepackage[british]{babel}
\usepackage{hyphenat}
\usepackage{graphicx}
\usepackage[mathscr]{eucal}
\usepackage{wasysym}
\usepackage{etex}
\usepackage{ragged2e}

\usepackage{amsfonts,amssymb,amsmath,bbm}
\usepackage[all]{xy}  
\usepackage{tikz}  

\newcommand{\phant}{\phantom{a}}

\numberwithin{equation}{section}

\usepackage{cite}
\usepackage{graphicx, import}
\usepackage{calc}
\usepackage[font=small, font+=it, format=hang, margin={1cm, 1cm}]{caption}
\graphicspath{{figures/}}

\title{Lecture notes on black hole astrophysics}

\begin{document}

\begin{titlepage}

\begin{center}{\huge{\bf Lecture notes on \vspace{0.7mm}\\ black hole binary astrophysics}}
\end{center}
\vspace{3mm}
\centerline{\normalsize{\bf{M.~Celoria\footnote{marco.celoria@gssi.infn.it}}, \bf{R.~Oliveri\footnote{roliveri@ulb.ac.be}}, 
    \bf{A.~Sesana\footnote{asesana@star.sr.bham.ac.uk}},
\bf{M.~Mapelli\footnote{michela.mapelli@uibk.ac.at}}}}
\vspace{2em}
\begin{center}
\textit{
$\phant^{1}$Gran Sasso Science Institute (INFN),\\ Viale Francesco Crispi 7,  I-67100 L'Aquila, Italy\\~\\
$\phant^{2}$Service de Physique Th{\'e}orique et Math{\'e}matique,\\ Universit{\'e} Libre de Bruxelles and International Solvay Institutes,\\ Campus de la Plaine, CP 231, B-1050 Brussels, Belgium\\~\\
$\phant^{3}$School of Physics and Astronomy and Institute of Gravitational Wave Astronomy, University of Birmingham,\\
  Edgbaston, Birmingham B15 2TT, United Kingdom\\~\\
  $\phant^{4}$Institut f\"ur Astro- und Teilchenphysik, Universit\"at Innsbruck, Technikerstrasse 25/8, A--6020, Innsbruck, Austria
}
\end{center}

\vspace{5mm}
\normalsize
\textit{
}


\begin{abstract}
\noindent 
{We describe some key astrophysical processes driving the formation and evolution of black hole binaries of different nature, from stellar-mass to supermassive systems. In the first part, we focus on the mainstream channels proposed for the formation of stellar mass binaries relevant to ground-based gravitational wave detectors, namely the {\it field} and the {\it dynamical} scenarios. For the field scenario, we highlight the relevant steps in the evolution of the binary, including mass transfer, supernovae explosions and kicks, common envelope and gravitational wave emission. For the dynamical scenario, we describe the main physical processes involved in the formation of star clusters and the segregation of black holes in their centres. We then identify the dynamical processes leading to binary formation, including three-body capture, exchanges and hardening. The second part of the notes is devoted to massive black hole formation and evolution, including the physics leading to mass accretion and binary formation. Throughout the notes, we provide several step-by-step pedagogical derivations, that should be particularly suited to undergraduates and PhD students, but also to gravitational wave physicists interested in approaching the subject of gravitational wave sources from an astrophysical perspective.
}
\end{abstract}


\end{titlepage}

\tableofcontents


\newpage

\section*{Physical constants}
Here is a list of the physical constants in the notes:
\begin{subequations}
\begin{align}
c &= 3,00 \times 10^{10} \mbox{cm s$^{-1}$} \qquad &&\mbox{speed of light}, \nonumber \\
G &= 6.67 \times 10^{-8} \mbox{cm$^{3}$ g$^{-1}$ s$^{-2}$} \qquad &&\mbox{gravitational constant}, \nonumber \\
K_B &= 1.38 \times 10^{-16} \mbox{erg~K$^{-1}$} \qquad &&\mbox{Boltzmann constant}, \nonumber \\
\sigma&=5.67 \times 10^{-5} \mbox{erg cm$^{-2}$ s$^{-1}$ K$^{-4}$} \qquad &&\mbox{Stefan-Boltzmann constant}, \nonumber \\
M_{\astrosun} &= 1.99 \times 10^{33} \mbox{g} \qquad &&\mbox{solar mass}, \nonumber\\
R_{\astrosun} &= 6.95 \times 10^{10} \mbox{cm} \qquad &&\mbox{solar radius}, \nonumber\\
m_{H} &= 1.67 \times 10^{-24}\mbox{g} \qquad &&\mbox{hydrogen mass}, \nonumber\\
t_{Hubble} &= 14.4 \times 10^{9} \mbox{yr} \qquad &&\mbox{Hubble time}. \nonumber
\end{align}
\end{subequations}

\newpage

\section{Astrophysical black holes}

Astrophysical black holes can be classified according to their mass. 
It is common to identify three classes of black holes: 
\begin{itemize}

\item[1.] \emph{Stellar-mass Black Holes} (SBHs) with masses from few to tens of solar masses, $M_\odot \lesssim M_{\rm SBH} \lesssim 10^2 M_\odot$. They are the natural relics of stars with $M\gtrsim{}20 M_\odot$. They have been observed  either in X-ray binary systems, where a SBH is coupled to a companion star (about twenty X-ray binaries host dynamically confirmed SBHs \cite{2006ARA&A..44...49R}), or as gravitational wave (GW) sources in the LIGO band (five binaries detected to date \cite{2016PhRvX...6d1015A,2017PhRvL.118v1101A,2017PhRvL.119n1101A,2017ApJ...851L..35A}).

\item[2.] \emph{Intermediate Mass Black Holes} (IMBHs) with masses from hundreds to hundreds of thousands of solar masses, $10^2 M_\odot \lesssim M_{\rm IMBH} \lesssim 10^6  M_\odot $. This is the most mysterious class, because there is only sparse evidence of these object up until now. IMBHs can form at high redshift either as population III star remnants \cite{2001ApJ...551L..27M} or via direct collapse from a marginally stable protogalactic disc or quasistar (\emph{e.g.}, \cite{2006MNRAS.370..289B,2012MNRAS.425.2854A}), or at lower redshifts via dynamical processes in massive star clusters \cite{2002ApJ...576..899P}. Although statistically significant samples of black holes with masses down to $\approx 10^5 M_\odot$ have been detected in dwarf galaxy nuclei \cite{2013ApJ...775..116R,2018arXiv180201567M}, still there is no unambiguous detection of IMBHs in the mass range $10^2 M_\odot \lesssim M \lesssim 10^5  M_\odot $.

\item[3.] \emph{(Super)-Massive Black Holes} (MBHs or SMBHs) with masses from few millions to billions of solar masses, $ 10^6 M_\odot \lesssim M_{\rm SMBH} \lesssim 10^{10} M_\odot $.
They are observed at the centre of almost all galaxies \cite{1998AJ....115.2285M} and power active galactic nuclei (AGN, \cite{1969Natur.223..690L}) and quasars already at $z\approx 7.5$ \cite{2018Natur.553..473B}. The MBH hosted at the centre of the Milky Way is called Sagittarius A$^{\star}$ and has a mass of $\approx 4\times10^6  M_\odot$ \cite{2009ApJ...692.1075G}.
\end{itemize}

\begin{figure} [h!]
  \centering
    \includegraphics[width=0.9\textwidth]{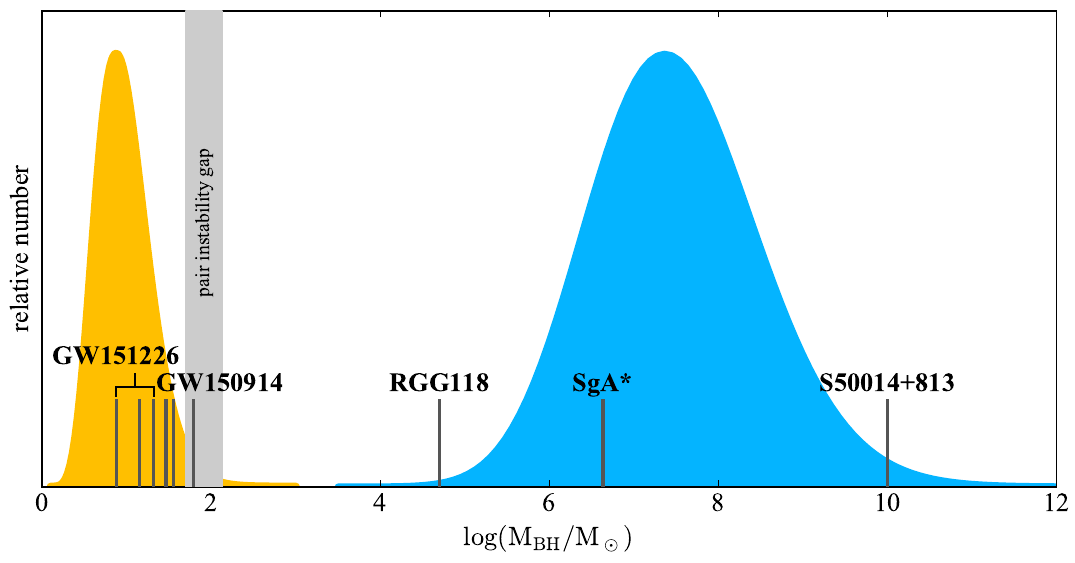}
    \caption{Cartoon of the black hole mass function (modelled as a log-normal distribution), from SBHs (orange distribution on the left) to SMBHs (blue distribution on the right). Vertical lines denote the black hole masses in GW150914 and GW151226 (including the mass of the new black hole, result of the merger) and the smallest and largest mass of the two black holes known as of today in galaxies. The grey vertical strip denotes the region of pair instability Supernovae that leave no remnant. The ``gap'' in the range $10^2 M_\odot \lesssim M \lesssim 10^5  M_\odot $, also known as BH desert, might either be real or due to selection effects, that make it particularly difficult to detect IMBHs. The next generation of ground based detectors, jointly with LISA in space, will shed light into this poorly sampled mass range. Figure taken from \cite{2017ogw..book...43C}.}
\label{BH_landscape}
\end{figure}

{\it In these notes we will use the acronyms BH and BHB to refer to general black holes and binaries, we will use SBH (SBHB) for stellar black hole (binaries), IMBH (IMBHBs) for intermediate mass black hole (binaries) and MBH (MBHB) for massive (or supermassive) black hole (binaries).} A pictorial representation of the mass ranges of known black holes is given in Fig.~\ref{BH_landscape}, which highlights the current difficulties in pinning down the IMBH range.

\section{Binary systems: basics}

Stars, neutron stars (NS) and BHs are typically observed in binary systems. For example, half of the total amount of stars and, more interesting, $70\%$ of the stars with mass greater than ten solar masses, that are relevant to the formation of compact objects, are in binaries \cite{2012Sci...337..444S}.

Remarkably, GWs emitted during the merger of two compact objects in a binary system have been recently observed by GWs detectors, advanced LIGO and Virgo, opening a new window to astrophysics. At the time of writing, five SBH and one NS-NS merging events have been announced by the LIGO/Virgo collaboration (LVC) \cite{2016PhRvX...6d1015A,2017PhRvL.118v1101A,2017PhRvL.119n1101A,2017ApJ...851L..35A,2017PhRvL.119p1101A}, and more are expected to come when detectors operations will be resumed in early 2019. From these observations, we have learned that SBH binaries exist, that they can be quite massive,\footnote{SBHs observed in X-ray binaries were characterized by $M\lesssim20 M_\odot$, much lighter than, \emph{e.g.}, the first GW detection, GW150914.} and that they can merge within a Hubble time.

Remarkably, in our galaxy we have observed at least 15 neutron star binaries, and 6 of them will merge in a Hubble time \cite{2008LRR....11....8L}. The observation of GW170817 has confirmed the existence of a large population of merging NS binaries beyond the Milky Way (MW), with an estimated merger rate $300< {\cal R}_{\rm NSB} <5000$ Gpc$^{-3}$yr$^{-1}$ \cite{2017PhRvL.119p1101A}.
Moreover, the large number of LIGO/Virgo detections -- implying a local rate $6< {\cal R}_{\rm BHB} <220$ Gpc$^{-3}$yr$^{-1}$ \cite{2016PhRvX...6d1015A} -- indicate that BHB formation is a relatively common phenomenon in Nature.

This section is devoted to the description of binary systems. We outline main physical features of such systems, like the coalescence phase and the emission of GWs. In particular, we  start by considering the Keplerian motion of the binary, where two compact objects can be regarded as point masses and, as long as the orbital velocities are small compared to the speed of light $c$, we can use the Newtonian gravitation theory as a good approximation. Then, we sketch the derivation of the change of the binary orbit caused by the emission of GWs and we present the relation between the initial separation of the two objects and the coalescence time. 

\subsection{Keplerian motion}
\label{sec:kepler}

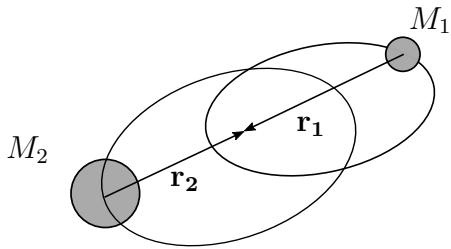
\begin{figure}[h!]
\centering
\def\svgscale{0.8}
\input{binsystem.pdf_tex}
\caption{Binary system governed by Newtonian gravity.}
\label{binarysyst}
\end{figure}

Let us consider two point masses $M_1$ and $M_2$ moving in elliptical orbits around the centre of mass (CoM), as depicted in Fig.~\ref{binarysyst}. Let $\vec{r}_i = \vec{r}_i(t)$ be the distance from the mass $M_{i}$ to the CoM and $M$ be the total mass of the system $M = M_1+M_2$.

Setting the origin of the coordinates in the CoM, the positions of the objects from the CoM are given by
\begin{equation}
\vec{r}_1 = \frac{M_2}{M}\vec{r}, \qquad \vec{r}_2 = - \frac{M_1}{M}\vec{r},
\end{equation}  
where we defined the relative distance between the two masses to be $\vec{r} = \vec{r}_1 - \vec{r}_2$. Assuming, for the moment, that the two masses are constant, the velocities relative to the CoM read as
\begin{equation}
\vec{v}_1 = \dot{\vec{r}}_1 =  \frac{M_2}{M}\dot{\vec{r}}=- \frac{M_2}{M_1}\dot{\vec{r}}_2=- \frac{M_2}{M_1} \vec{v}_2 \, .
\end{equation} 
Therefore, total energy of the system is
\begin{equation}
E = \frac{M_1 v_1^2}{2} + \frac{M_2 {v}_2^2}{2} - 
  \frac{GM_1M_2}{r} =
  \frac{\mu {v}^2}{2} - \frac{GM \mu}{r} =  -\frac{GM \mu}{2a} \, ,
  \label{B:E}
\end{equation}
where we introduced the relative velocity between the two bodies $\vec{v}= \vec{v}_1-\vec{v}_2$ and the reduced mass of the system $\mu = M_1 M_2/M$.
In the last step, we have used the conservation of energy along the orbit and wrote the energy in terms of the semi-major axis $a$ ($a=|\vec{r}|=r_c$ in the limit of circular orbits). 
Hence, the binary system is equivalent to considering a single body with mass $\mu$ moving in an effective external gravitational potential \cite{landau1976mechanics}.
The motion is in an elliptic orbit with eccentricity $e$, semi-major axis $a$, orbital period $P$ and orbital frequency $\omega=2\pi/P$ satisfying Kepler's third law
\begin{equation}
  \omega^2 = \left( \frac{2\pi}{P} \right)^2 = \frac{GM}{a^3}\, .
  \label{B:3Kepl}
\end{equation}
The orbital angular momentum vector,  perpendicular to the orbital plane, is given by
\begin{equation}
  \vec{J}_{\rm orb}=
  M_1\vec{v}_1\times\vec{r}_1 + M_2\vec{v}_2\times\vec{r}_2 =
  \mu\vec{v}\times\vec{r}
  \label{B:vecJ}
\end{equation}
and its  absolute value is
\begin{equation}
J_{\rm orb}= \mu\sqrt{GMa(1-e^2)} \, .
  \label{B:Je}
\end{equation}
For circular binaries, we have $r(t)=r_{c}$ and $e=0$, so the angular momentum reduces to
\begin{equation}
\begin{aligned}
J_{\rm orb} = \mu\sqrt{GMr_c}
\end{aligned}
\end{equation}
and, from $v=\omega r_c$, we get the circular velocity of the binary 
\begin{equation}
v= \sqrt{\frac{GM}{r_c}} \, .
\end{equation}

\subsection{Gravitational radiation from a binary system}

In the previous subsection, we considered the binary system in Newtonian approximation. However, in order to describe the gravitational radiation emitted by a binary, we need a general relativistic framework.

A full description of binary dynamics and GW emission in General relativity is beyond the scope of these notes, and a full treatment can be found in \cite{2014LRR....17....2B}. Here, we just mention that the BHB modifies the geometry of spacetime in a time varying fashion, generating GWs that propagate at the speed of light. In the weak-field approximation, sometimes also known as Isaacson short-wave approximation, the spacetime metric can be written as the sum of the background spacetime metric $\bar{g}_{\mu\nu}$ and a small perturbation metric $h_{\mu\nu}$ describing the GWs
\begin{equation}
ds^2=g_{\mu\nu}dx^\mu dx^\nu=(\bar{g}_{\mu\nu}+h_{\mu\nu})dx^\mu dx^\nu.
\end{equation}
The approximation is valid as long as the perturbative scale of the waves $h_{\mu\nu}$ is much smaller than the curvature scale of the background metric $\bar{g}_{\mu\nu}$. For any astrophysical purposes, the background metric is the Minkowski metric $\bar{g}_{\mu\nu}=\eta_{\mu\nu}$. Among the ten independent components of the linearised field $h_{\mu\nu}$, only two of them describe the dynamical propagating degrees of freedom. To eliminate any gauge redundancy of the theory, we fix the harmonic or de Donder gauge and impose the transverse and traceless (TT) conditions \cite{misner1973gravitation}.

For Newtonian sources, \emph{i.e.}, non relativistic sources for which $v_{\rm source}\ll c$, localised in a compact region of space, the gravitational power (energy per unit time) radiated is governed by the Einstein quadrupole formula given by\footnote{The mass monopole term has to vanish because of the mass-energy conservation, while the mass dipole term vanish because of the conservation of the linear and angular momentum.}
\begin{equation}
\frac{dE_{\rm rad}}{dt} = \frac{G}{5}\left<  \dddot{I}_{ij}\dddot{I}_{ji}\right>,
\end{equation}
where the brackets stand for average over the solid angle and the tensor $I_{ij}$ is the mass quadrupole moment given by the following integral of the Newtonian mass density over the compact region of the source
\begin{equation}
I_{ij}(t) = \int_{\rm source}d^{3}r \left(r_i r_j - \frac{1}{3}\delta_{ij}r^2 \right)T_{00}(t, \vec{r}),
\end{equation}
and $T_{00}$ is the 00 component of the source stress energy tensor \cite{misner1973gravitation}. For binary systems, it is easy to show that the averaged power emitted is given by \cite{PhysRev.131.435,1964PhRv..136.1224P} 
\begin{equation}\label{GWpower}
\frac{dE_{\rm rad}}{dt} =  \frac{32}{5}\frac{G^4}{c^5}\frac{\mu^2 M^3}{a^5}F(e) = \frac{32}{5}\frac{G^4}{c^5}\frac{M_1^2 M_2^2(M_1+M_2)}{a^5}F(e),
\end{equation}
where the factor
\begin{equation}\label{eq:Fe}
  F(e)=\left(1-e^2\right)^{-7/2}\left(1+\frac{73}{24}e^2+\frac{37}{96}e^4\right)
\end{equation}
depends on eccentricity only and shows that highly eccentric binaries are much more efficient at radiating away energy in form of GWs. The averaged angular momentum flux reads
\begin{equation}\label{GWangularmomentum}
\frac{dJ_{\rm rad}}{dt} 
=\frac{32}{5}\frac{G^{7/2}}{c^5}\frac{M_1^2M_2^2(M_1+M_2)^{1/2}}{a^{7/2}}\left(1-e^2\right)^{-2}\left(1+\frac{7}{8}e^2\right).
\end{equation}
In other words, gravitational waves extract energy and momentum out of the orbit and, as a consequence, the masses spiral together around each other until they eventually merge. From Eqs.~\eqref{GWpower}-\eqref{GWangularmomentum} it is possible to show that  the separation between the two bodies decreases according to
\begin{equation}
  \frac{da}{dt} = -\frac{64}{5}\frac{G^3}{c^5}\frac{M_1 M_2(M_1+M_2)}{a^3}F(e),
  \label{eq:dadtgw}
\end{equation}
whereas the eccentricity evolves according to
\begin{equation}
  \frac{de}{dt} = -\frac{304}{15}\frac{G^3}{c^5}\frac{M_1 M_2(M_1+M_2)}{a^4}e\left(1-e^2\right)^{-5/2}\left(1+\frac{121}{304}e^2\right),
  \label{eq:dedtgw}
\end{equation}
from which we see that GWs drive the binary towards circularization along the inspiral. By integrating Eq.~\eqref{eq:dadtgw} and neglecting that the eccentricity $e$ changes in time, we estimate the time of coalescence of the binary system
\begin{equation}\label{tcoal}
t_{\rm coal}=\frac{5}{256}\frac{c^5}{G^3}\frac{a_0^4}{M_1M_2 (M_1+M_2)}\frac{1}{F(e)}.
\end{equation}
It is instructive to solve for the initial separation $a_{0}$ and rewrite the expression as
\begin{equation}
  a_{0}= 1.6~ R_{\astrosun}\left(\frac{M_1}{M_{\astrosun}}\right)^{3/4}\left[q(1+q) F(e)\left(\frac{t_{\rm coal}}{1 {\rm Gyr}}\right)\right]^{1/4},
\label{a0}
\end{equation}
where we introduced the binary mass ratio $q=M_2/M_1$. From Eq.~\eqref{a0} we immediately see that the initial separation $a_0$ required for two suns to merge within an Hubble time is $a_0\approx 10^{11}$ cm $\approx 0.01$ AU. This constitutes a major problem to the naive concept of creating SBH binaries (SBHBs) from pre-existing stellar binaries. In fact, during their evolution, stars undergo a giant phase characterised by  $R_G \approx 10^{14}$ cm. Therefore, stars with initial separations $a_0<R_G$ during their life on the main sequence, would simply engulf each other during their giant phase, thus merging in a single star {\it before} forming the binaries of compact objects observed by LIGO/Virgo. Only binaries with separations bigger then $1$~AU, \emph{i.e.}, $100$ times bigger than $a_{0}$, would survive the giant phase intact, thus leaving behind remnants that would merge in $t_{\rm coal}\approx 10^8 ~t_{\rm Hubble}$! One of the primary challenges of stellar population/evolution modellers is to envisage mechanisms to bring the two objects close enough to efficiently emit GWs and merge in a Hubble time.

\section{The common evolution channel of SBHB formation}

In the previous section, we have considered the relation between the initial separation $a_0$ of two objects in the binary and the coalescence time $t_{\rm coal}$; see Eq.~\eqref{tcoal}. Importantly, we have seen that in order to have an efficient GW emission and merge in a Hubble time, the initial separation of a solar mass binary should be $a_0\approx 10^{-2}$ AU. However, binaries should have  $a_0 \gtrsim 1$ AU to survive the giant phase of the component stars without resulting in a stellar merger. A mechanism to bring the two objects close enough to form the SBHBs observed by LIGO is therefore needed. 
Two formation channels consistent with the first gravitational-wave observations of SBHB mergers have been proposed: the \emph{common evolution of field binaries} and the \emph{dynamical capture in dense environments}. We refer the reader to \cite{Barack:2018yly} and references therein for an updated review of the topic.

The \emph{common evolution} is the astrophysical scenario in which the two stars, eventually producing the SBHB, form as a stellar binary system and evolve together through the different phases of stellar evolution. In particular, four key ingredients are needed to get a general understanding of the different evolutionary phases of the system:
\begin{enumerate}
\item[1.] the gravitational potential of a binary system;
\item[2.] the mass transfer between the two stars forming the binary;
\item[3.] the physics of supernov\ae~explosions and the role of supernov\ae~kicks;
\item[4.] the formation and evolution of a common envelope.
\end{enumerate}

\subsection{The gravitational potential of a binary system}

As in section \ref{sec:kepler}, we consider two objects of masses $M_1$ and $M_2$, approximated by two point masses, separated by $a$ and moving in circular orbits about their common CoM with velocities $\vec{v}_1$ and $\vec{v}_2$. Let $\vec{r}_1$ and $\vec{r}_2$ be the distances between the centre of mass and the two objects, the angular velocity is $\omega=v_1/r_1=v_2/r_2$. Working in the co-rotating coordinate system with the centre of mass at the origin,  we consider a test mass $m$ at a distance $r_{\rm CM}$ from the CoM and distance $s_i$ from the masses $M_i$.

\begin{figure}[h!]
  \centering
    \includegraphics[width=0.6\textwidth]{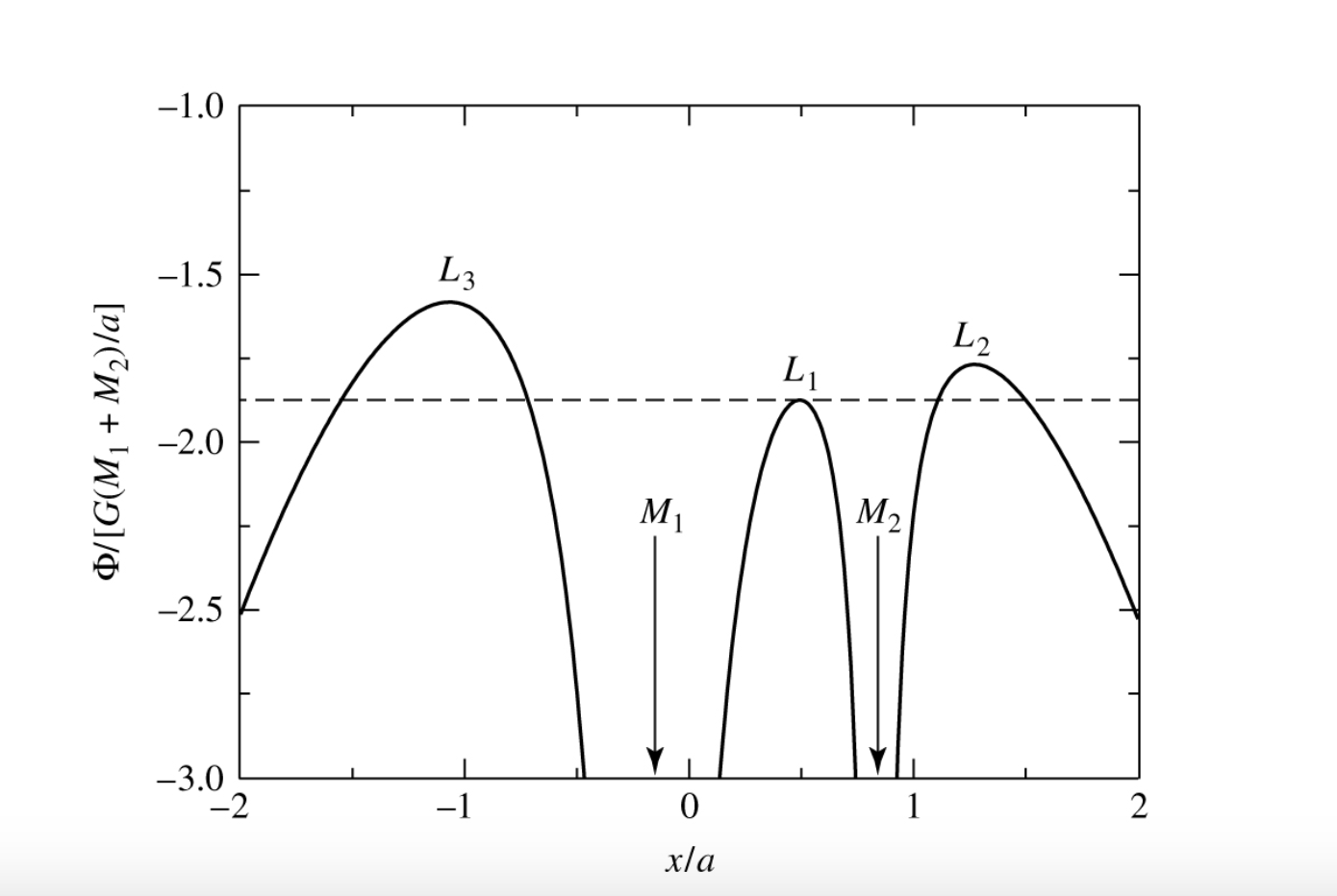}
    \caption{The effective gravitational potential $\Phi$ along the $x$-axis for two objects of masses $M_1=0.85 M_\odot$ and $M_2=0.17 M_\odot$, separated by $a=0.718 R_\odot$. Figure taken from~\cite{carroll2017introduction}. }
       \label{fig:Roche2}
\end{figure}

The energy potential $U$ of the system is 
\begin{equation}
U=-\frac{Gm(M_1+M_2)}{r_{\rm CM}}-\frac{1}{2}m\omega^2 r_{\rm CM}^2=-Gm\left(\frac{M_1}{s_1}+\frac{M_2}{s_2}\right)-\frac{1}{2}m\omega^2 r_{\rm CM}^2.
\end{equation}
Let $U \equiv m \Phi$, where $\Phi$ is the effective gravitational potential plotted in Fig.~\ref{fig:Roche2}. 

\begin{figure} [h!]
  \centering
    \includegraphics[width=0.6\textwidth]{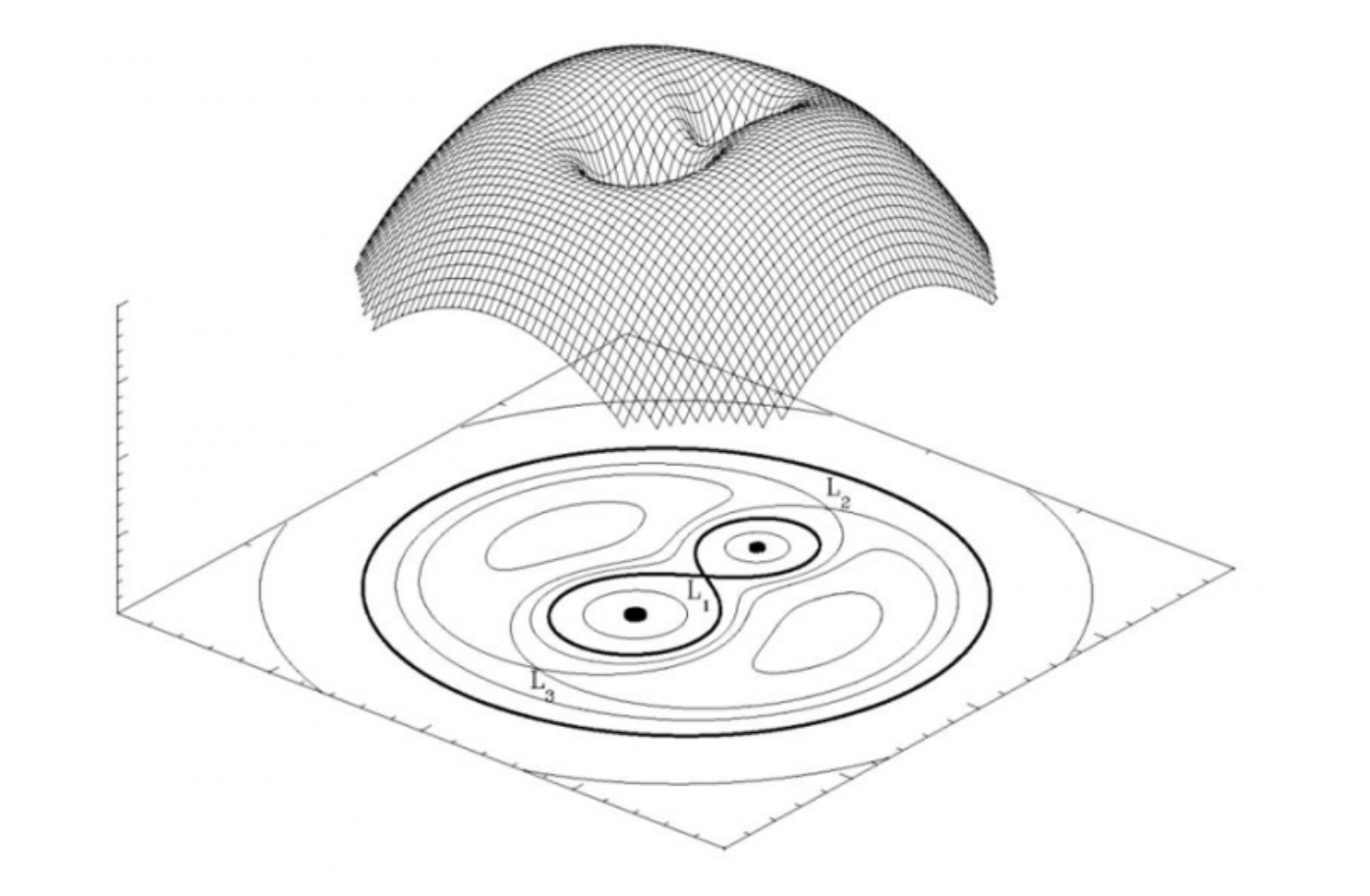}
    \caption{A three-dimensional representation of the effective potential in a binary star with a mass ratio of $2$, in the co-rotating frame. The projection at the bottom represents the shape of the equipotential surfaces, with the Roche lobes marked in thick. The three Lagrangian points $L_1, L_2, L_3$ are indicated. From \cite{Postnov2014}.}
    \label{fig:Roche}
\end{figure}

The equilibrium points (Lagrangian points) satisfy
\begin{equation}
\vec{F}=-m \vec{\nabla}\Phi=0.
\end{equation}
Close to each object, surfaces of equal gravitational potential, known as equipotential surfaces of the binary system, are approximately spherical and concentric with the nearer object. Far from the binary system, the equipotential surfaces are approximately ellipsoidal and elongated parallel to the axis joining the centres. 
There exists a critical equipotential surface, forming a two-lobed figure-of-eight, with one of the two object at the centre of each lobe and intersecting itself at the $L_1$ Lagrangian point, as shown in Fig.~\ref{fig:Roche}. This critical equipotential surface defines the \emph{Roche lobes}.
The precise shape of the Roche lobes depend on the mass ratio $q$, and must be evaluated numerically. However, an approximate formula (up to 1\% accuracy) was derived by Eggleton \cite{1983ApJ...268..368E}
\begin{equation}\label{ApproxRoche}
\frac {R_{1}}{a}=\frac {0.49q^{2/3}}{0.6q^{2/3}+\ln(1+q^{1/3})},
\end{equation}
where $R_1$ is the radius of the Roche lobe around $M_1$.

\subsection{Mass transfer}

Let us assume that one of the two objects in the binary, say $M_2$, fills its Roche lobe. This will happen at some point in the evolution of the system, when the most massive star, upon fuel exhaustion, will start expanding ascending the giant branch in the Hertzsprung-Russel diagram. Then, matter (represented by the test mass $m$) may flow to $M_1$ through $L_1$, without any need of energy exchange. This mechanism thus allows mass transfer from one body to another in the binary system.

Suppose that $M_2$ loses matter at a rate $\dot{M}_2 <0$ and let $\beta \in [0,1]$ be the fraction of the ejected matter which leaves the system, \emph{i.e.} $\dot{M}_1 = -(1-\beta)\dot{M}_2  \ge  0$. If $\beta=0$, all the mass lost by $M_1$ is captured by $M_2$ and the mass transfer is fully conservative. By keeping $\beta$ as free parameter, we are considering the more general case in which a fraction of the mass can be lost, \emph{e.g.}, by stellar winds. Differentiating the angular momentum of the system
\begin{equation}
J=\mu \sqrt{GMa}
\end{equation}
with respect to time and using $\dot{M}_1+\dot{M}_2\equiv \dot{M}=\beta \dot{M}_2$, we have
\begin{equation}
\begin{aligned}
\frac{\dot{a} }{a}&=2\frac{\dot{J}}{J}-2\frac{\dot{\mu}}{\mu}-\frac{   \beta \dot{M}_2 }{ M}
\\
&=2\frac{\dot{J}}{J}-2\left[ (\beta-1)\frac{M_{2}}{M_1}+1\right]\frac{\dot{M}_2 }{M_2}.
\end{aligned}
\end{equation}
If there is no leak of matter from the system, then $\beta=0$. and the angular momentum is conserved ($\dot{J}=0$). The total mass is also conserved ($\dot{M}=0$) and it easy to derive that
\begin{equation}
\frac{\dot{a} }{a}= -2 \left(1-\frac{M_2}{M_1} \right)\frac{\dot{M}_2}{M_2}.
\end{equation}
Therefore, the orbit expands if $M_1 > M_2$, otherwise it shrinks.
Moreover from the Kepler's third law, we have
\begin{equation}
\frac{\dot{\omega}}{\omega}=-\frac{3}{2}\frac{\dot{a}}{a},
\end{equation}
and thus the angular frequency increases as the orbit shrinks.

As a side note, non-conservative mechanisms of mass transfer with $\beta\neq 0$ exist and they are known as Jeans (fast winds) model and isotropic re-emission model. We refer the reader to \cite{Postnov2014} and references therein for a more detailed account of the topic. 


\subsection{Supernova kicks}
\label{sec:kicks}
Suppose that the two objects in the binary system are two stars separated by $a \approx 1$ AU. At the end of all the subsequent stages of nuclear burning, the most massive star can undergo a supernova (SN) explosion, expelling  much of the stellar material and leaving a compact remnant, usually a NS or a SBH. The mass loss is practically instantaneous as the typical timescale for the explosion is much shorter than the orbital period. In general, the collapse is not perfectly symmetric nor isotropic. As a result, the SN imprints a kick to the object characterised by a recoil velocity $v_{\rm kick}$. Since mass ejection decreases the total mass of the binary, also the gravitational potential changes and, if enough mass is ejected, the SN explosion can unbind the binary. Moreover,   $v_{\rm kick}$ is  typically much greater than the orbital velocity $v_{\rm kick}  \gg  v_{\rm orb}$. As a result,  kicks might destroy most of the  binaries.

To describe the effect of a SN explosion in a binary system, consider  two stars  in a circular orbit with  initial separation $a_i$ and masses $M_1$ and $M_2$.  The pre-SN relative velocity is 
\begin{equation}
v_i=\sqrt{\frac{G(M_2+M_1)}{a_i}}.
\end{equation}
After the  SN explosion of the giant star $M_1$,  the compact remnant is characterized by $M_c<M_1$, and $\Delta M = M_1-M_c$ is ejected. Right after the instantaneous explosion, the position  of the exploded star $M_1$ has not changed, but the final reduced mass of the system  is $\mu_{f}=M_cM_2/(M_c+M_2)$ and the final velocity is $\vec{v}_f=\vec{v}_i+\vec{v}_{\rm kick}$. Thus, the final energy is
\begin{equation}
E_{f}=\frac{1}{2}\mu_f v_{f}^2-\frac{GM_{c}M_2}{a_i}.
\end{equation}  
The system remains bound if the final velocity is smaller than the escape velocity
\begin{equation}
v_f \le v_{e}=\sqrt{\frac{2G(M_2+M_c)}{a_i}}.
\end{equation}
In the case of a spherically symmetric SN explosion, $\vec{v}_{\rm kick}=0$, \emph{i.e.} $\vec{v}_{i}=\vec{v}_f$, and the condition  $v_f \le v_{e}$ implies
\begin{equation}
\Delta M  \le \frac{M_1+M_2}{2}.
\end{equation}
Therefore, even in absence of kicks, a binary can be disrupted due to mass loss only, if the SN explosion ejects more than half of the initial mass of the binary system.

The case of asymmetric SN explosion is more complicated and usually characterized by $\vec{v}_{\rm kick} \approx 10^2$ km/s. NSs experience such large kicks because the collapsing outer shells of the pre-existing stellar nucleus eventually bounce against the hard surface of the new-born NS core, sustained by the strong degeneracy pressure of the neutrons. Since the process is likely asymmetric, conservation of linear momentum can result in large kicks. Specifically, from the measurement of proper motions of the NSs in the MW, typical supernovae kicks have been estimated to be of the order of  $v_{\rm kick} \approx 250$ km/s \cite{2005MNRAS.360..974H}. More precisely, the natal kick distribution is typically modelled by a Maxwellian probability distribution with velocity dispersion $\sigma_v = 190$ km/s  \cite{Hansen} 
\begin{equation}\label{maxwellian_kick}
f(v_{\rm kick})=\sqrt{\frac{1}{\pi}}\frac{v_{\rm kick}^2}{\sigma_v^3}e^{-\frac{v_{\rm kick}^2}{2\sigma_v^2}}\, .
\end{equation}
Note that the typical circular velocity of a  binary of massive stars with $a_i \approx 1$ AU is $v_i \approx 10-100$ km/s.
 Therefore, $v_i  \lesssim  v_{\rm kick}$ and  kicks are expected to destroy the vast majority of the systems that would otherwise result in NS binaries.  This is one of the main problem in the formation of NS binaries emitting GWs. 
 
 However, we mention the possibility of a bimodal or dichotomous   distribution of neutron star kick speeds, proposed in order to solve  the retention problem in globular clusters.
 Since the escape velocity for the most of the globular clusters in the Milky Way is under $50$ km/s, we would expect less than $6$\%  of all NSs  to survive and  remain in a globular cluster. 
 On the other hand,  observational and theoretical evidences indicate that some of the massive globular clusters in our Galaxy contain $\sim 1000$ NSs, much more than we would expect from the Maxwellian probability distribution given by equation \eqref{maxwellian_kick}.
 A possible solution to this discrepancy, proposed in \cite{Podsiadlowski}, is to model the natal kick distribution as the superposition of two Maxwellian functions, one peaked at lower velocities, the other peaked at higher ones. The physical reason for this is sought in different mechanisms of SN explosion and NS formation.


Finally, for SBHs the situation is less clear and the problem of binary disruption due to SN kicks might be less severe. In this case, there is no hard surface to bounce onto and, as a consequence, the kicks might be smaller -- $v_{\rm kick}\approx 50$ km/s -- with respect to the NS case. However, the situation is controversial.
According to   \cite{Repetto},  the velocity distribution of black hole natal kicks seems similar to that of neutron-star kick velocities, probably as a consequence  of the large-scale asymmetries created in the supernova ejecta by the explosion mechanism (see \cite{Janka:2013hfa} for details). Moreover, observations from binary black hole GW151226 indicate a non-zero spin for the most massive black hole, misaligned from the binary's orbital angular momentum.
If the black holes were formed through isolated binary evolution from a binary star, all angular momenta would be initially parallel (because of tidal interactions within the binary).
Then, the most likely processes that can misalign their spin angular momenta are the linear momentum recoils imparted by  SN kicks.
In this case, kinematic arguments can be used to constrain the characteristic magnitude of the natal kick \cite{OShaughnessy:2017eks} to be    $v_{\rm kick} \gtrsim 50$ km/s in order to be consistent  with the  misalignment measured in  GW151226, if no processes act to realign stellar spins. Significantly larger natal kicks, with one-dimensional velocity dispersion $\simeq 200$ km/s, are required if stellar spins efficiently realign prior to the second BH's birth \cite{Wysocki:2017isg}.

\subsection{Common envelope}
Giant stars are composed of a core and an envelope, thus $M_{\rm gs}=M_{\rm core}+M_{\rm env}$. In the core, hydrogen has been converted into helium and nuclear reactions stop, causing the core to shrink under the action of gravity. Hence, the helium core and the hydrogen envelope are well separated objects. When the giant star overfills its Roche lobe, mass transfer is allowed, as we have seen before. Depending on the mass of the two stars, the orbit shrinks, $\dot{a}<0$, causing even more material to overflow the Roche lobe. This eventually leads to the run-away process of dynamically unstable mass transfer. It is therefore possible that the mass transfer rate from the mass-losing star is so high that the SBH (or NS, but we focus on SBHBs here) cannot accommodate all the accreting matter. In this situation, the envelope continues to expand engulfing the companion SBH, leading to the formation of a \emph{common envelope}. The common envelope can extract energy from the orbit of the binary system, formed by the SBH and the core of the massive star, via dynamical friction, eventually unbinding itself from the system.

A proper understanding of the details of common envelope formation and evolution has to rely on 3D hydrodynamic simulations, including nuclear reactions. At present, we do not have a  clear solution to the problem, although important advances have been made recently \cite{2013A&ARv..21...59I}. However, simple energy balance arguments can provide a general understanding of the problem. The initial binding energy of the stellar envelope is given by
\begin{equation}
E_{\rm{env}, i}=-\frac{GM_{\rm gs} M_{\rm env}}{\lambda R_L},
\end{equation}
where $R_L \approx 1$ AU is the Roche lobe radius of the star, that can be approximated by Eq.~\eqref{ApproxRoche}, while $\lambda$ is the concentration parameter, that depends on the density profile of the envelope, which is in general denser closer to the core. Usually the density of the envelope is described as a power law $\rho(r)\propto r^{-\gamma}$, with $\gamma>0$; the bigger is $\gamma$, the more concentrated is the envelope and the smaller is $\lambda$.
If we assume that, at the end of the common envelope stage, the envelope unbinds formally reaching infinity with zero velocity, that is $E_{\rm env, f}=0$, we have 
\begin{equation}
\Delta E_{\rm env}=-E_{\rm env, i}=\frac{GM_{\rm gs} M_{\rm env}}{\lambda R_L}.
\end{equation}
Note that  the more concentrated the envelope is, \emph{i.e.}, the smaller is $\lambda$, the more binding energy is possible to extract from it. As a consequence, the separation between the SBH and the core of the giant star decreases. By computing the variation of the binary orbital energy, we obtain
\begin{equation}
\Delta E_{\rm orb}=E_{\rm orb,f}-E_{\rm orb,i}=\alpha_{\rm ce} \left[-\frac{GM_{\rm core}M_{\rm BH}}{2a_f}-\left(-\frac{G(M_{\rm core}+M_{\rm env})M_{\rm BH}}{2a_i}\right)\right],
\end{equation}
where $\alpha_{\rm ce}$ is the common envelope parameter which describes the efficiency of expenditure of orbital energy on expulsion of the envelope. Since 
\begin{equation}
\Delta E_{\rm orb}+\Delta E_{\rm env}=0,
\end{equation}
after using $M_{\rm gs}=M_{\rm core}+M_{\rm env}$, we get
\begin{equation}
\alpha_{\rm ce} \left(\frac{GM_{\rm core}M_{\rm BH}}{2a_f}-\frac{GM_{\rm gs}M_{\rm BH}}{2a_i}\right)=\frac{GM_{\rm gs} M_{\rm env}}{\lambda R_L}.
\end{equation}
Therefore, the ratio between the initial and final semi-major axes is given by
\begin{equation}
\frac{a_f}{a_i}=\frac{M_{\rm core}}{M_{\rm gs}}\left( 1+\frac{2}{\lambda \alpha_{\rm ce}} \frac{a_{i}}{R_{L}}\frac{M_{\rm env}}{M_{\rm BH}}\right)^{-1}=\frac{M_{\rm core}}{M_{\rm gs}}\left( \frac{M_{\rm BH}}{M_{\rm BH}+\frac{2 M_{\rm env}}{\lambda \alpha_{\rm ce}}\frac{a_i}{R_L}}\right).
\end{equation}
The evaluation of $\lambda$ (the measure of the binding energy of the envelope to the core prior to the mass transfer in a binary system) and $\alpha_{\rm ce}$ (the common envelope efficiency) suffer from large physical uncertainties. Nevertheless, it is possible to estimate the product $\alpha_{\rm ce}\lambda$ by modelling specific systems or well-defined samples of objects corrected for observational selection effects. In general, the envelope is quite concentrated and typical values are of the order  $\alpha_{\rm ce}\lambda \ll 1$. A crude approximation is
\begin{equation}
\frac{M_{\rm core}}{M_{\rm gs}}\approx 0.2-0.3, \qquad \frac{M_{\rm BH}}{M_{\rm BH}+\frac{2 M_{\rm env}}{\lambda \alpha_{\rm ce}}\frac{a_i}{R_L}} \approx  0.05 - 0.01.
\end{equation}
Thus, the order-of-magnitude estimation is
\begin{equation}
\frac{a_f}{a_i}\sim 10^{-3}-10^{-2}.
\end{equation}
This process takes about $10^2-10^3$ years and upon its completion the system has shrunk to $a_f \approx R_\odot$, close enough to merge in a Hubble time due to GW emission.

 \subsubsection*{Summary}
 For the sake of clarity, we list the main steps in the formation of binaries by common evolution. The following is just a general evolutionary scheme, several variations of it exist depending on the exact masses of the two objects, initial separations, and detailed assumptions of mass transfer and common envelope. As we gather more information from GW observations, we will be able to effectively constrain the relevant physical processes.
 \begin{enumerate}
 \item Two massive stars with masses $M \approx 50 M_\odot$ are initially bound in a wide binary system with semi-major axis $a_i \approx 1$ AU.
 \item The more massive of them becomes a giant star, fills its Roche lobe and transfers mass to the other star, changing the mass ratio and the separation of the binary.
 \item The most massive star undergoes a SN explosion. The compact remnant (\emph{e.g.}, a BH) gets the SN kick which might destroy the binary.
 \item If the binary survives, the other star also evolves and becomes a giant star, fills its Roche lobe and  transfers mass to the BH. The mass transfer becomes unstable and a common envelope can eventually form.
 \item The common envelope unbinds and the system shrinks by a factor $10^2-10^3$. The final separation between the BH and the helium core is  $a_f\sim R_\odot$. 
 \item At this point the second star undergoes SN explosion getting another kick. However, in this case $a_f\sim R_\odot$, the circular velocity is 
 \begin{equation}
 v_c \approx \sqrt{\frac{GM}{a_f}} \sim 10^3 \ \text{km/s},
 \end{equation}
 so the chances of disrupting the binary with the second SN kick are small (probably most binaries survive the second SN kick).
 \item Finally, we are left with a SBHB with separation $a_f\sim R_\odot$ allowing merger in a Hubble time (formation of BH-NS and NS-NS binaries proceeds through similar steps).
 \end{enumerate}

Note that, before the first GW detections, the only way to probe SBHs was via electromagnetic observations of accreting binaries. During the evolutionary phases described above, the SBH can accrete mass from the stellar companion, thus shining as an X-ray source. We observe SBHs in binaries where a massive companion emits strong stellar winds that are partly captured and accreted by the SBH. Those systems are labelled High Mass X-Ray Binaries (HMXRB), and we observe plenty of them in the MW. Similarly, SBHs can accrete material from degenerate helium cores in several Low Mass X-Ray Binaries (LMXRB) and we also observe plenty of them in our galaxy. Although the LMXRB we can observe in the MW are not BHB progenitors (the stellar companion is too light to become a SBH), the observation of such close interacting binaries support the evolutionary path described above for massive binaries.

Before closing, we note that an alternative common evolution channel has been identified in {\it chemically homogeneous systems} \cite{2016MNRAS.458.2634M,2016A&A...588A..50M}. Without entering in details, in massive contact binaries, tides can induce high spins. The fast rotation mixes the interior of the stars, preventing a clear separation between a core and an envelope. As a result, the stars do not expand to become giants at the end of their life on the main sequence and they can directly evolve to form a SBHB that has already a separation of few $R_\odot$, without the need of any common envelope phase.

\subsection{A back of the envelope estimate of BHB merger rate from common evolution}
A simple estimate of the expected SBHB merger rate in the universe from the field channel can be easily obtained by combining general arguments about the stellar content of the Universe.

The average today's stellar mass density in the Universe is $\rho_*\approx 3\times 10^8~M_\odot$ Mpc$^{-3}$, which means $3\times 10^8$ stars, assuming an average of 1\,$M_\odot$ per star. For a Salpeter initial mass function (IMF), only about 0.3\% of these stars have $M>30\,M_\odot$, thus leaving behind a SBH as relic. Since 70\% of the massive stars are observed to be in binaries, and we need two stars to form a binary system, we can estimate about $3\times 10^5$ massive binary stars (\emph{i.e.}, SBHB progenitors) per Mpc$^{-3}$.

We can make the extreme assumption that those binaries are produced in a continuous, steady state star formation process over about 10 Gyr, thus resulting in a formation rate of  $3\times 10^{-5}$ yr$^{-1}$ Mpc$^{-3}$. By assuming that all those massive binaries give rise to SBHBs that merge in a short time-scale, then we get a SBHB merger rate of  $3\times 10^{4}$ yr$^{-1}$ Gpc$^{-3}$, which is about two orders of magnitude higher than the upper end of the 90\% confidence interval estimated by LIGO.

Our estimate is a sort of an ``upper limit'' and two-three extra orders of magnitude can be accommodated by playing around with the physical processes outlined above. For example, it is very likely that the vast majority of these binaries are disrupted by the first SN explosion, which can easily decrease the rate by more than an order of magnitude to, say $10^3$ yr$^{-1}$ Gpc$^{-3}$. Moreover, the details of common envelope are poorly understood. If the process is too efficient, then the first SBH can merge with the core of the secondary star before it undergoes SN, thus preventing the formation of a SBHB. Therefore, it is reasonable that only a fraction of systems surviving the first SN explosion actually produce a SBHB. Finally, we assumed a constant star formation rate across cosmic time. It is known that the cosmic star formation rate peaks at $z\approx 1.5$, being about an order of magnitude higher than today. LIGO, so far, measured the local SBHB merger rate, so even without invoking SN disruption and failed common envelope evolution, it is likely that the number provided above is an overestimate of the local rate by a factor of a few.

In general, population  synthesis models are able to match current observations, perhaps being on the high side of the measured rates; see, \emph{e.g.}, \cite{2015ApJ...806..263D,2018MNRAS.479.4391M}.

\section{Dynamical processes of SBHB formation}
Another channel to form SBHBs is via dynamical processes. This channel encompasses a number of physical mechanisms, including dynamical capture, three body hardening, hierarchical triples, Kozai-Lidov oscillations, \emph{etc}. 
This scenario is closely related to the fact that most stars are observed to form in clusters and associations. However, the vast majority of the stars we observe today in the MW are field stars, \emph{i.e.} they do not belong to stellar associations (groups of more than $10^3$ stars, like globular clusters, young massive star clusters and open clusters \cite{2007ARA&A..45..565M}). In order to understand the dynamical formation scenario, it is therefore important to start from the physics of star cluster formation and evolution.

This Section makes use of several concepts and derivations performed in the course {\it Lecture on collisional dynamics} by Prof. Michela Mapelli\footnote{The notes are collected in five pdf presentations available online at\\ \url{http//web.pd.astro.it/mapelli/2014colldyn1.pdf}\\
  \url{http//web.pd.astro.it/mapelli/2014colldyn2.pdf}\\
\url{http//web.pd.astro.it/mapelli/2014colldyn3.pdf}\\
 \url{http//web.pd.astro.it/mapelli/2014colldyn4.pdf}\\
\url{http//web.pd.astro.it/mapelli/2014colldyn5.pdf}},   {\it Structure and Evolution of Stars} by Prof. Max Pettini
\footnote{The notes are available at \\ \url{https://www.ast.cam.ac.uk/~pettini/STARS/}} and {\it Stellar Dynamics and Structure of Galaxies} by Dr.  Vasily Belokurov 
\footnote{The notes are available at \\
\url{https://www.ast.cam.ac.uk/~vasily/Lectures/SDSG/}}.
\subsection{Star cluster formation}
Stars in cluster form from the fragmentation of a giant molecular cloud (GMC) (sometimes called stellar nursery), \emph{i.e.}, a type of interstellar cloud, the density and size of which permit the formation of molecules, most commonly molecular hydrogen, H$_2$. From observations, we know that in the Milky Way there  are GMCs of mass $ \approx 10^6 M_\odot$, with temperature of $T \approx 10-100 $ K and radius $\approx 10$ pc. However, in gas rich star forming galaxies, GMCs can have masses up to $ \approx 10^8-10^9 M_\odot$ \cite{2012MNRAS.427..688L}.

\subsubsection{Jeans mass}
Consider a cloud of gas and denote with $K$ and $U$ the kinetic energy (related to the gas pressure) and potential (gravitational) energy, respectively. The virial theorem states that, in order to maintain  hydrostatic equilibrium, the gravitational potential energy must equal twice the internal thermal energy:
\begin{equation}
-2\left<K\right> =\left<U\right>,
\end{equation}
where the brackets denote time averages. However, if the gas pressure is insufficient to sustain gravity, the cloud will  collapse. The condition for gravitational collapse is
\begin{equation}\label{gravinstability}
2K < \left|U\right|.
\end{equation}
Assuming spherical symmetry, the gravitational force acting on a  thin shell of thickness $dr$ and mass $dm$ at a distance $r$ from the centre of the cloud is
 \begin{equation}
 dF_{g}=\frac{GM(r)dm}{r^2},
 \end{equation}
 where $M(r)$ is the mass contained within $r$.  The corresponding gravitational potential energy of the test mass is:
 \begin{equation}
 dU_{g}=-\frac{GM(r)dm}{r}=-GM(r) 4\pi r\rho dr,
 \end{equation}
where we have considered $dm=4\pi r^2 \rho dr$. 
As an approximation,  we assume  that the cloud is a sphere of constant density 
\begin{equation}
M(r) = \frac{4\pi}{3} r^3\rho_0,
\end{equation}
and the mass of a GMC of radius $R$ is $M = (4/3)\pi R^3\rho_0$. With this approximation, we can integrate the gravitational potential energy to obtain
\begin{equation} \label{U}
U_g=-4\pi G \int_0^RM(r)\rho r dr = -\frac{16\pi^2}{15}G\rho_0^2R^5=-\frac{3}{5}\frac{GM^2}{R}.
\end{equation}     
The total internal kinetic energy of the cloud is 
\begin{equation} \label{K}
K = \frac{3}{2}Nk_BT ,
\end{equation}
where $N$ is the total number of particles. If we define the mean molecular weight
\begin{equation}
\mu\equiv \frac{\langle{}m\rangle{}}{m_H},
\end{equation}
where $\langle{}m\rangle{}$ is  the average mass of the particles (atoms, ions, or molecules) in the gas and $m_H$ the mass of an hydrogen atom, we can write $N$ as
\begin{equation}
N=\frac{M}{\mu m_H}.
\end{equation}
The condition given in Eq.~\eqref{gravinstability} can be rewritten as
\begin{equation}\label{gravinstability2}
\frac{3Mk_BT}{\mu m_H}<\frac{3}{5}\frac{GM^2}{R}.
\end{equation}
Solving for $M$ and substituting the radius of a spherical and constant distribution of matter, we obtain
\begin{equation}
\begin{split}
  M>M_J &=  \left[\frac{375}{4\pi} \left(\frac{K_B}{G m_{H}}\right)^3 \frac{T^3}{\mu^3 \rho_0} \right]^{1/2}
  \\
  & \approx 3 \times 10^{3} M_\odot  \left[\frac{1}{\mu^4}\left(\frac{T}{100 \mbox{K}}\right)^3 \left(\frac{10^{3}~\mbox{cm}^{-3}}{n} \right)\right]^{1/2},
  \label{eq:mj}
  \end{split}
\end{equation}
where we used $\rho_0 = n \langle{}m\rangle{} = n \mu m_{H}$.
If the mass of the cloud exceeds the Jeans mass, the cloud will be unstable against gravitational collapse.
Assuming the cloud to consist of atomic neutral hydrogen ($\mu=1$) with a typical value of temperature $T \approx 100 $ K and a number of hydrogen atoms per volume of $n_{H} \approx 10\, \text{cm}^{-3}$, then $M_{J} \approx 3\times 10^{4} M_\odot$.

\paragraph{Free-fall timescale.}
An equivalent way to calculate the Jeans mass is by comparing two relevant timescales: the free-fall timescale $t_{\rm ff}$ -- \emph{i.e.}, the collapsing time of the cloud -- and the time for transferring information within the cloud $t_{\rm sound}$. The starting point of the calculation is the  equation of motion for a point particle under the influence of the gravitational field of $M(r)$:
\begin{equation}
\ddot{r}=-\frac{GM(r)}{r^2},
\end{equation}
where as usual $M(r)$ denotes the mass enclosed within radius $r$. By using the hypothesis that the mass distribution is spherical and constant, \emph{i.e.}, $M(r)=(4/3)\pi r_0^3\rho_0$, the equation of motion becomes
\begin{equation}
\ddot{r}=-\left(\frac{4\pi}{3}G\rho_0r_0^3\right)\frac{1}{r^2}.
\end{equation}
By multiplying both sides by $\dot{r}$ and integrating with respect to time, we get
\begin{equation}\label{freefall1}
\dot{r}=- \left[\frac{8\pi}{3}G\rho_0 r_0^2\left(\frac{r_0}{r}-1\right)\right]^{1/2},
\end{equation}
where we have chosen the negative root because the cloud is collapsing. The above differential equation might be written as
\begin{equation}\label{freefall2}
\frac{\dot{r}/r_{0}}{{\sqrt{r_0/r-1}}}=-\sqrt{\frac{8\pi}{3}G\rho_0}.
\end{equation}
Since $r/r_0 \in [0,1]$, it is more convenient to parametrise the ratio by $r/r_0 = \cos^2 \xi$, with $\xi \in [0, \pi/2]$. Hence, Eq.\eqref{freefall2} reads as
\begin{equation}
\dot{\xi}  \cos^2 \xi = \sqrt{\frac{2\pi}{3}G\rho_0},
\end{equation}
which can be integrated with respect to time to give
\begin{equation}
\frac{\xi}{2}+\frac{1}{4}\sin(2\xi)=\sqrt{\frac{2\pi}{3}G\rho_0}~(t-t_{0}).
\end{equation}
Here $t_0 = t_{r =r_0}$ corresponds to $\xi_{0} = 0$.
By definition, the  free-fall timescale is the time taken by a cloud in free-fall to collapse from the initial radius $r = r_0$ to $r = 0$, \emph{i.e.}
\begin{equation}
  t_{\rm ff} \equiv t_{r=0}-t_{r=r_0}=\sqrt{\frac{3\pi}{32}\frac{1}{G\rho_0}} \approx (1.63~\mbox{Myr}) \left(\frac{10^{3}{\mbox{cm}}^{{-3}}}{\mu n}\right)^{{1/2}},
  \label{eq:tff}
\end{equation}
where, in the approximation step, we used the relation $\rho_{0} = n \langle{}m\rangle{} = n \mu m_{H} $. The free-fall timescale depends only on the initial density $\rho_0$.
In other words, in a spherical molecular cloud of uniform density  all parts of the cloud will take the same time to collapse and the density will increase at the same rate everywhere within the cloud. 

Now, in order to derive the Jeans mass, we need to compare the free-fall timescale $t_{\rm ff}$ with the time for transferring information within the cloud  $t_{\rm sound}$. Suppose that the gas is slightly compressed. Then, the time needed for the sound waves to re-establish the system in pressure balance is
\begin{equation}
  t_{\rm sound}={\frac  {R}{c_{s}}} \approx (0.5 {~\mbox{Myr}})\left({\frac  {R}{0.1{\mbox{ pc}}}}\right)\left(\frac{0.2 ~\mbox{km s$^{-1}$}}{c_{s}}\right),
  \label{eq:tsound}
\end{equation}
where $1$ pc $\approx 3.08 \times 10^{18}$ cm.
If $t_{\rm ff}>t_{\rm sound}$, the system returns to a stable equilibrium as the pressure forces can overcome gravity. Conversely, for $t_{\rm ff}<t_{\rm sound}$, gravitational collapse takes place. It is easy to show that equating expressions (\ref{eq:tff}) and (\ref{eq:tsound}) and solving for $M$, yields the same condition of Eq.~(\ref{eq:mj}).

The Jeans length $\lambda _{J}$ is defined by equating twice the total internal kinetic energy \eqref{K} and the gravitational potential energy \eqref{U} and solving for the radius of the cloud. This yields 
\begin{equation}
\lambda _{J} = \sqrt{\frac{15}{4\pi} \frac{K_B}{G m^2_H} \frac{T}{\mu^2 n}} = \sqrt{\frac{15}{4\pi} \frac{c_s^2}{\gamma G \rho_0}},
\end{equation}
where $c^2_s = \gamma K_B T/ \langle{}m\rangle{}$ is the sound speed for an ideal gas with adiabatic index $\gamma$. The Jeans mass  $M_{J}$ is the mass contained in a sphere of radius  $\lambda_{J}$.

\subsubsection{Cloud fragmentation}
\label{sec:fragment}
The Jeans instability  plays a very important role in star formation, because it is responsible for the fragmentation of the molecular cloud.
In the previous sections, we have shown that the Jeans mass depends on the temperature and the density (see Eq.~\eqref{eq:mj})
\begin{equation}\label{JeansMass}
  M_{J} \propto \left(\frac{T^3}{\rho_0}\right)^{1/2}.
\end{equation}
Suppose that  the cloud is big and massive enough that potential energy overcomes internal energy and the cloud starts to collapse. During the collapse, $R$ decreases and $T$ increases. The process can be either \emph{isothermal} or \emph{adiabatic}, depending on whether the gas can efficiently cool or not, which is ultimately related to the metallicity content of the gas as we now describe.

GMCs and protostellar clouds are mostly composed by hydrogen (H), helium (He) and other heavier atomic elements, which are commonly referred to as ``metals''. We call \emph{metallicity}, $Z$, the mass fraction of the cloud that is neither hydrogen nor helium. As a reference, the solar metallicity is $Z_{\odot} =1.34 \times 10^{-2}$, \emph{i.e.}, only a mere 1.34\% of the Sun's mass is composed by metals. Whether a cloud can cool efficiently or not depends on its metallicity. This is because at typical GMCs temperatures ($T\approx 100$ K), collisions in the gas cannot excite the atomic levels of H and He ($T> 10^3$ K is needed). Heavier elements (\emph{i.e.}, metals), however, have much smaller energy gaps between atomic states. Therefore, also at low temperatures, collisions in the gas excite atomic/molecular levels that return to the ground state by emitting photons that can escape, cooling the cloud. Such efficient cooling is possible if $Z>Z_{\rm crit}$, with $10^{-4}\lesssim Z_{\rm crit}/Z_\odot \lesssim 10^{-3} $.


\paragraph{Adiabatic collapse.}
If the metallicity  of the cloud is $Z<Z_{\rm crit}$ the cloud cannot cool efficiently and the collapse is approximately {\it adiabatic}. This means that
\begin{equation}
PV^\gamma= const,
\end{equation}
being $P$ the gas pressure, $V$ the volume of the cloud and $\gamma$ the adiabatic index of the gas. It follows that 
\begin{equation}
P = K\rho^\gamma,
\end{equation}
where  $K$ is a constant. From the equation of state for an ideal gas
\begin{equation}
P = \frac{k_B }{\mu m_H}\rho T,
\end{equation}
we have the relation between the temperature and density:
\begin{equation}\label{Trho}
T \propto \rho^{\gamma-1}.
\end{equation}
Substituting Eq.~(\ref{Trho}) in Eq.~(\ref{JeansMass}) we get
\begin{equation}
M_J \propto  \rho ^{(3\gamma-4)/2}.
\end{equation}
For atomic hydrogen $\gamma = 5/3$, and we obtain $M_J \propto \rho^{1/2}$, so that the Jeans mass increases as the density increases (\emph{i.e.}, when the cloud is shrinking). As a consequence,  the cloud cannot fragment and the collapse is approximately {\it monolithic}. In other words, a cloud  with $Z<Z_{\rm crit}$ contracts adiabatically without fragmenting, thus potentially forming a single massive protostar of the order of $10^3 M_\odot$. The reason for the upper limit of $10^3 M_\odot$ comes from the fact that, when the cloud collapses adiabatically, the temperature increases up to $T \approx 10^2$ K, corresponding to $M_J \approx 10^3 M_\odot$. At this point, molecular hydrogen transitions start to cool the cloud, and the Jeans mass does not increase anymore.

Adiabatic collapse is advocated as one of the main formation channels of IMBHs. Remarkably, early on in structure formation, at redshift $20<z<15$, the gas metallicity was very low\footnote{The metallicity of primordial gas is $Z\approx 10^{-4}Z_\odot$. Metals are mostly synthesized in stellar cores and the average metallicity content of the universe increases with time.}, so the first protostellar clouds likely collapsed with little fragmentation, leaving behind a first generation of stars (known as population III or popIII stars) that were very massive and could potentially evolve into IMBHs characterized by mass of the order of $ 10^2-10^3 M_\odot$ \cite{2001ApJ...551L..27M}. Some of those 'seeds' could have later grown through gas accretion and mergers with other black holes to become the SMBHs we see in galactic nuclei today \cite{2003ApJ...582..559V}.

At later stages, supernovae explosions progressively enriched the interstellar medium with metals, so that subsequent star formation episodes would occur at $Z>Z_{\rm crit}$, thus leading to significant fragmentation. For example, the protostars that form today are characterized by masses of the order of $0.2-0.5 M_\odot$, as we now discuss.

\paragraph{Isothermal collapse.}
For $Z>Z_{\rm crit}$ collapse is essentially {\it isothermal}. Temperature remains approximately constant and Eq.~(\ref{JeansMass}) implies that the mass limit for instability decreases when the density of the cloud increases as $M_j\propto \rho^{-1/2}$. As a consequence, any initial density inhomogeneities will cause individual regions within the cloud to cross the instability threshold independently and collapse locally, forming  a large number of smaller objects: isothermal collapse naturally leads to fragmentation. Those clumps eventually turn into protostars, and this is essentially the reason why stars are mostly formed in stellar associations. The typical time scale for star formation is $t\sim 10^5$ years.

It is however obvious that stars cannot directly form as a result of isothermal collapse, if anything simply because temperatures around $10^7$ K are needed to turn on nuclear reactions. Protostellar clumps have to heat up at some point. This happens when the clump density becomes high enough that gas becomes opaque  to infrared photons (corresponding to the typical energy range of the photons emitted by cooling metals). At this point, the cloud cannot cool efficiently, the isothermal approximation breaks down and further evolution of the clump is approximately adiabatic. As a consequence, temperature increases and so does $M_j$, as we just saw, thus leading to a minimum fragment size into which the cloud can break up. This is know as the opacity limit. It is the lower mass limit of the fragmentation process, and we now compute its value.


The starting point is the energy released during the collapse of the protostellar cloud. From the virial theorem
\begin{equation}
-2\left<K\right> =\left<U\right>,
\end{equation}
we have
\begin{equation}
\left< E\right>=\left<K\right> +\left< U\right>=\frac{\left< U\right>}{2},
\end{equation}
\emph{i.e.}, only half of the change in gravitational potential energy is available to be radiated away as the protostar collapses.
Since
\begin{equation}
U=-\frac{3}{5}\frac{GM^2}{R},
\end{equation}
the energy released during the collapse is
\begin{equation}
\Delta E_g = \frac{3}{10}\frac{GM^2}{R}.
\end{equation}
Using $\rho_0 =3M/(4\pi R^3)$, the emitted average luminosity over the free-fall time is therefore
\begin{equation}
L_{\rm ff}=\frac{\Delta E_g}{t_{\rm ff}} = \frac{3}{10}\frac{GM^2}{R}\left(\frac{3\pi}{32}\frac{1}{G\rho_0}\right)^{-1/2} = \frac{3\sqrt{2}}{5\pi} G^{3/2}\left( \frac{M}{R}\right)^{5/2},
\end{equation}
This is the power that, in the opacity limit, is absorbed by the gas. In the following, we assume that the gas emits radiation as a grey-body, thus the cloud can radiate away energy at a rate given by
\begin{equation}
L_{\rm rad} = 4\pi R^2e_f\sigma T^4,
\end{equation}
where we have introduced the efficiency factor $e_f$ ($0 < e_f < 1$), because the collapsing cloud is not in thermodynamic equilibrium. The gas will start to heat up (becoming adiabatic) as soon as the cooling becomes slower than the heating rate, \emph{i.e.}, when
\begin{equation}
L_{\rm ff} \gtrsim L_{\rm rad}.
\end{equation}
In other words, the  isothermal approximation breaks down when $L_{\rm ff} = L_{\rm rad}$. This condition amounts to
\begin{equation}
M^{5} \lesssim M^5_{\rm crit}=\frac{200}{9}\frac{\pi^4 \sigma^2}{G^3}e_f^2 R^{9} T^8.
\end{equation}
Since the mass must be bigger than the Jeans mass $M_J$ to collapse gravitationally, we have the constraint $M_{J} \leq M \leq M_{c\rm rit}$. The opacity limit is reached when $M_J = M_{\rm crit}$. This happens when the clumps reach the mass limit
\begin{equation}
M_{\rm frag} = 0.01 \left(\frac{T/\mbox{K}}{\mu ^{9}~e_f^{2}}\right)^{1/4}M_\odot.
\end{equation}
Typical values of $\mu = 1$, $e_f = 0.1$, and $T = 1000$ K lead to $M_{\rm frag} \approx  0.2 M_\odot$. Hence, fragmentation ceases when individual fragments are approximately solar-mass objects. See  Lecture 11 of the series {\it Structure and Evolution of Stars} by Prof. Max Pettini
\footnote{\url{https://www.ast.cam.ac.uk/~pettini/STARS/Lecture11.pdf}} for details. 

\subsection{Star cluster evolution}


We now discuss the main physical processes at play in the evolution of a star cluster. These processes determine whether the cluster is going to evaporate or to form a segregated core of massive objects that can be the nurse of SBHBs of the type of those observed by LIGO/Virgo. 
We refer to \cite{1987Spitzer, 1987ElsonHutInagaki, binney2011galactic} for more details.

\subsubsection{Two-body relaxation timescale}\label{subsec:Relaxation_timescale}
Stars in clusters can reach equilibrium through mutual interactions. This process is called \emph{two-body relaxation} and is analogous to \emph{thermalization}. Therefore, a very important timescale in collisional dynamics in clusters is the two-body relaxation time, \emph{i.e.}, the time for a star to completely lose memory of its initial velocity, by means of gravitational encounters.
\begin{figure}[h!]
  \centering
    \includegraphics[width=0.6\textwidth]{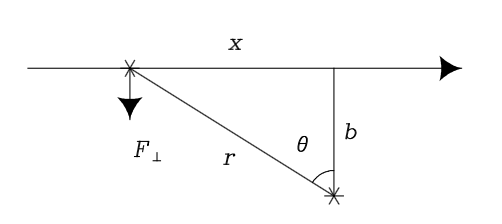}
    \caption{A field star approaches the subject star at speed $v$ and with impact parameter $b$. From \cite{binney2011galactic}. }
       \label{fig:Relaxation}
\end{figure}
Following \cite{binney2011galactic}, we consider an idealized cluster of size $R$, consisting of $N$ identical stars with mass $m$ uniformly distributed.
In the cluster, we focus on a single star that passes close to a field star at relative velocity $v$ and impact parameter $b$ as drawn in Fig.~\ref{fig:Relaxation}. The gravitational force is
\begin{equation}
\vec{F}=-\frac{Gm^2}{r^3}\vec{r}=\vec{F}_{\parallel} +\vec{F}_{\perp},
\end{equation}
where $F_{\perp}=F\cos\theta$, or explicitly, by using the Pythagoras theorem
\begin{equation}
  F_{\perp}=-\frac{Gm^2}{r^2}\frac{b}{r}=-G\frac{m^2}{b^2}\left[1+\left(\frac{vt}{b}\right)^2\right]^{-3/2}=-m\dot{v}_\perp.
  \label{eq:fperprelax}
\end{equation}
The change in velocity integrated over one entire encounter is
\begin{equation}
\delta v_\perp =\int_{-\infty }^{\infty}\dot{v}_\perp dt=\frac{Gm}{b^2}\int_{-\infty }^{\infty}\left[1+\left(\frac{vt}{b}\right)^2\right]^{-3/2}dt.
\end{equation}
Note that, for symmetry reasons, $\delta v_\parallel=0$. Defining $s=vt/b$, we have 
\begin{equation}
  \delta v_\perp =\frac{Gm}{bv}\int_{-\infty }^{\infty}\frac{ds}{\left(1+s^2\right)^{3/2}}=\frac{2Gm}{bv}.
  \label{eq:vperp}
\end{equation}
Taking into  account all stars in the system, the surface density of stars in an idealized cluster is ${N}/{\pi R^2}$ and the number of interactions per unit element is
\begin{equation}
\delta n= \frac{N}{\pi R^2} d(\pi b^2)= \frac{2 b N}{ R^2} db.
\end{equation}
Defining $\delta v^2_{\rm tot}=\int \delta v_\perp ^2 \delta n$, where the integral is performed over all possible impact parameters from $b_{\rm min}$ to $b_{\rm max}$, we get
\begin{equation}
\begin{aligned}
\delta v^2_{\rm tot}&=8N\left(\frac{Gm}{Rv}\right)^2\int_{b_{\rm min}}^{b_{\rm max}}\frac{db}{b}=8N\left(\frac{Gm}{Rv}\right)^2\log\left(\frac{b_{\rm max}}{b_{\rm min}}\right)
\\
& \equiv 8N\left(\frac{Gm}{Rv}\right)^2\log\Lambda\, .
\end{aligned}
\end{equation}
The integration limit $b_{\rm max}$ is of the order of the size $R$ of the system and $b_{\rm min}$ corresponds to the smallest $b$ to avoid stellar collisions. This is roughly the impact parameter for which $\delta v=v$, and can be readily estimated form Eq.~(\ref{eq:vperp}) as
\begin{equation}
b_{\rm min}=\frac{2Gm}{v^2},
\end{equation}
thus leading to
\begin{equation}
  \delta v^2_{\rm tot}=8N\left(\frac{Gm}{Rv}\right)^2\log\left(\frac{Rv^2}{2Gm}\right).
  \label{eq:vtot1}
\end{equation}
Now, the typical speed of a star in a virialized system is given by
\begin{equation}
Nmv^2=\frac{G(Nm)^2}{R},
\end{equation}
and reads
\begin{equation}\label{v2GNmR}
v^2=\frac{GNm}{R}.
\end{equation}
Replacing $v$ in Eq.~(\ref{eq:vtot1}), we get
\begin{equation}
\delta v^2_{\rm tot}=8\frac{Gm}{R}\log\left(\frac{N}{2}\right)=v^2\frac{8}{N}\log\left(\frac{N}{2}\right).
\end{equation}
The number of crossings of the system for which  $\delta v^2_{\rm tot}/v^2\approx 1$ -- \emph{i.e.}, for which the star has changed its initial velocity completely thus losing memory of its initial conditions -- is given by 
\begin{equation}
n_{\rm cross} \approx \frac{N}{8}\frac{1}{\log\left(N/2\right)}.
\end{equation}
The time needed to cross the system, $t_{\rm cross}$ is given by
\begin{equation}\label{tcrossdef}
t_{\rm cross}=\frac{R}{v}=\sqrt{\frac{R^3}{GNm}} \propto \frac{1}{\sqrt{G\rho}} \propto t_{\rm ff}.
\end{equation}
Finally, the time necessary for stars in a system to lose completely the memory of their initial velocity, called the relaxation time, is defined by
\begin{equation}\label{trelaxdef}
t_{\rm rlx}=n_{\rm cross}t_{\rm cross}= \frac{N}{8}\frac{1}{\log\left(N/2\right)}\frac{R}{v} \approx 10^7-10^{10}\text{ yr}.
\end{equation}
With more accurate calculations, based on diffusion coefficients, we have \cite{1971ApJ...164..399S}
\begin{equation}
\begin{aligned} \label{eq:relaxdef2}
t_{\rm rlx}&= 0.34 \frac{\sigma^3}{G^2m \rho\log \Lambda}
\\
&=\frac{1.8 \times 10^{10} \text{yr}}{\log\Lambda}\left( \frac{\sigma}{10 \ \text{ km s}^{-1}}\right)^3\left(\frac{1 \ M_\odot	}{m}\right)\left( \frac{10^3 M_\odot \ \text{pc}^{-3}}{\rho}\right) \, ,
\end{aligned}
\end{equation}
where $\sigma^2$ is the mean-square velocity in any directions.
\\
Note that, using $\sigma\approx \sqrt{M/R}$, $\rho\approx M/R^3$  and $\Lambda\approx N$, we have \cite{1971ApJ...164..399S}
\begin{equation}\label{eq:relaxdef3}
t_{\rm rlx}\approx \frac{10^8 \ \text{yr}}{\log{N}} \left( \frac{M}{10^5 M_\odot} \right)\left(\frac{R}{1\ \text{pc}} \right)^{3/2}\left(\frac{1 \ M_\odot	}{m}\right)\, .
\end{equation}
Importantly,  Eq.~(\ref{eq:relaxdef2}) draws the line between collisional and collisionless systems. Collisional systems have $t_{\rm rlx}\ll t_{\rm Hubble}$ and are therefore 'relaxed' by two body interactions to a state that does not retain memory of their initial conditions. Examples are:
\begin{itemize}
\item open clusters, with  $M\approx 10^3 M_\odot$ and $R\approx 1$\,pc, resulting in $t_{\rm rlx}\approx 10^7$\, yr; 
\item globular and young massive star clusters, with $M\approx 10^5-10^6 M_\odot$ and $R\approx 1-10$\,pc, resulting in $t_{\rm rlx}\approx 10^8-10^9$\,yr;
\item the Galactic centre, with with $M\approx 10^6 M_\odot$ and $R\approx 1$\,pc, resulting in $t_{\rm rlx}\approx 10^8$ yr.  
\end{itemize}
Conversely  collisionless systems have $t_{\rm rlx}\gg t_{\rm Hubble}$ and therefore two body interactions are not important in their dynamical evolution. Examples are:
\begin{itemize}
\item galaxies at large, with $M\approx 10^{11} M_\odot$ and $R\approx 10$\,kpc, resulting in $t_{\rm rlx}\approx 10^{15}$\, yr;
\item galaxy clusters, with $M\approx 10^{15} M_\odot$, $R\approx 1$\,Mpc and $m\approx 10^{11} M_\odot$, resulting in $t_{\rm rlx}\approx 10^{12}$\, yr.  
\end{itemize}

\subsubsection{Infant mortality}
As we already mentioned, most stars form in clusters, but today we observe only a minority of stars in clusters and associations. This is because most clusters dissolve with time due to several physical processes. One of them is \emph{infant mortality}, as first discussed in \cite{1980ApJ...235..986H}, who performed the derivation reported below.

Newborn clusters (or protoclusters) are bound systems of stars and gas and, after the initial evolution of the most massive stars, SN explosions and stellar winds blow away the internal gas of the cluster. In other words, the gas not turned into stars is lost within  a few Myrs of the cluster formation. As a consequence,  the mass of the cluster -- generating the gravitational potential holding the stars together -- decreases and stars can escape in a runaway process that can completely dissolve the cluster itself. Specifically, for a  cluster with mass $M=M_{g}+M_*$,  if the mass of the gas getting lost is $M_g>M_*$, then the cluster dissolves. From the virial theorem, the velocity dispersion  of the stars before gas removal is estimated as
\begin{equation}
\sigma_0^2=\frac{G(M_g+M_*)}{R_0},
\end{equation}
where $R_0$ is the initial size of the cluster. Assuming instantaneous gas removal, the energy soon after gas is lost is
\begin{equation}
E_f=\frac{1}{2}M_*\sigma_0^2 -\frac{GM_*^2}{R_0} = -\frac{GM_*^2}{2R_f},
\end{equation}
 so that
 \begin{equation}
 E_f= -\frac{GM_*^2}{2R_f}=\frac{1}{2}M_*\sigma_0^2 -\frac{GM_*^2}{R_0}=\frac{1}{2}M_*\frac{G(M_g+M_*)}{R_0} -\frac{GM_*^2}{R_0}.
 \end{equation}
This implies that 
 \begin{equation}
  -\frac{M_*}{R_f}=\frac{1}{2}\frac{(M_g+M_*)}{R_0} -\frac{M_*}{R_0}.
 \end{equation}
 By solving for $R_f$, we get
 \begin{equation}
 \frac{R_f}{R_0}= \left[ 1- \frac{1}{2}\left(1+\frac{M_g}{M_*}\right)\right]^{-1},
 \end{equation}
 and
\begin{equation}
 R_f>0 \Longleftrightarrow M_g < M_*.
\end{equation}
Therefore, the cluster survives only if the expelled gaseous mass is smaller than the total mass in stars. This result is a strong lower limit to the maximum mass of gas which can be expelled without destroying the star cluster, because we assume instantaneous expulsion of the gas component. More accurate calculations, assuming a realistic timescale for gas expulsion, find $M_g\lesssim{}4\,{}M_*$ (e.g. \cite{baumgardt2007}).

Most clusters might not survive the first $10$ Myrs for this reason. On the other hand, the more massive is the cluster the more difficult is to blow it apart, so the most massive clusters tend to survive. 

\subsubsection{Evaporation}
\begin{figure}[h!]
  \centering
  \includegraphics[width=0.6\textwidth]{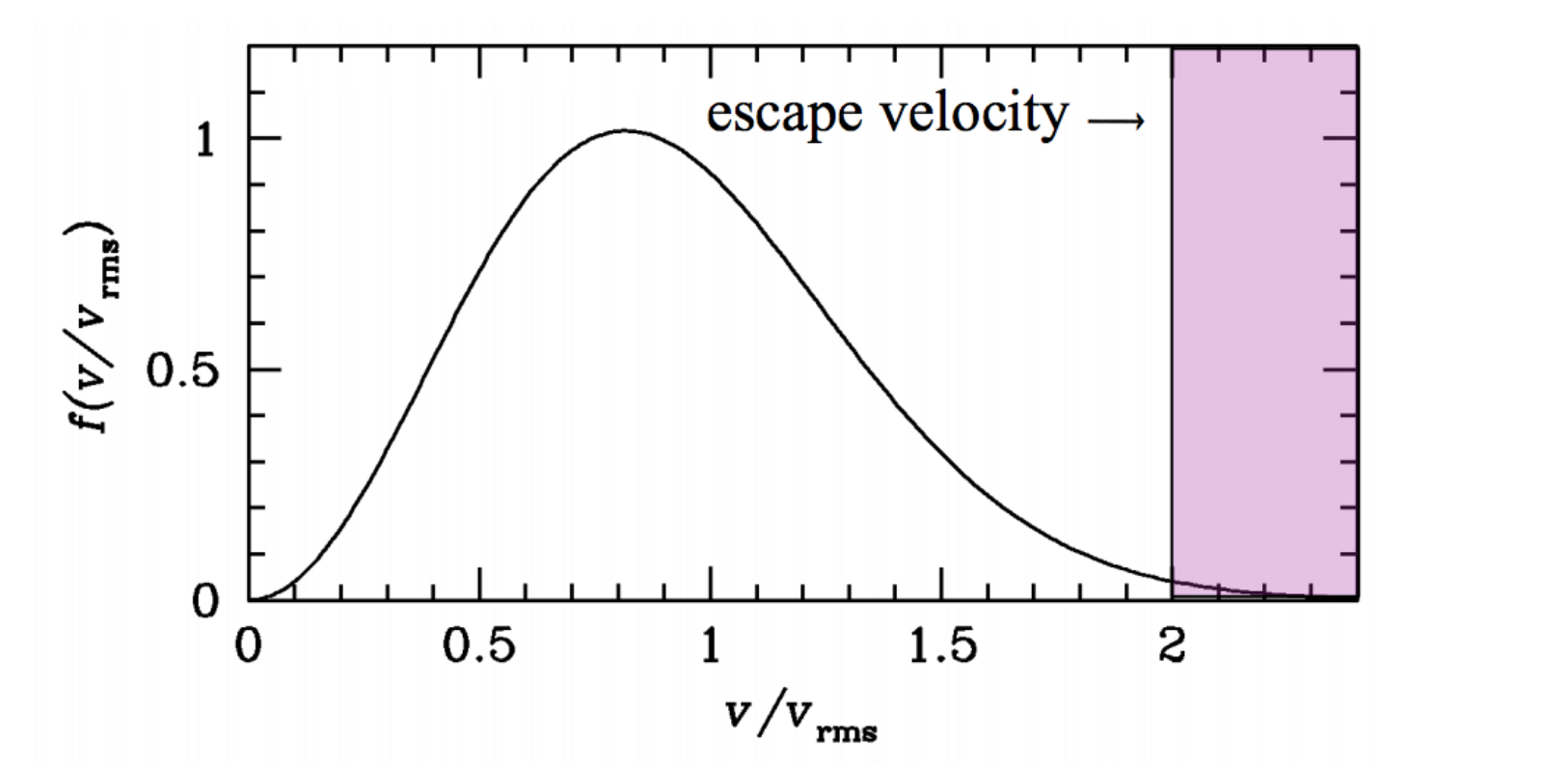}
  \caption{Maxwell velocity distribution normalized to $v_{\rm rms}$. The shaded area highlights the tail of unbound stars with $v>2v_{\rm rms}$}\label{Maxwell distr}
\end{figure}

Following the derivation in Lecture 7 of {\it Stellar Dynamics and Structure of Galaxies}\footnote{\url{https://www.ast.cam.ac.uk/~vasily/Lectures/SDSG/sdsg_7_clusters.pdf}} by Dr.  Vasily Belokurov, a star cluster is a self gravitating system for which we can define an escape velocity at any position $r$ as 
\begin{equation}
v_e^2(r)=-2\phi(r),
\end{equation}
where $\phi$ is the gravitational potential.
The mean square escape velocity is then obtained by averaging over the mass density distribution $\rho(r)$ (we assume spherical symmetry) as
\begin{equation}
\left< v_e^2\right>=\frac{\int\rho (r) v_e^2(r)d^3r}{\int \rho(r)d^3r}=-\frac{2}{M}\int\rho (r) \phi(r)d^3r=-\frac{4E_g}{M},
\end{equation}
where $E_g$ is the potential self-energy and $M$ is the total mass. From the  virial theorem:  $-E_g = 2E_K$ with $E_K=\frac{1}{2}M\left<v^2\right>$ being the kinetic energy, thus leading to
\begin{equation}
\left<v_e^2\right> =4\left<v^2\right>.
\end{equation}
Therefore, stars with velocities exceeding twice the root-mean-square (RMS) velocity of the distribution are unbound. For a typical Maxwellian velocity distribution, this amounts to a fraction $\epsilon= 7.4 \times 10^{-3}$ of all stars; see Fig.~\ref{Maxwell distr}. 
Roughly, evaporation removes $dN= -\epsilon N$ stars on a timescale $dt=t_{\rm rlx}$ and the system gets slightly hotter and contracts. Meanwhile, the velocity distribution adjusts to another Maxwellian distribution,  and so in every relaxation time  $\epsilon N$ stars are removed
\begin{equation}
\frac{dN}{dt}=-\frac{\epsilon N}{t_{\rm rlx}}=-\frac{N}{t_{\rm evap}},
\end{equation}
where we have defined
\begin{equation}
t_{\rm evap}=\frac{t_{\rm rlx}}{\epsilon}.
\end{equation}
Note that  there are other mechanisms for stars to escape such as close encounters and SNs.  The first is  rare, but can lead to stars escaping with $v \gg v_e$. The second is related to SN kicks that result in NSs (and perhaps BHs) with high velocities, as we discussed in Section \ref{sec:kicks}. 


\subsubsection{Core Collapse}
We now see how evaporation leads to core collapse \cite{1987Spitzer}, closely following the derivation in the second lecture of the series {\it lectures on collisional dynamics}.\footnote{\url{http://web.pd.astro.it/mapelli/2014colldyn2.pdf}}
In the previous section, we have seen that, at a typical position within the cluster, stars with velocity $v_e \geq 2\left<v\right> $ are unbound, in other words  $0.74 \%$ of the stellar population, if the velocity distribution were exactly Maxwellian, evaporates. Generally,  the evaporation is  not  a  steady-state process\footnote{In a steady-state process the state variables which define the behaviour of the process, such as the density, the velocity dispersion, \emph{etc}\dots do not change in time}. However, we can search for a self-similar solution that can describe evaporation with sufficient accuracy (over some range in spatial and temporal scales). Self-similar solutions are homologous solutions in which the radial variations of density, potential and other physical factors remain invariant with time except for time-dependent scale factors \cite{1987Spitzer}.
In this case, the spatial distributions of the state variables  at different times can be obtained from one another by a scale transformation, \emph{i.e.}, a rescaling of the axes. In the self-similar regime, we expect a constant rate of mass loss, \emph{i.e.},  \cite{1987Spitzer}
\begin{equation}\label{dMdt}
\frac{dM}{dt}=-\xi_e\frac{M(t)}{t_{\rm rlx}(t)} \, .
\end{equation}
From Eq.~\eqref{eq:relaxdef3}, neglecting changes in the Coulomb logarithm $\log \Lambda\sim \log N$, the relaxation time is proportional to
\begin{equation}\label{relaxdef4}
t_{\rm rlx}\propto \sqrt{M}R^{3/2},
\end{equation}
so that
\begin{equation}\label{eqtrelax}
t_{\rm rlx}(t)=t_{\rm rlx}(0)\left(\frac{R(t)}{R(0)}\right)^{3/2}\left(\frac{M(t)}{M(0)}\right)^{1/2}.
\end{equation}
Replacing Eq.~\eqref{eqtrelax} in Eq.~\eqref{dMdt}, we get
\begin{equation}\label{massloss}
\frac{dM}{dt}=-\xi_e\frac{M(0)}{t_{\rm rlx}(0)}\left(\frac{R(t)}{R(0)}\right)^{-3/2}\left(\frac{M(t)}{M(0)}\right)^{1/2}.
\end{equation}
Each star escaping from the cluster carries away a certain kinetic energy per unit mass
\begin{equation}
\frac{dE_{\rm TOT}}{dM} = \zeta E_m = \zeta \frac{E_{\rm TOT}}{M},
\end{equation}
where $E_m$ is the mean energy per unit mass of the cluster.
As a consequence, the total cluster energy changes by
\begin{equation}
\frac{dE_{\rm TOT}}{dt}=\frac{dE_{\rm TOT}}{dM} \frac{dM}{dt}=\zeta E_m\frac{dM}{dt}=\zeta  \frac{E_{\rm TOT}}{M} \frac{dM}{dt}.
\end{equation}
Because 
\begin{equation}\label{ETOT}
E_{\rm TOT}\propto -\frac{M^2}{R},
\end{equation}
we have
\begin{equation}
\zeta  \frac{E_{\rm TOT}}{M} \frac{dM}{dt}\propto -\zeta \frac{M}{R}\frac{dM}{dt}.
\end{equation}
On the other hand, from Eq.~\eqref{ETOT}
\begin{equation}
E_{\rm TOT}\propto -\frac{M^2}{R}\qquad \Longrightarrow \qquad \frac{dE_{\rm TOT}}{dt}\propto -\frac{2M}{R}\frac{dM}{dt}+\frac{M^2}{R^2}\frac{dR}{dt},
\end{equation}
and we conclude that
\begin{equation}
\left(2-\zeta\right) \frac{dM}{M}=\frac{dR}{R}.
\end{equation}
Integrating this expression, we have
\begin{equation}\label{massloss2}
\frac{R(t)}{R(0)}=\left(\frac{M(t)}{M(0)}\right)^{2-\zeta}\qquad \Longrightarrow \qquad \frac{\rho(t)}{\rho(0)}\propto \left(\frac{M(t)}{M(0)} \right)^{3\zeta -5}.
\end{equation}
For realistic clusters $\zeta<1$. This can be understood by considering that most of the stars are ejected just above $v_e$ so that their velocity at infinity is generally $v_{\infty} < \sigma$. Since the typical energy per unit mass of particles in the clusters is $\approx \sigma^2$ we have $v^2_{\infty} < \sigma^2$ so that $\zeta<1$. It follows that, when mass is lost due to evaporation, the cluster radius contracts and the density increases. Note that $\rho(t)\rightarrow \infty$ as $M(t)\rightarrow 0$. The cluster formally collapses in a finite time into a point mass of infinite density.
Substituting Eq.~(\ref{massloss2}) into Eq.~(\ref{massloss}), we get
\begin{equation}
\frac{dM}{dt}=-\xi_e\frac{M(0)}{t_{\rm rlx}(0)}\left(\frac{M(t)}{M(0)}\right)^{\frac{5-\zeta}{2}},
\end{equation}
which can be integrated, and the result is
\begin{equation}
M(t)=M(0)\left[1-\frac{\xi_e(7-3\zeta) }{2}\frac{t}{t_{\rm rlx}(0)}\right]^{\frac{2}{7-3\zeta}}\equiv M(0)\left[1-\frac{t}{t_{0}}\right]^{\frac{2}{7-3\zeta}},
\end{equation}
where $t_{0}$ is the collapse time, satisfying
\begin{equation}
M(t_{0})=R(t_{0})=0.
\end{equation}
For a cluster composed of equal-mass stars $t_0\gtrsim 10 ~t_{\rm rlx}$.  

\subsubsection{Post Core-Collapse phase}
In the previous section, we have seen that mass loss due to evaporation leads to collapse, a  runaway  process called gravothermal instability. If  the system contracts, it becomes denser and  the two-body encounter rate increases along with the evaporation rate. As a consequence, the core of the cluster loses energy (the kinetic energy of the evaporated stars) to the halo, $dE_{\rm core}<0$. Since  any bound finite system in which the dominant force is gravity exhibits negative heat capacity, defined as 
\begin{equation}
C\equiv \frac{dE}{dT}<0 \, ,
\end{equation} 
the temperature increases $dT_{\rm core}>0$. Therefore, stars exchange more energy and become dynamically hotter, and faster stars tend to evaporate at higher rate. This runaway  process leads to the unphysical situation of star clusters with infinite core density.

To avoid this catastrophe, we consider the possibility that the core collapse is reversed by  an external source injecting kinetic energy into the core ($dE_{\rm core}>0$), thus cooling it ($dT_{\rm core}<0$) until the temperature gradient declines to zero, halting the collapse. Possible sources of extra kinetic energy are \cite{binney2011galactic}
\begin{itemize}
\item mass loss by stellar winds and supernovae (if massive star evolution lifetime is similar to core collapse timescale, \cite{mapelli2013,trani2014});
\item formation of binary systems; 
\item three-body encounters between single stars and binaries, extracting kinetic energy from the internal energy of the binary systems (see Section~\ref{sec:Hardening} below).
\end{itemize}
Note that, with the increasing density in  the core during the collapse, the probability to form binary systems also increases. In order to understand how the formation of binaries can  inject kinetic energy into the core, consider a three-body interaction of three stars with initial kinetic energies $K_i$, $i = 1,2,3$. Suppose that after the 3-body interaction, stars $1$ and $2$ form a binary, thus  \cite{binney2011galactic}
\begin{itemize}
\item  the kinetic energy of the centre of mass of the binary is $K_{\rm bin}$,
\item  the internal energy of the binary is $E_{\rm bin} < 0$,
\item the kinetic energy of the third star is $K_3'$.
\end{itemize}
Energy conservation implies
\begin{equation}
K_1+K_2+K_3=K_{\rm bin}+E_{\rm bin}+K_3' \qquad \Longrightarrow  \qquad K_{\rm bin}+K_3'>K_1+K_2+K_3,
\end{equation}
from which we see that the kinetic energy after the interaction, \emph{i.e.}, the kinetic energy stored in the centres of mass of the single star and the binary, is larger than the initial kinetic energy of the three stars.
\\
Therefore, the formation of binaries can pump kinetic energy into single stars crossing the core of the cluster (where most of the binaries form). Those stars then share the acquired extra kinetic energy with other stars through two-body relaxation, heating up the cluster.

As before, we suppose that a spherical cluster evolves self-similarly as a result of relaxation. 
The evolution is described by two functions $M(t)$ and $R(t)$, the mass and characteristic radius as functions of time, satisfying Eq.~\eqref{dMdt} and \cite{binney2011galactic, 1993Goodman}
\begin{equation}\label{dRdt}
\frac{dR}{dt}=-\xi_r\frac{R(t)}{t_{\rm rlx}(t)} \, .
\end{equation}
where $\xi_r$ is a constant of order unity. Moreover, recall from Eq.~\eqref{relaxdef4} that the two-body relaxation time is given by $t_{\rm rlx} \propto \sqrt{M}R^{3/2}$.

After core collapse, the kinetic energy injection by binaries in the cluster core causes an increase of the total kinetic  energy of the cluster, but for an isolated cluster the mass $M$ remains approximately constant.
In this case, we have
\begin{equation}
\sqrt{R(t)}\frac{dR}{dt}=-\frac{\xi_r}{\sqrt{M}}=const.
\end{equation}
with solution
\begin{equation}
R(t)\sim (t-t_0)^{2/3}\, ,
\end{equation}
where $t_0$ is roughly the time of core collapse.
As a consequence  the halo expands.
\\
However, the situation is more complicated as the self-similar post-collapse evolution is unstable, leading to gravothermal oscillations, characterized by a series of core contractions/re-expansions.

%
%

\subsubsection{Mass Segregation and Spitzer's instability}

All the processes described so far work for any stellar population, even for a population of equal-mass stars. Stars in clusters, however,  have a mass spectrum ranging from $\sim 0.5 M_{\rm \odot}$ to $\sim 150 M_{\odot}$. In the following, we briefly discuss how a realistic mass spectrum affects the processes we described so far.

First of all, stars more massive than the average stellar mass are expected to undergo the process called ``dynamical friction'' (see Section~\ref{sec:dynfric}). This means that a massive star walking through a sea of lighter stars feels a drag force, which decelerates its motion. The timescale of dynamical friction for a star of mass $M$ is approximately
\begin{equation}
t_{\rm segr}(M)=\frac {\langle{}m\rangle{}}{M}t_{\rm rlx} ,
\end{equation}
where $\langle{}m\rangle{}$ is the average star mass and $t_{\rm rlx}$ is the two-body relaxation timescale. For a star with mass $M=25$ M$_\odot$ (assuming a typical value of $\langle{}m\rangle{}=1$ M$_\odot$), $t_{\rm segr}(M)=0.04\,{}t_{\rm rlx}$, which implies that dynamical friction is much more efficient than two-body relaxation. The effect of dynamical friction is that the most massive stars in a star cluster lose kinetic energy in favour of the light stars and segregate toward the centre of the star cluster. This generates the phenomenon called {\emph{mass segregation}}: the radial distribution of massive stars tends to be more centrally concentrated than the average stellar distribution in dense star clusters.

The main effect is that core collapse occurs much faster ($t_0\sim{}0.2\,{}t_{\rm rlx}$, \cite{fujii2014}) in a stellar population with a realistic mass function than in a cluster of equal-mass stars, because mass segregation increases the central density faster and accelerates the runaway collapse. 

Mass segregation also favours the formation of very massive binaries in the core of the cluster, which might become progenitors of massive compact-object binaries.

Moreover, we know from the equipartition theorem of statistical mechanics (Boltzmann 1876) that in gas systems at thermal equilibrium energy is shared equally by all particles. For analogy with gas, we expect that in a two-body relaxed stellar system the kinetic energy of a star $i$ is {\emph{locally}} the same as that of the star $j$
\begin{equation}\label{equipartition}
\frac{1}{2}m_i\left< v_i^2\right>=\frac{1}{2}m_{j}\left<v_{j}^2\right>,
\end{equation}
If the velocities of all stars are initially drawn from the same distribution, massive stars are thus expected to transfer kinetic energy to lighter stars and slow down, till they reach equipartition. The condition of equipartition requires that $v(m)\propto{}m^{-0.5}$.

Does the fact that star clusters are mass segregated also implies that they reach equipartition in a two-body relaxation timescale?

Spitzer (1969, \cite{1969ApJ...158L.139S}) demonstrated through an analytic calculation that there is at least one case in which equipartition cannot be reached by a stellar system (known as Spitzer's instability or mass stratification instability). The main assumption of Spitzer's calculation is that the cluster is a two-component system\footnote{The only analytic generalization of Spitzer's calculation to a star cluster with a realistic mass function was done by Vishniac (1978, \cite{vishniac1978}). However, \cite{vishniac1978} assume similar density profiles between various stellar mass groups, which is another strong (and quite unrealistic) assumption. On the other hand, recent numerical models \cite{trenti2013,spera2016} have shown that Spitzer's instability is very common in star clusters with a realistic mass function.}, \emph{i.e.}, there are  $N_1$ stars with mass $m_1$ (the total mass of the lighter population is $M_1=N_1 m_1$) and $N_2$ stars with mass $m_2$ (the total mass of the heavier population is $M_2=N_2m_2$). We further assume that $m_2 \gg{} m_1$.
 


Under these assumptions, Spitzer demonstrated that a star cluster can reach equipartition only if
\begin{equation}
\frac{M_2}{M_1}\left( \frac{m_2}{m_1}\right)^{3/2} <0.16.
\end{equation}
If this condition is not satisfied, the heavy population forms a cluster within the cluster, \emph{i.e.}, a sub-cluster at the centre of the cluster, dynamically decoupled from the rest of the cluster. Since the system cannot reach equipartition, the core of massive stars continues to contract until  most of the massive stars bind into binary systems and/or eject each-other from the cluster by 3-body encounters, or when most of the massive stars collapse into a single object.

We now give a proof of the Spitzer condition; see, \emph{e.g.}, \cite{2013iesb.book.....B}. Let  $\rho_i$ be the local density of stars of mass $m_i$, $r_{h,i}$ the half-mass radius of population $i=1,2$, and $M_i(r)$  the total mass of population $i$ contained within radius $r$.  From the virial theorem, we have
\begin{equation}
\begin{aligned}\label{Spitzer_virial1}
\left<v_1^2\right>&=\frac{\alpha G M_1}{r_{h,1}}+\frac{G}{M_1}\int_0^{\infty}\frac{\rho_1 M_2(r)}{r}4\pi r^2dr,
\\
\left<v_2^2\right>&=\frac{\alpha G M_2}{r_{h,2}}+\frac{G}{M_2}\int_0^{\infty}\frac{\rho_2 M_1(r)}{r}4\pi r^2dr,
\end{aligned}
\end{equation}
where  $\alpha$ is a parameter that describes the density distribution throughout the cluster. The first term on the right-hand side represents the self-gravity of the population and
the second term on the right-hand side corresponds to the gravitational energy of one population due to the other. As before, we assume that $m_2\gg m_1$. 
Finally, as a consequence of segregation, the more massive stars become centrally concentrated compared to the distribution of low-mass stars. We thus assume that the density of lighter stars is constant $\rho_1(r)\approx \rho_1(0)$ throughout the region occupied by the heavier stars, \emph{i.e.},
\begin{equation}
M_1(r) = \frac{4\pi}{3} r^3 \rho_{c1},
\end{equation}
where $\rho_{\rm c1}$ is the central density of stars of mass $m_1$. With these assumptions, Eq.~\eqref{Spitzer_virial1} can be approximated by
\begin{equation}\label{Spitzer_virial2}
\begin{aligned}
\left<v_1^2\right>&=\frac{\alpha G M_1}{r_{h,1}},
\\
\left<v_2^2\right>&=\frac{\alpha G M_2}{r_{h,2}}+\frac{4\pi G}{3}\rho_{c,1}r_{s,2}^2,
\end{aligned}
\end{equation}
where 
\begin{equation}
r_{s,2}^2= \frac{1}{M_2}\int_0^{\infty}r^2 \rho_2 ~(4\pi r^2)dr.
\end{equation}
Defining the mean density of stars of each type within their half-mass radius 
\begin{equation}
\rho_{m,i}\equiv\frac{3}{4\pi r_{h,i}^3}\frac{M_i}{2},
\end{equation}
using this equation to express $r_{h,i}$ in terms of $\rho_{m,i}$  and substituting Eq.~\eqref{Spitzer_virial2} into the equipartition condition \eqref{equipartition}, we obtain, after some algebra,
\begin{equation}\label{Spitzer_virial3}
\chi\equiv \frac{M_2}{M_1}\left( \frac{m_2}{m_1}\right)^{3/2}=\frac{\left( \rho_{m,1}/\rho_{m,2}\right)^{1/2}}{\left[ 1+\beta \left( \rho_{m,1}/\rho_{m,2}\right)\right]^{3/2}},
\end{equation}
where we have defined
\begin{equation}
\beta=\frac{\rho_{c,1}}{\rho_{m,1}}\frac{1}{2\alpha} \left( \frac{r_{s,2}}{r_{h,2}}\right)^2.
\end{equation}
Note that for $\rho_{m,1}/\rho_{m,2}=(2\beta)^{-1}$,  Eq.~\eqref{Spitzer_virial3}  has the maximum value 
\begin{equation}
\chi_{\rm max}= \sqrt{\frac{4}{27\beta}}.
\end{equation}
For realistic values of $\beta$, one has $\chi_{\rm max}=0.16$. We conclude that  the condition for equipartition in equilibrium is
\begin{equation}
\chi<\chi_{\rm max}=0.16.
\end{equation}
Let us see what this means for a typical astrophysical situation. For a Salpeter IMF \cite{1955ApJ...121..161S}, about 0.3\% of the stars have $m>30 M_\odot$. Let assume that those stars evolve into SBHs with mass $m_2=10 M_\odot$. The rest of the cluster is approximately composed of solar mass stars on the main sequence. We therefore have: $m_2\approx10 m_1$ and $M_2\approx 0.03M_1$. Thus $\chi=0.03\times 10^{3/2}\approx 1 > \chi_{\rm max}$. Therefore, typical clusters likely undergo Spitzer instability, resulting in a dense core of SBHs, prone to the formation of tight SBHBs via capture and other physical processes.

\subsection{Black hole binaries: hardening and gravitational waves}\label{sec:Hardening}

In this section, we consider the hardening mechanism via 3-body interactions and its relation to GW emission from SBHBs formed by dynamical capture.  For more details, we refer to, \emph{e.g.}, \cite{colpi2009physics}.
\subsubsection{Three body encounters}
We start by summarizing the dynamics of 3-body encounters. Recall that the internal energy of a binary  (i.e. the total energy of a binary after subtracting the kinetic energy of its centre of mass) is
\begin{equation}
E_{\rm bin}=-\frac{G\,{}M_1\,{}M_2}{2\,{}a}=-E_b,
\end{equation}
where $a$ is semi-major axis, $M_1$ and $M_2$ the masses of the two objects and $E_b$ is the binding energy. 

Consider a 3-body  interaction between an object $m_3$ and a binary system, formed by two masses $M_1$ and $M_2$ (we use capital $M$ for objects forming the initial binary and $m$ for the intruder). If the original binary is preserved in the encounter, there are two possibilities: 
\begin{itemize}
\item the single body extracts internal energy from the binary, so that  the final kinetic energy $K_f$ of the CoM of the single object and of the binary is higher than the initial one $K_i$, \emph{i.e.}, $K_f>K_i$.
\item the single body loses a fraction of its kinetic energy, which is converted into internal energy of the binary.
\end{itemize}
In the first case, the object and the binary acquire recoil velocity and the binding energy increases (as the binary becomes more bound). In other words, since
\begin{equation}
K_i-E_{b,i}=K_f-E_{b,f} \qquad \Longrightarrow \qquad E_{b,f} -E_{b,i}=K_f- K_i,
\end{equation}
for $K_f>K_i$ we have 
\begin{equation}
E_{b,f}=\frac{GM_1M_2}{2a_f}>\frac{GM_1M_2}{2a_i}=E_{b,i} \qquad \Longrightarrow \qquad a_f<a_i.
\end{equation}
The opposite happens in the second case.

Another possibility for the binary to increase the binding energy during a  3-body  interaction is the exchange, \emph{i.e.}, the single object $m_3$ replaces one of the members of the binary. This usually happens when $M_2<m_3<M_1$, in which case, after the exchange the binary is formed by $M_1$ and $m_3$ and
\begin{equation}
E_{b,f}=\frac{GM_1m_3}{2a}>\frac{GM_1M_2}{2a} =E_{b,i}.
\end{equation}

The final binary can also becomes less bound and can even be ionized  if its velocity at infinity exceeds the critical velocity $v_c$ \cite{1983ApJ...268..319H}. In fact, from
\begin{equation}
E_f=\frac{1}{2}\frac{m_3(M_1+M_2)}{(M_1+M_2+m_3)}v^2-\frac{G M_1M_2}{2a},
\end{equation}
the system is unbound if $E_f=0$, so that
\begin{equation}
v_c=\sqrt{\frac{GM_1M_2(M_1+M_2+m_3)}{am_3(M_1+M_2)}}.
\end{equation} 

\paragraph{Hard and Soft binaries.}
We define {\it hard binaries} those characterised by
\begin{equation}
  E_b>\frac{1}{2}\langle{}m\rangle{}\sigma^2,
\end{equation}
where $\sigma{}$ is the average velocity of the stars and $\langle{}m\rangle{}$ is the average mass of a star. 
Conversely, {\it soft binaries} satisfy
\begin{equation}
E_b<\frac{1}{2}\langle{}m\rangle{}\sigma^2.
\end{equation}
Heggie's law  \cite{1975MNRAS.173..729H} states  that  statistically, during three-body interactions, hard binaries tend to become harder  whereas soft binaries tend to become softer. 

\paragraph{Cross Section for 3-body encounters}
To define the cross section for 3-body encounters, let us consider the maximum impact parameter  $b_{\rm max}$ for a non-zero energy exchange between the single object $m_3$ and the binary (formed by $M_1$ and $M_2$). To estimate $b_{\rm max}$, we need to consider gravitational focusing, \emph{i.e.}, the fact that the trajectory of $m_3$ is significantly deflected by the presence of the binary, thus approaching it with an effective pericentre $p$ much smaller than the formal impact parameter $b$ at infinity. From energy conservation we can write
\begin{equation}
\Delta E=\frac{1}{2}\frac{m_3(M_1+M_2)}{(M_1+M_2+m_3)}(v_f^2-v_i^2)+Gm_3(M_1+M_2)\left(\frac{1}{D}-\frac{1}{p}\right)=0,
\end{equation}
where  $D$ is the initial distance between the single object and the binary, and for the initial velocity of the single object we consider $v_i=\sigma$.  Assuming $D\rightarrow \infty$, we get
\begin{equation}\label{energycons}
\frac{1}{2}\frac{\sigma^2}{(M_1+M_2+m_3)}=\frac{1}{2}\frac{v_f^2}{(M_1+M_2+m_3)}-\frac{G}{p}.
\end{equation}
On the other hand, angular momentum conservation imposes
\begin{equation}\label{momentumcons}
\Delta J =(pv_f-b\sigma)\frac{m_3(M_1+M_2)}{(M_1+M_2+m_3)}=0,
\end{equation}
so that
\begin{equation}
b\sigma=pv_f.
\end{equation}
Combining Eq.~(\ref{energycons}) and Eq.~(\ref{momentumcons}), we get
\begin{equation}
\frac{1}{2}\frac{\sigma^2}{(M_1+M_2+m_3)}-\frac{1}{2}\left(\frac{b\sigma}{p}\right)^2\frac{1}{(M_1+M_2+m_3)}
+\frac{G}{p}=0,
\end{equation}
which can be solved for $p$, leading to \cite{sigurdsson1993}
\begin{equation}
p=\frac{G(M_1+M_2+m_3)}{\sigma^2}\left[\sqrt{1+\frac{b^2\sigma^4}{G^2(M_1+M_2+m_3)^2}}-1\right].
\end{equation}
Finally, Taylor expansion of the right-hand-side for $\frac{b\sigma^2}{G(M_1+M_2+m_3)}\ll 1$ (which holds when $M_1,M_2\gg m_3$ and for hard binaries in general) gives
\begin{equation}
p\simeq \frac{b^2\sigma^2}{2G(M_1+M_2+m_3)}.
\end{equation}
The  3-body cross section is defined as
\begin{equation}
\Sigma=\pi b_{\rm max}^2\simeq \pi \left[\frac{2G(M_1+M_2+m_3) }{\sigma^2} \right] p_{\rm max}\simeq  
\frac{2\pi G(M_1+M_2+m_3) a}{\sigma^2} ,
\end{equation}
where we approximated $p_{\rm max}\simeq a$ (which is correct only for very energetic three-body encounters).

\subsubsection{Three body hardening}
\label{sec:3body}
Once the interaction cross section is determined, the interaction rate can be readily estimated as 
\begin{equation}\label{collision_rate}
\frac{dN}{dt}=n\Sigma \sigma=\frac{2\pi G(M_1+M_2+m_3) na}{\sigma}. 
\end{equation}
We now make a series of simplifying assumptions that characterise those binaries that will eventually become GW sources. Importantly, those assumptions are relevant for SBHs and SMBHs alike, thus providing a useful description to the dynamics of SBHBs and MBHBs. We assume that
\begin{enumerate}
\item the binary is hard;
\item the effective pericentre satisfies $p\lesssim 2a$;
\item  the mass of the single object is small with respect to binary mass, $m_3\ll M_1,M_2$.
\end{enumerate}
The average binding energy variation per encounter can be estimated by
\begin{equation}
\left<\Delta E_b\right>=\xi \frac{m_3}{(M_1+M_2)}E_b=\xi \frac{m_3}{(M_1+M_2)}\frac{GM_1M_2}{2a},
\end{equation}
where $\xi\approx 0.2 - 1$ 
is a parameter that can be extracted from 3-body scattering experiments \cite{1992MNRAS.259..115M,1996NewA....1...35Q}. The rate of binding energy exchange for a hard binary is\footnote{Note the extra factor of $2$, because we assumed $p\lesssim 2a$ instead of $p\lesssim a$.}
\begin{equation}
\frac{dE_b}{dt}=\left<\Delta E_b\right>\frac{dN}{dt}=2\pi \xi \frac{M_1M_2m_3(M_1+M_2+m_3)}{(M_1+M_2)}\frac{  G^2 n}{\sigma},
\end{equation}
where we have used (\ref{collision_rate}). Supposing a single mass population of intruders characterized by $m_3=\langle{}m\rangle{}$, we can write the rate of binding energy exchange in terms of  the local mass density  $\rho=\langle{}m\rangle{}n$. By exploiting the condition $m_3\ll M_1,M_2$ we obtain
\begin{equation}
\frac{dE_b}{dt}=\frac{2\pi \xi G^2 M_1M_2 \rho}{\sigma}.
\end{equation}
Therefore, hard binaries harden at a constant rate. Finally, expressing $a$ in terms of $E_b$, the hardening rate is given by
\begin{equation}
  \frac{d}{dt}\left(\frac{1}{a}\right)=\frac{2}{GM_1M_2}\frac{dE_b}{dt}=4\pi G\xi\frac{\rho}{\sigma}= \frac{GH\rho}{\sigma^2},
  \label{eq:starhardening}
\end{equation}
which can be written as 
\begin{equation}
\frac{da}{dt}=-\frac{GH\rho}{\sigma}a^2,
\end{equation}
where $H\approx 15-20$ is an dimensionless hardening rate (as introduced in \cite{1996NewA....1...35Q}).

\subsubsection{Hardening and gravitational waves}
\label{sec:3body_gw}
From Eq.~(\ref{eq:starhardening}) we can see that hardening in a given stellar background proceeds at a constant rate that is solely determined by the properties of the stellar background, in particular the density $\rho$ and velocity dispersion $\sigma$. Note that $da/dt|_{\rm 3b} \propto a^2$, whereas $da/dt|_{\rm GW} \propto a^{-3}$ (see Eq.~(\ref{eq:dadtgw})).
The evolution of the semimajor axis can therefore be written as \cite{2003ApJ...599.1260C}
\begin{equation}
\frac{da}{dt}=\frac{da}{dt}\Big{|}_{3b}+\frac{da}{dt}\Big{|}_{\rm gw}=-Aa^2-\frac{B}{a^3},
\label{aevtot}
\end{equation}
where
\begin{equation}
  A=\frac{GH\rho}{\sigma},  \,\,\,\, B=\frac{64}{5}\frac{G^3}{c^5} M_1M_2(M_1+M_2)F(e),
  \label{eq:AB}
\end{equation}
and $F(e)$, given by Eq.~(\ref{eq:Fe}), takes into account for the accelerated GW evolution of eccentric binaries.\footnote{Conversely, binary evolution from three body scattering is largely insensitive to the BHB eccentricity, with the dimensionless rate $H$ increasing modestly from $\approx 15$ for circular binaries to $\approx 20$ for very eccentric ones \cite{1996NewA....1...35Q}.}

Since the stellar hardening is $\propto a^2$ and the GW hardening is $\propto a^{-3}$, binaries spend most of their time at the transition separation obtained by imposing $(da/dt)_{3b}=(da/dt)_{\rm gw}$:
\begin{equation}
  \begin{aligned}
    a_{*/{\rm gw}}= & \left[\frac{64G^2\sigma M_1M_2MF(e)}{5c^5H\rho}\right]^{1/5}\\
    & \approx 0.15 {\rm AU}\left(\frac{M}{60{\rm M}_\odot}\right)^{3/5}\frac{q^{1/5}}{(1+q)^{2/5}}\left(\frac{\sigma}{10{\rm km\,s}^{-1}}\right)^{1/5}\left(\frac{\rho}{10^5 M_\odot{\rm pc}^{-3}}\right)^{-1/5}F(e)^{-1/5},
\label{atrans}
\end{aligned}
\end{equation}
and their lifetime can be written as
\begin{equation}
  t(a_{*/{\rm gw}})=\frac{\sigma}{GH\rho a_{*/{\rm gw}}} \approx 3\,{\rm Gyr}\left(\frac{\sigma}{10{\rm km s}^{-1}}\right)\left(\frac{\rho}{10^5 M_\odot{\rm pc}^{-3}}\frac{a_{*/{\rm gw}}}{0.15 {\rm AU}}\right)^{-1}.
\label{ttrans}
\end{equation}

Eq.~(\ref{atrans}) and Eq.~(\ref{ttrans}) have been normalised for a massive SBHB of 30$+$30 $M_\odot$ in a typical cluster with $\sigma=10$km s$^{-1}$ and core density $\rho=10^5 M_\odot{\rm pc}^{-3}$. From this we see that relatively massive SBHB (such as GW150914) can harden via dynamical processes in $t<t_{\rm Hubble}$, resulting in a GW driven merger. Moreover, the mechanism efficiency increases with the BHB mass. If IMBHs (with $M\approx 10^3 M_\odot$) can indeed form in star clusters, stellar hardening provides an efficient mechanism to merge them with SBHs or with a companion IMBH. Since we do not focus on IMBH formation in these lectures, we refer the reader to \cite{2002ApJ...576..899P,giersz2015,mapelli2016} for more details. 

Note that this framework also applies to MBHBs in galactic nuclei. One uncertainty here is that the density is usually a function of $r$ so that it is not obvious what number to pick. It has been shown \cite{2015MNRAS.454L..66S,2015ApJ...810...49V} that in the limit of efficient loss cone refilling, the set of equation given above is valid if $\rho$ and $\sigma$ are evaluated at the {\it influence radius} $r_{\rm inf}$ of the MBHB, where $r_{\rm inf}$ is defined as the distance to the centre of mass of the binary (also assumed to be the centre of the stellar distribution) enclosing twice the mass of the binary in stars. For a typical LISA event with $M_1=M_2=10^6 M_\odot$ residing in a nucleus with $\sigma=100$km s$^{-1}$ and $\rho(r_{\rm inf})=10^4 M_\odot{\rm pc}^{-3}$, $a_{*/{\rm gw}}\approx 0.001$pc and $t(a_{*/{\rm gw}})\approx 0.3$ Gyr. This shows that stellar hardening is also an effective mechanism to drive MBHBs to merger.


The number of 3-body interactions before the GW regime can be calculated as
\begin{equation}
N_{\rm int}=\int_0^t\frac{dN}{dt}dt=\int_0^t \frac{4\pi G(M_1+M_2+m_3) na}{\sigma} dt.
\end{equation}
Since
\begin{equation}
\frac{da}{dt}=-\frac{GH\rho}{\sigma}a^2,
\end{equation}
for $\rho=n\langle{}m\rangle{}$ and $m_3\ll M_1,M_2$, we have
\begin{equation}
N_{\rm int}=-\int_{a_i}^{a_f} \frac{ (M_1+M_2+m_3) }{\xi \langle{}m\rangle{} a}da\simeq \frac{ 4\pi (M_1+M_2) }{H \langle{}m\rangle{}}\log\left(\frac{a_i}{a_f}\right).
\end{equation}
Therefore, the binary has to interact with a mass in stars of the order of its own mass in order to shrink by an e-fold. This mechanism is thought to be important in the shaping of galactic nuclei. It has been in fact shown \cite{2006RPPh...69.2513M} that the low density cores in massive galaxies can be explained by merging MBHBs ejecting few times their own mass (\emph{i.e.}, up to several billion solar masses) in stars during the hardening process. 

\subsubsection{Other dynamical processes and merger rates}
We have shown that  hardening is a viable mechanism to form SBHBs separated by $a\sim R_\odot$, \emph{i.e.}, potential GW sources. There are, however, other competitive dynamical channels that have been put forward to form SBHBs, involving exchanges, ejections and interactions in triple and multiple systems.

{\bf Exchanges}. First, most SBHBs in star clusters tend to form in exchange interactions whereby a binary composed of a SBH and a star interacts with another SBH. The intruder, during the 3-body interaction, replaces the star in the binary. The final result is a SBHB and a single star. BHs are particularly efficient in acquiring companions through dynamical exchanges, because the probability of an exchange is maximized if the intruder is more massive than one of the members of the binary \cite{hills1980} and BHs are among the most massive objects in a star cluster (through N-body simulations, \cite{2014MNRAS.441.3703Z} find that 90\% BH-BH binaries in young star clusters form by exchange).

{\bf Ejections}. During three body interactions between a SBHB and a single star (or a single SBH), a part of the internal energy of the binary is extracted from the binary and converted  into kinetic energy  of the intruder and of the CoM of the binary. Both the intruder and the SBHB might experience a significant recoil and can be ejected from the star cluster. During the hardening process, the binary can also acquire a significant eccentricity, so that its coalescence timescale can be shorter than the Hubble time even if the SBHB is ejected from the cluster and does not experience any further interaction \cite{2016PhRvD..93h4029R}.

{\bf Hierarchical triples}. SBHBs can also be found in hierarchical triples with a companion on a wide orbit that does not interact strongly with the individual members of the binary,  thus preventing significant energy exchanges. In this case, however, angular momentum can be efficiently exchanged between the inner and the outer orbit of the triple. In particular, inclination of the outer orbit can be traded for eccentricity of the inner orbit via Kozai-Lidov oscillations \cite{1962AJ.....67..591K,1962P&SS....9..719L}. Depending on the relative inclination of the inner and outer orbits, the mechanism can be extremely efficient in driving the inner SBHB to $e>0.99$, at which points it swiftly merges due to GW emission [essentially because of the factor $F(e)$ in Eq.~(\ref{eq:AB})]. Antonini et al. (2016, \cite{2016ApJ...816...65A}) estimate that up to 10\% of dynamically formed SBHBs can merge in this way. A distinctive signature of such binaries is the extremely high eccentricity that will certainly be measurable by LISA and maybe also by ground-based detectors \cite{2016ApJ...816...65A,2017MNRAS.465.4375N}.

A different flavour of this scenario has been proposed in \cite{2018ApJ...856..140H}. Here SBHBs orbiting around a MBH undergo Kozai-Lidov oscillations because of the perturbation driven by the former. In practice we have an inner SBHB with a perturber MBH. This process can be extremely efficient in galactic nuclei and also results in extremely eccentric SBHB mergers.

In general, the estimated merger rate of SBHBs in globular clusters sits around ${\cal R}_{\rm SBHB} \approx 10$ Gpc$^{-3}$yr$^{-1}$ \cite{rodriguez2016,askar2016}, and up to ${\cal R}_{\rm SBHB} \approx 1-3$ Gpc$^{-3}$yr$^{-1}$ events might be due to hierarchical triples. These numbers set to the lower end of the LIGO estimated rate, $6< {\cal R}_{\rm SBHB} < 220$ Gpc$^{-3}$yr$^{-1}$, but are highly uncertain. 


Finally, we note that SBHBs are characterized not only by the orbital angular momentum but also by the spins of the individual BHs. Generally, systems formed through dynamical interactions among compact objects (dynamical formation channel) are expected to have isotropic spin orientations. Conversely,  binaries formed from the isolated binary evolution channel are more likely to have spins aligned with the binary orbital angular momentum, although this is an active area of research and several alternatives have been proposed to this naive picture (see, \emph{e.g.}, \cite{2017arXiv170607053B}). Therefore,  in principle, gravitational wave measurements of the binary spins (together with their eccentricity) may shed light on the formation of SBHBs. 


\newpage

\section{Supermassive black holes} \label{SMBH}
We now turn to discuss some relevant astrophysical aspects of \emph{(super)-massive black holes}, hereafter abbreviated as (S-)MBHs. It is customary to use the term MBH for BHs in the range $10^6 - 10^{10}M_{\astrosun}$. Those objects has been observed at the centre of massive galaxies, and inhabit virtually all nuclei of galaxies with $M_*>10^{11}M_{\astrosun}$, whereas their ubiquity in lighter galaxies is much debated. Our own galaxy, the Milky Way, hosts a MBH named Sagittarius A$^{\star}$ with $M\approx 4\times10^6M_{\astrosun}$.

\begin{figure}[h!]
  \centering
    \includegraphics[width=0.7\textwidth]{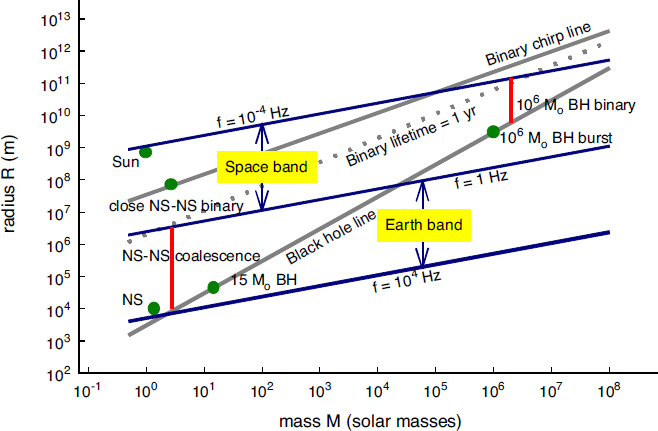}
    \caption{Characteristic frequencies estimated from Eq.~\eqref{frequency estimation} for  various gravitational waves sources of radius $R$ and mass $M$. From \cite{Sathyaprakash2009}.}
    \label{frequency band}
\end{figure}

To put things in the GW detection context, we start our discussion with a rough (Newtonian) order-of-magnitude estimate of the \emph{characteristic frequency} $f_c$ associated to a BH of radius $r_s$ and mass $M$ and defined as \cite{Sathyaprakash2009}
\begin{equation} \label{frequency estimation}
f_c \equiv \frac{\omega_c}{2\pi} = \frac{1}{2\pi} \sqrt{\frac{G M}{r_s^3}} \approx 1 \mbox{kHz} \left(\frac{10 M_{\astrosun}}{M}\right),
\end{equation}
where  $r_s  = 2 G M/ c^2$ is the Schwarzschild radius. In simple words, the characteristic frequency is inversely proportional to the mass of the objects. This means that MBHs are expected to be GW sources in a frequency range that is well below the ground-based frequency at $1 - 10^4$ Hz and space-based detectors are needed to detect them, as shown in Fig.~\ref{frequency band}.

\subsection{The first observation of a supermassive black hole}
The first observational evidence of a MBH dates back to 1963  \cite{1963Natur.197.1040S} with the identification of the quasar 3C 273 located at redshift $z=0.158$ with luminosity $L \approx 10^{46}$ erg s$^{-1}$. A quasar, or quasi-stellar-object (QSO), is an active galactic nucleus (AGN) consisting of a MBH powered by an accretion disk of gas, and it is an extremely bright object, usually at cosmological distance far from Earth.

Right after the discovery, astronomers (wrongly) assumed that 3C 273 was a star. It was soon clear that this assumption could not be correct. Indeed, assuming 3C 273 to be a massive star, the mass-luminosity power-law relation $L \propto M^{3.5}$ would imply a stellar mass in the range $10^4-10^5M_{\astrosun}$, which is well above the mass of any other observed star. Moreover, a star powered by nuclear reactions, thus shining at a luminosity $L=\eta \dot{M}c^2$ with $\eta=0.007$ would result in $\dot{M}\approx 10 M_{\astrosun}$ yr$^{-1}$. Therefore assuming that 10\% of the stellar mass is processed by nuclear reactions in the core, such star would survive about 100 yr.
Another puzzling property of 3C 273 was its spectral energy distribution (SED). It was not that of a black body, which is a good approximation for a star, but it was pretty flat (from radio to $\gamma$-ray frequencies, as we know today). Having discarded the star-like nature of 3C 273, it was noticed that its luminosity was compatible with that of a galaxy. Indeed, it was about hundreds of times the luminosity of the Milky Way $L_{\rm MW} = 10^{11} L_{\astrosun} = 10^{44}$ erg s$^{-1}$. But even this proposal was rejected because of the variability on a timescale of days, which was incompatible with that of a galaxy.

The solution to the puzzle of the nature of 3C 273 came from the variability scales of its luminosity. Let $\delta t_{\rm var} \approx $ days be the variability of the luminosity of 3C 273; then the characteristic size of the emitting object is $r_c < c~\delta t_{\rm var} \approx 0.01$ light-year, which is comparable with the size of the solar system. In other words, 3C 273 is an object with a luminosity of hundreds of time that of the Milky Way and with a size of our solar system or, more astonishingly, it is like an object 'burning' $10^{13}$ suns inside the solar system.

In order to explain the total luminosity, in 1969 Lynden-Bell proposed a model consisting of a MBH, located at the centre of the host galaxy, accreting the surrounding matter \cite{LyndenBell:1969yx} . The model describes a mechanism, the accretion process onto the MBH, in which the accreting matter forms a disk-like object - the accretion disk - where loss of angular momentum due to viscosity effects heats the gas that radiates away efficiently its gravitational energy, eventually vanishing into the MBH horizon. The model can accommodate both the energetic and the time variability of the source. For a $M=10^9M_{\astrosun}$, $R_S=3\times 10^{14}$ cm, since most of the luminosity comes from the inner regions of the accretion disk, within $10-100 R_S$, it is reasonable to expect variability on a timescale $t=(10-100 R_S)/c\approx 1-10$ days.

\subsection{Basics concepts of accretion}

\subsubsection{Bondi accretion}
In order to accrete, a MBH needs to capture gas from its surroundings at a sufficient rate. The problem of accretion onto a compact object was tackled by Hoyle, Lyttleton and Bondi in 1940s and then refined by Bondi in 1952 \cite{Bondi:1952ni}. We refer to this latter work in the following.


The model assumes an object (a black hole in our case) of mass $M$ surrounded by an infinite cloud of gas, accreting with stationary and spherically symmetric motion. The model neglects any self-gravity effects of the cloud, magnetic fields, angular momentum and viscosity due to the accretion mechanism. The gas is assumed to be perfect and polytropic, namely
\begin{equation} \label{perfect gas}
\frac{p}{p_{\infty}} = \left( \frac{\rho}{\rho_{\infty}}\right)^{\gamma},
\end{equation}
where $ 1\leq\gamma\leq 5/3$ is the polytropic index, while $p_{\infty}$ and $\rho_{\infty}$ are, respectively, the constant pressure and the constant density of the gas far from the black hole. Because of the stationarity and spherical symmetry, the conservation of rest-mass $\nabla_{\mu} (\rho u^{\mu}) =0$ implies 
\begin{equation} \label{continuity}
\dot{M} = \mbox{const} = 4 \pi r^2 \rho v,
\end{equation}
where $r$ is the radial coordinate, $v = - u^{r}$ is the inward velocity of the gas, and $\dot{M}$ is the constant of integration, which is the \emph{accretion rate}.
The conservation of energy, or the Bernulli's equation for compressible fluid, reads as
\begin{equation} \label{Bernoulli}
\frac{v^2}{2} + \int_{p_{\infty}}^{p}\frac{dp'}{\rho(p')} - \frac{GM}{r} =\mbox{const} = 0.
\end{equation}
The constant is set to zero if the gas is at rest far from the black hole. By inserting Eq.~\eqref{perfect gas} into Eq.~\eqref{Bernoulli}, the Bernoulli's equation becomes
\begin{equation} \label{Bernoulli2}
\frac{v^2}{2} + \int_{p_{\infty}}^{p}\frac{dp'}{\rho(p')} - \frac{GM}{r} =
\begin{cases}
\frac{v^2}{2} +  \frac{\gamma}{\gamma -1} \frac{p_{\infty}}{\rho_{\infty}}\left[\left(\frac{\rho}{\rho_{\infty}}\right)^{\gamma - 1} -1\right] - \frac{GM}{r} =0 &\qquad \gamma \neq1 ,\\
\frac{v^2}{2} + \frac{p_{\infty}}{\rho_{\infty}}\ln \left(\frac{\rho}{\rho_{\infty}}\right)  - \frac{GM}{r} =0, &\qquad \gamma = 1.
\end{cases}
\end{equation}
To simplify the two governing equations \eqref{continuity} and \eqref{Bernoulli2}, we introduce the speed of sound in the gas at infinity
\begin{equation}
c^2_{s} = \gamma \frac{p_{\infty}}{\rho_{\infty}},
\end{equation}
the characteristic length-scale of the problem, the \emph{Bondi radius}
\begin{equation}
r_{B} = \frac{GM}{c^{2}_{s}},
\end{equation}
and we rescale the variables
\begin{align}
r &= x r_B,\\
v &= y c_{s}, \\
\rho &= z \rho_{\infty}.
\end{align}
The variable $y$ plays the role of the Mach number.
Simple algebra transforms Eqs.~\eqref{continuity} and \eqref{Bernoulli2} for $\gamma \neq 1$ to
\begin{equation} \label{Bondi system}
\begin{cases}
x^2yz = \lambda,\\
\frac{y^2}{2} + \frac{z^{\gamma -1} -1}{\gamma -1} - \frac{1}{x}=0,
\end{cases}
\end{equation}
where the dimensionless accretion rate parameter $\lambda$ is given by
\begin{equation} \label{accretion parameter}
\lambda  = \frac{\dot{M}_{B}}{4\pi r^2_{B}c_s \rho_{\infty}} = \frac{\dot{M}_{B}c_{s}^3}{4\pi G^2 M^2 \rho_{\infty}}. 
\end{equation}
Notice that for $\lambda =1$, we get the Bondi's accretion rate
\begin{equation}
\dot{M}_{B} = \frac{4\pi G^2 M^2 \rho_{\infty}}{c_{s}^3} = 2.5\times 10^{2}\left(\frac{c_s}{100 {\rm km\, s}^{-1}}\right)^{-3}\left(\frac{M}{10^8 M_{\astrosun}}\right)^2\, {\rm M}_{\astrosun}\,{\rm yr}^{-1}.
\end{equation}

Therefore, accretion from an uniform distribution of gas is proportional to $M^2$ and inversely proportional to the third power of $c_s$. Note that $\dot{M}\propto M^2$ implies that the mass of the compact object diverges to infinity in a finite amount of time for an infinite fuel supply. The Bondi accretion rate has a large variety of applications from HMXRB, to planet formation, and it also provides a good estimate of the rate at which a MBH can capture gas in a low density environment. However, it assumes perfectly spherical geometry and does not take into account for any feedback due to the radiation emitted by the accretion flow. The latter, in particular, is of capital importance as it imposes a maximum rate at which mass can be accreted that uniquely depends on the accretor mass and not on the properties of the accreted gas, as we now see. 



\subsubsection{The Eddington limit}
The radiation reaction onto the accretion flow is the physical rationale behind the \emph{Eddington accretion limit}, and sets the maximum luminosity $L$ that an AGN (or, in fact, any astrophysical object) can emit when the radiation force acting outward equals the gravitational force acting inward. Beyond this limit, the radiation force overwhelms the gravitational force and the accretion process is considerably softened or halted.

Let assume that the gas around the MBH consists of a spherically distributed cloud of fully ionized hydrogen, so that photons and electrons interact among each other via Thomson scattering. At a given radial distance $r$ from the accretor, the flux of energy through the spherical surface of radius $r$ is given by
\begin{equation}
\Phi = \frac{L}{4 \pi r^2},
\end{equation}
and, since for a photon $L = dE/dt = c~dp/dt $, the momentum flux is
\begin{equation}
p_{\rm rad} = \frac{\Phi}{c} = \frac{L /c}{4 \pi r^2 }.
\end{equation}
Therefore, the force exerted by the radiation on a single electron is given by
\begin{equation}
F_{\rm rad} = p_{\rm rad}~\sigma_T = \frac{L /c}{4 \pi r^2 } \sigma_T, \quad \sigma_{T}=\frac{8\pi}{3}\left( \frac{e^2}{m_e c^2}\right)^2 = 6.65 \times 10^{-25} \mbox{cm}^2,
\end{equation}
where $\sigma_{T}$ is the Thomson cross-section for electrons. The dependence of $\sigma_{T}$ on the particle mass justifies neglecting photon-proton scattering, since the cross section is suppressed by a factor $10^6$ compared to photon-electron interactions. However, because of electrostatic forces between electrons and protons, the latter will be carried away along with electrons. The gravitational force, on the other hand, is
\begin{equation}
F_{\rm grav} = \frac{G M (m_e + m_p)}{r^2} \approx \frac{G M m_p}{r^2}.
\end{equation}
By equating the two opposite forces and by solving for $L$, we get
\begin{equation} \label{Edd luminosity}
L_{\rm Edd} = \frac{4\pi G m_{p} c}{\sigma_T} M = 1.26\times10^{38}\left(\frac{M}{M_{\astrosun}}\right)\,{\rm erg\,s}^{-1}.
\end{equation}
Hence, it is immediate to infer that if the luminosity of an AGN is $L \approx 10^{46}$ erg s$^{-1}$, then the mass of the accretor must be $M > 10^8$ M$_{\astrosun}$, supporting the MBH accretion interpretation of QSOs and AGNs.

Now, assume that the accretion process occurs at a given rate $\dot{M} = dM/dt$, which is the \emph{mass accretion rate}, and that a fraction $\epsilon$ of the rest mass energy of the accreted matter is radiated away. Then, the luminosity can be expressed as a fraction of the rest-mass accretion rate
\begin{equation} \label{luminosity}
L = \epsilon \dot{M} c^2.
\end{equation}
This relation implies a limit on the accretion rate, namely the \emph{Eddington accretion rate}, which is defined by
\begin{equation} \label{Edd rate}
L_{\rm Edd} = \epsilon \dot{M}_{\rm Edd} c^2 \Longrightarrow \dot{M}_{\rm Edd}= \frac{4\pi G m_{p} }{\sigma_T c}  \frac{M}{\epsilon} = 2.2\times 10^{-8}\left(\frac{\epsilon}{0.1}\right)^{-1}\left(\frac{M}{M_{\astrosun}}\right)\, {\rm M}_{\astrosun}\,{\rm yr}^{-1}.
\end{equation}

\subsubsection{Radiative efficiency and emission processes}
The fraction $\epsilon$ is the \emph{radiative efficiency} of the accretion process and can be estimated from the energy loss of the accreted material. Without entering in too much details, accreted gas usually has a finite amount of angular momentum, causing it to settle into a disk-like geometry. A detailed treatment of accretion disk theory is well beyond the scope of these lectures, and an excellent pedagogic introduction can be found in \cite{1981ARA&A..19..137P,Abramowicz:2011xu}. For our purposes it is sufficient to say that a gas element joining the disk at large radii, loses energy and angular momentum through viscous processes, spiralling inwards until the inner rim of the disk. Viscosity heats up the disk and the heat is dissipated locally as a blackbody spectrum (for optically thick disks). For a gas element in a wide initial orbit we can safely assume $E_i=0$, but when the gas reaches the inner rim of the disk in an approximately circular orbit, its energy is given by $E_{\rm rim}=-GmM/(2r_{\rm rim})$, where $M$ is the accretor mass, $m$ is the gas element mass and $r_{\rm rim}$ is the radius of the inner rim of the disk. Therefore,
\begin{equation}
  \Delta{E}=-\frac{GmM}{2r_{\rm rim}}  \Longrightarrow L=-\frac{dE}{dt}=\frac{G\dot{M}M}{2r_{\rm rim}},
  \label{eq:rrim}
\end{equation}  
where $\dot{M}=-dm/dt$ is the mass accretion rate. From this we can estimate the efficiency $\epsilon$ by writing $r_{\rm rim}=\beta \left(2GM/c^2\right)$ so that
\begin{equation}
L=\frac{1}{4\beta}\dot{M}c^2=\epsilon \dot{M}c^2,
\end{equation}  
where we set $\epsilon=4\beta$. If the accretor is a MBH, then $\beta=3$ for a Schwarzschild BH and $\beta=1$ for a maximally spinning BH, resulting in $\epsilon=0.08$ and $\epsilon=0.25$ respectively. This is a very rough estimate based on Newtonian physics, the full GR calculation gives $\epsilon=0.057$ for a  Schwarzschild BH  and $\epsilon=0.42$ for a maximally rotating Kerr BH.

\begin{figure} [h!]
  \centering
    \includegraphics[width=0.6\textwidth]{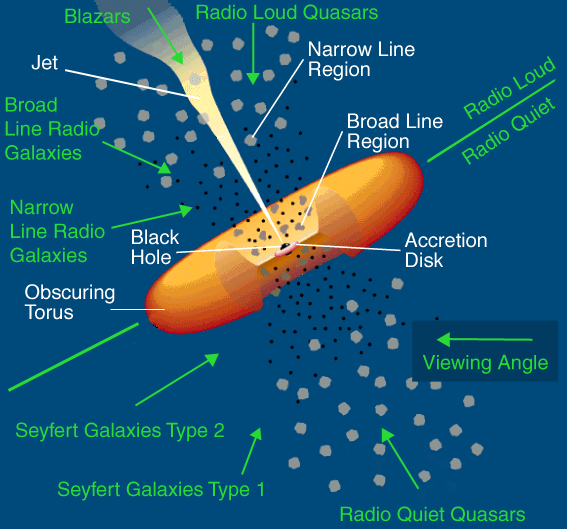}
    \caption{Cartoon showing the unified model of QSOs and AGNs. From \cite{1995PASP..107..803U}.}
    \label{fig:unified}
\end{figure}

Energy in each disc annulus is radiated as a black-body with a temperature
\begin{equation}
  T\approx 2\times 10^7\left(\frac{M}{M_{\astrosun}}\right)^{-1/4}r^{-3/4} {\rm K},
\end{equation}
where $r$ is the annulus radius normalised to the Schwarzschild radius $r_s$. From this we see that the inner rims (which are the more luminous) of SBHs emit at $T\approx 10^7$ K in the X-ray band. However, because of the mass dependence, MBHs emit the bulk of their luminosity at $T\approx 10^5$ K in the UV-optical band. The emitted radiation is partly re-processed across the electromagnetic spectrum by several mechanisms. Inverse Compton scattering against free electrons in the ionised corona (or atmosphere) surrounding the inner portions of the disk re-processes UV photons at higher energies, causing a non-thermal X-ray emission component \cite{1991ApJ...380L..51H}. A dust torus surrounding the accreting MBH, on the other hand, is responsible for re-processing part of the photons in the infrared \cite{1992ApJ...401...99P}. In the unified model of QSOs and AGNs \cite{1993ARA&A..31..473A}, depending on the orientation of the torus, radiation from the inner disk can either be blocked or travel to the observer. In the latter case we have a Type I AGN, characterised by continuum emission at all wavelengths (including optical and UV) and several broad emission lines; in the former case, we have a Type II AGN, characterised by the absence of broad lines or optical continuum, and by the prominent hard X-ray emission (only hard X-rays can penetrate the torus travelling to the observer). Finally, accretion can power jets \cite{1977MNRAS.179..433B} observed either in radio or X/$\gamma$-rays and winds \cite{2010A&A...521A..57T} that can exert a significant feedback on the host galaxy \cite{2012ARA&A..50..455F}. The cartoon in Fig.~\ref{fig:unified} shows the different accretion-powered emission processes operating in AGNs. 


\subsection{MBH mass measurements}
Besides powering QSOs and AGNs in general, ``dormant'' MBHs have also been detected at the centre of massive galaxies in the local Universe. This has led to the notion that virtually every galaxy hosts a MBH, and QSOs and AGNs represent only a particular phase in the lives of those MBHs, as we will see later.
In order to understand the difficulties encountered in MBH mass measurements, let us consider the length-scales of the system. Given a MBH of mass $M$ residing in the centre of a galaxy, we can consider several characteristic scales:
\begin{itemize}
\item The Schwarzschild radius
\begin{equation}
r_{s} = \frac{2GM}{c^2} \approx 3\times10^{5} \left(\frac{M}{M_{\astrosun}}\right) \mbox{cm}.
\end{equation}
As a remark, the Schwarzschild radius is not a well-defined geometrical quantity. Nevertheless, we define the Schwarzschild radius from the area of the event horizon, which is a well-defined quantity, by assuming a spherical geometry $A = 4 \pi r^2_{s}$. 
\item The influence radius, which is defined as that radial distance where the gravitational potential well of the object of mass $M$ overwhelms the internal energy of the surrounding stellar distribution
\begin{equation}
\frac{1}{2}\sigma^{2} = \frac{GM}{r_{i}} \Longrightarrow r_{i} = \frac{2GM}{\sigma^2} = r_{s}\left(\frac{c}{\sigma}\right)^2 \approx 10^6 r_s,
\end{equation}
where $\sigma$ is the velocity dispersion of the stellar distribution. For $M=10^9$M$_\odot$, $r_{i}\approx 30$ pc.
\item The Bondi radius, which is defined by a similar condition
\begin{equation}
\frac{1}{2}c_s^{2} = \frac{GM}{r_{i}} \Longrightarrow r_{i} = \frac{2GM}{c_s^2} = r_{s}\left(\frac{c}{c_s}\right)^2 \approx 10^6 r_s,
\end{equation}
where $c_s$ is the sound speed of the gas in the galactic nucleus, which is comparable to $\sigma$.
\item The typical radius of the galaxy $r_{\rm gal} \approx 100$ kpc.
\end{itemize}
From this four characteristic length-scales, it is evident that the effects of a MBH of $10^6 M_{\astrosun}\lesssim M \lesssim 10^9 M_{\astrosun}$ are measurable only in a tiny portion of the galaxy, from $10^{-6}$ to $10^{-3}$ of its typical radius. Therefore, the effects of a MBH are very difficult to observe in galaxies, unless it is located very close to us. To put things in perspective, consider that the best angular resolution achievable with HST is of the order of $1$ mas. This corresponds to about 1 pc spatial separation at a distance of 100 Mpc. It follows that the influence radius of MBHs is only well resolved within $\approx 100$ Mpc. For a local galaxy density of 0.01 Mpc$^{-3}$, there are only about $10^4$ galaxies in the local universe for which we can hope to resolve the MBH influence radius. The vast majority of them, however, is not massive enough, still $r_i$ can be resolved for $\approx 100$ galaxies, as we will see below.

MBH mass measurements may be classified in the following categories (see, \emph{e.g.}, the reviews \cite{2013ARA&A..51..511K,peterson2014measuring} and references therein for more details):
\begin{enumerate}
\item test particle measurements, either a) stars or b) megamasers;
\item ensemble motions , either a) spatially resolved of b) temporally resolved;
\item virial mass estimates.
\end{enumerate}

\subsubsection{Test particle measurement}
\paragraph{Resolved stellar motions.} The ``easiest'' way to measure a MBH mass is by resolving the Keplerian motion of a star in orbit around it. This technique is currently applicable to one case only, that of SgrA$^{*}$, the MBH at the centre of our galaxy. By imaging the orbit of stars in the MW centre and taking their spectra, we can reconstruct the projected 2D position and 2D velocity (from imaging), and the radial velocity (from the spectra). The six degree of freedom is then fixed by matching the position of SgrA$^{*}$ with the focus of the ellipse and the orbit is fully reconstructed. The star S2 in the MW centre has a period of about 15 yr, and sits in an orbit with periastron $r_p\approx 2000 r_s$ and $e=0.88$. The full determination of its orbit (along with measurements from other stars) has led to a robust estimate of $M \approx 4 \times 10^{6}$ M$_{\astrosun}$ for the SgrA$^{*}$ mass; see, \emph{e.g.}, \cite{2009ApJ...692.1075G}.

\paragraph{Megamasers.} The word maser derives from the acronym MASER, which stands for Microwave Amplification by Stimulated Emission of Radiation. An atom or molecule may absorb a photon and move to a higher energy level, or the photon may stimulate emission of another photon of the same energy causing a transition to a lower energy level. Producing a maser requires population inversion, \emph{i.e.}, a system with more members in a higher energy level relative to a lower energy level. Water maser emission is observed primarily at 22 GHz, due to a transition between rotational energy levels in the water molecule. In practice, photons with the right frequency $\nu$ stimulate emission of photons at the same frequency causing an exponential amplification of intensity
\begin{equation}
  I_\nu=I^0_\nu ~e^{-\int_0^Lk_\nu dL},
\end{equation}
where $k_\nu<0$ is a negative absorption coefficient due to stimulated emission and $L$ is the coherency length of the process, \emph{i.e.}, the length over which the energy of the photons matches the difference of the water molecule energy levels involved in the stimulated emission.

\begin{figure} [h!]
  \centering
    \includegraphics[width=0.7\textwidth]{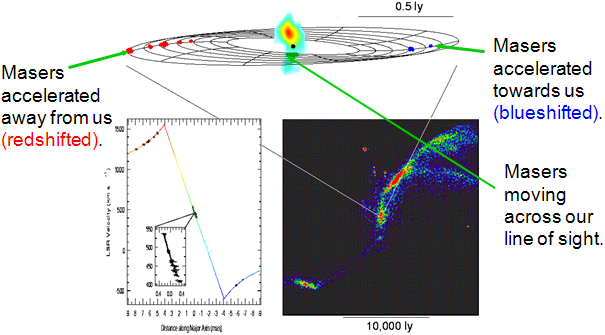}
    \caption{from http://astronomy.swin.edu.au/cosmos/V/VLBI. Measured maser rotation curve in NGC 4258.}
    \label{fig:maser}
\end{figure}
The phenomenon is observed in edge disks hosting molecular clouds with water molecules. The rest frame frequency of the maser line is 22 GHz and by measuring the Doppler shift due to the Keplerian motion of the clouds, the rotation curve of the disk can be measured, allowing the estimate of the MBH mass. Note that maser emission occurs only along the major axes (at 90$^0$ with respect to the observer line of sight) of the edge on disk and the nearest semiminor axis. This is because along those axes geometry favours a longer coherency length $L$. In fact, if a photon  at 22 GHz is emitted at an angle with respect to the observer, it encounters molecules with different radial velocities along the line of sight to the observer and Doppler shifts prevent maser amplification to occur. Fig.~\ref{fig:maser} shows the rotation megamaser curve measurement of NGC 4258, at 7.6 Mpc from us. The system is resolved with sub-arcsec precision and the mass of the MBH is estimated as $M \approx 4 \times 10^{7} M_{\astrosun}$ at a few \% level \cite{1995Natur.373..127M}. Megamasers have been exploited in about 10 MBH mass measurements.

\subsubsection{Ensemble motions}
\paragraph{Spatially resolved.} Whenever the resolution of the measurement is enough to observe the velocity dispersion of the stellar distribution $\sigma$ down to the sphere of influence of the MBH, we can infer the presence of a compact object in the centre of the galaxy as follows. Assume a spherical symmetric distribution of stars in virial equilibrium. Then, the motion of a star at distance $r_{*}$ from the centre of the distribution is dictated by the total mass inside the sphere of radius $r_{*}$, let us say $M_{r < r_{*}}$. The stellar velocity dispersion is thus given by
\begin{equation}
\sigma(r; r_{*}) = \sqrt{\frac{2 G M_{r < r_{*}}}{r}} \underset{r \rightarrow 0} {\longrightarrow}
\begin{cases}
0 \quad &\mbox{if} \quad M_{r < r_{*}} = M_{\rm galaxy} \rightarrow 0,\\
r^{-1/2}\quad &\mbox{if} \quad M_{r < r_{*}} = M_{\rm galaxy} +M_{\rm BH} \rightarrow M_{\rm BH}.
\end{cases}
\end{equation}
In the first case, the mass inside the sphere of radius $r_{*}$ is simply given by the mass of the stellar distribution, which goes to zero as r approaches the centre of the distribution. The exact limit deepens on the density profile, but for typical stellar distributions with $\rho(r)\propto r^{-\gamma}$ with $\gamma<2$, $M(r)\rightarrow 0$ faster than $r$, and consequently the velocity dispersion vanishes. An observation of this behaviour implies that there is no compact object or MBH inside the stellar distribution. In the second case, due to the presence of a MBH, the velocity dispersion shows a power law behaviour in the radial profile. The $r^{-1/2}$ behaviour starts to be evident for radii $r < r_{i}$ smaller than the radius of influence of the central object. The same technique can be used to measure the rotation curve of gas sitting in circumnuclear disk (if present). The masses of about 70 MBHs in local galaxies at distance $<100$ Mpc have been obtained in this way.

\begin{figure} [h!]
  \centering
    \includegraphics[width=0.6\textwidth]{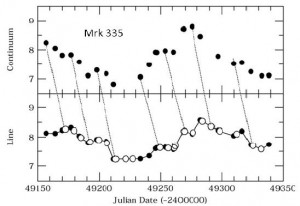}
    \caption{Delayed line variability with respect to the continuum floor in the lightcurve of Mrk 335. Figure taken from \cite{2001sac..conf....3P}.}
    \label{fig:reverberation}
\end{figure}

\paragraph{Temporally resolved.} This type of measurement uses the technique of {\it reverberation mapping}
\cite{1993PASP..105..247P}. The technique requires the measurement of both continuum and line emission and it is therefore applicable only to type I AGNs. The intensity of emission of both components is taken at a series of epochs, thus constructing a light-curve as the one shown in Fig.~\ref{fig:reverberation}. The emission line variability shows a time lag with respect to the continuum. This is because lines are emitted from atomic gas in the so called ``broad line region'' (BLR), which sits at some distance from the central engine, whereas the ionised gas responsible for the continuum emission is located at few $r_s$. Line intensity varies in response to the continuum variability, with a time lag $\Delta{t}$ that can be associated to the distance of the BLR from the central engine via $R_{\rm BLR}=c\Delta t$. From the full width half maximum (FWHM) of the broad lines the velocity $v$ of the gas can be inferred. Therefore, assuming virial equilibrium, the MBH mass is simply obtained as
\begin{equation}
  M = \frac{f R_{\rm BLR} v}{G},
  \label{eq:rev}
\end{equation}
where the factor $f$ accounts for the unknown geometry of the BLR. For example, for a spherically symmetric distribution of gas $f=\sqrt{3}/2$, whereas for a disk inclined by an angle $\theta$ with respect to the line of sight $f=1/(2\sin\theta)$. 

\subsubsection{Virial estimates and the MBH mass ladder}
Reverberation mapping is a powerful technique, in principle applicable to a large number of Type I AGNs. It is, however, extremely time consuming, requiring multiple spectra of the source to be taken at different epochs. The culprit is the evaluation of $R_{\rm BLR}$, which requires data-taking at multiple epochs in order to measure $\Delta{t}$. But once $R_{\rm BLR}$ is measured for a sizeable sample of objects, correlations can be found with quantities that are easier to measure. In particular, it was found that $R_{\rm BLR}\propto L^{0.5}$ \cite{2000ApJ...533..631K,2005ApJ...629...61K,2009ApJ...697..160B}, which opens the door to ``single epoch'' mass measurements: by taking a single spectrum, $v$ and $L$ are measured, from $L$ one can derive $R_{\rm BLR}$ and the MBH mass is readily measured via Eq.~(\ref{eq:rev}). Single epoch measurements allowed mass estimates for tens of thousands of objects, although particular care needs to be taken in addressing sources of errors and systematic biases \cite{2008ApJ...680..169S}.

All the techniques described thus far can be combined in constructing a MBH mass ladder that allows the estimate of MBH masses hosted by millions of galaxies, as we now sketch:
\begin{itemize}
\item The masses of $\approx 100$ objects within 200 Mpc can be {\it directly} estimated via spatially resolved measurements.
\item MBH masses were then found to correlate with easy-to-measure galaxy properties like the velocity dispersion $\sigma$ and the stellar bulge mass $M_*$. These are the notorious $M-\sigma$ and $M-M_*$ relations (see \cite{2013ARA&A..51..511K} and references therein). The relations suffer a small degree of intrinsic dispersion of about $0.3$ dex (\emph{i.e.}, a factor of two) and they can be used to perform an {\it indirect} measure of MBH masses in millions of quiescent galaxies.
\item Masses estimated via the $M-\sigma$ and $M-M_*$ can then be used to calibrate reverberation mass measurements, to pin down the geometric factor $f$. This was done in \cite{2004ApJ...615..645O} where e scatter of about 0.5 dex (\emph{i.e.}, a factor of three) was estimated for reverberation map measurements.
\item The $R_{\rm BLR}-L$ relation (intrinsic scatter 0.4 dex) can finally be anchored to the calibrated reverberation mass measurements to estimate the masses of millions of type I AGN \cite{2006ApJ...641..689V}. 
\end{itemize}
The ``ladder'' allows to measure masses of millions of MBHs in different environments (active and inactive galaxies), it should be noticed, however, that indirect measurements rely on calibrations that bear an intrinsic scatter of a factor of two or more, and are generally derived for local sources (i.e. at redshift zero). It is probably safe to assume that indirect measurements of individual MBHs are accurate within a factor of three at most.

\subsection{Unified model of MBH evolution}

Since the discovery of the first quiescent MBHs in galactic nuclei, a link had been made with QSOs and AGN. The idea is that MBHs indeed inhabit all galaxies and only sporadically accrete gas efficiently thus shining as quasars. 

\subsubsection{$M - \sigma$ relation: accretion feedback?}
\label{sec:msigma}
Some insights about the connection between quiescent and accreting MBHs come from the $M - \sigma$ relation, which puts in correlation the mass of MBHs and the velocity dispersion of the gas of stars in the host galaxy. The most popular explanation of the relation relies on accretion feedback regulating the growth of the MBH. The main idea is that MBHs accrete until their mass sits on the relation. At this point, the radiation output is strong enough to unbind the gas reservoir, thus halting accretion. If this physical interpretation holds, then the $M - \sigma$ relation would be evidence of a direct link between accreting and quiescent MBHs. The fact that MBH accretion has the potential of affecting the host galaxy at large can be simply understood with the following argument based on energetics. We have seen in Eq.~\eqref{luminosity} that the luminosity is a fraction of the rest-mass accretion rate $\dot{M}$, \emph{i.e.}, $L = \epsilon \dot{M}_{\rm acc} c^2$. Since a fraction $\epsilon$ of the mass accreted $M_{\rm acc}$ is radiated, the mass stored into the MBH is $\dot{M}_\bullet=(1-\epsilon)\dot{M}_{\rm acc}$. Hence, the total energy radiated during the accretion process can be related to the mass gained by the MBH via
\begin{equation}
  E  = L T = \frac{\epsilon}{1-\epsilon} M_{\bullet} c^2,
  \label{eq:energyacc}
\end{equation}
where $M_{\bullet}$ is the mass accumulated by the MBH via accretion and $T$ is the whole time in which the accretion process has taken place. On the other hand, the binding energy of the whole galaxy is of the order of $E_G=M\sigma_*^2$, where $M$ is the total galaxy mass and $\sigma_*$ is the velocity dispersion. By taking the ratio of the two, we get
\begin{equation}
\frac{E}{E_G} = \frac{\epsilon}{1-\epsilon} \frac{M_{\bullet}}{M} \left(\frac{\sigma_*}{c}\right)^{-2} \approx 250  \left(\frac{\sigma_*}{200\,{\rm km s}^{-1}}\right)^{-2},
\end{equation}
where in the last passage we have assumed that the MBH grows to $M_\bullet = 10^{-3}M$. Therefore, growing MBHs with masses typically seen today, injects plenty of energy into the parent galaxy to strongly affect its evolution. We need to see whether the enormous energy output of the MBH can efficiently couple with the host galaxy.

Here, we give a derivation of the $M - \sigma$ relation by following the original paper of Silk \& Rees \cite{Silk:1997xw} and the subsequent work of King \cite{King:2003ix}. We describe the spatial distribution of stars and clouds around the MBH as a singular isothermal sphere. In other words, we assume that the phase space distribution function is a Gaussian given by
\begin{equation}
  \rho(v; \sigma) = \frac{1}{(2 \pi \sigma^{2})^{3/2}} e^{-\left(\frac{v^2}{2}+\Phi\right)/\sigma^2},
  \label{eq:distfunc}
\end{equation}
where $\frac{v^2}{2}+\Phi$ is the total energy per unit mass, with $\Phi$ being the potential energy and $\sigma$ the velocity dispersion. In the following, we assume $\sigma$ to be a constant. By integrating over the all the three-dimensional velocities, we get the mass density of the stellar distribution
\begin{equation}
\rho_{\rm mass} \equiv \int_{0}^{\infty}\int_{0}^{2\pi}\int_{0}^{\pi}\rho(v; \sigma) v^2 \sin(\theta) dv d\theta d\phi = e^{-\Phi/\sigma^2}.
\end{equation}
This mass distribution is the source of the gravitational potential of the stellar distribution. Therefore, from the Poisson equation
\begin{equation}
\nabla^2 \Phi = 4 \pi G \rho_{\rm mass} \Longrightarrow \frac{d}{dr}\left(r^2 \frac{d \ln \rho_{\rm mass}}{dr}\right) =4 \pi G \frac{r^2 \rho_{\rm mass}}{\sigma^2},
\end{equation}
and assuming the power-law ansatz, we get the singular isothermal mass distribution
\begin{equation}
\rho_{\rm mass}(r) = \frac{1}{2\pi G} \frac{\sigma^2}{r^2}.
\end{equation}
Then, the total mass inside a sphere of radius $r$ is given by
\begin{equation} \label{M isothermal}
M(r) =4\pi \int_{0}^{r} \rho(r')_{\rm mass}r'^2dr'  = \frac{2\sigma^2 r}{G}.
\end{equation}
Note the unphysical properties of the isothermal mass distribution: first, it is singular in $r=0$ and, second, it leads to a divergent mass for $r\rightarrow \infty$. Nevertheless, those pathologies can be easily cured by modifying the distribution function to obtain a finite-density core and by introducing a cutoff at large radii. Overall the isothermal sphere is a reasonable approximation for typical stellar bulges (\emph{e.g.}, the MW bulge has a density profile $\rho(r)\propto r^{-1.8}$ \cite{2007A&A...469..125S}), so we keep using it because, despite its simplicity, it is able to catch the main physical aspects leading to the $M - \sigma$ relation.

The underlying idea behind the $M - \sigma$ relation is that
accretion onto the MBH can produce a sufficiently intense radiation outflow which, in turn, interacts with the accreting cloud of gas \cite{King:2003xf} unbinding it from the galaxy. If the radiation transfer its momentum to the gas, then assuming a constant outflow velocity, the momentum change of the outflow is given by
\begin{equation}
\dot{M}_{\rm out}v_{\rm out} \approx \frac{L_{\rm Edd}}{c}.
\end{equation}
Since $v_{\rm out}$ is constant, the shell's equation of motion takes the form 
\begin{equation}
\frac{d}{dt}\left[M_{\rm out}(r) v_{\rm out}\right] = \dot{M}_{\rm out}(r) v = \frac{L_{\rm Edd}}{c}.
\end{equation}
Note that $\dot{M}_{\rm out}(r) = 4\pi r^2 \rho(r) v^2_{\rm out}$ by conservation of rest-mass. Integrating both sides over time and setting the constant of integration equal to zero (because for relevant timescales it is negligible) one finds
\begin{equation}
M_{\rm out}(r) v_{\rm out} = \frac{L_{\rm Edd}}{c} t.
\end{equation}
By setting $M_{\rm out}(r) = f_g M(r)$ with $f_g\approx 0.16$ (the exact number is not important) being the gas fraction and $M(r)$ the total mass as computed by means of the isothermal mass distribution \eqref{M isothermal}, the above equation becomes
\begin{equation} \label{v wind}
v^2_{\rm out} = \frac{G L_{\rm Edd}}{2 c} \frac{1}{f_g\sigma^2},
\end{equation}
where we approximated $r/t \approx v_{\rm out}$.

We observe that the accretion process occurs very efficiently until it is quenched by the outflow radiation wind. This quenching effect starts to be important as soon as the outflow wind is not anymore bound to the MBH, namely when $v_{\rm out} = \sigma$.
Therefore, from Eq.~\eqref{v wind} with $v_{\rm out} = \sigma$ and Eq.~\eqref{Edd rate}, we solve for the mass of the MBH and find an $M - \sigma$ relation of the form
\begin{equation}
  M_{\rm BH} = \left(\frac{\sigma_T/ m_p}{2\pi G^2}\right) f_g\sigma^4 \approx 1.8 \times 10^{8} \sigma_{\rm 200} ^4 M_{\astrosun},
  \label{eq:king}
\end{equation}
where $\sigma_{200} = \sigma/(200 ~\mbox{km s$^{-1}$})$. Eq.~(\ref{eq:king}) is in reasonable agreement with relations inferred from observations (see a compilation in \cite{2018NatCo...9..573M}). Empirically, $M\propto \sigma^\alpha$ with $4<\alpha<5$. It is interesting to note that the derivation above, leading to $M\propto \sigma^4$ assumes a ``momentum driven'' outflow. In the limit of an ``energy driven'' outflow, as considered in \cite{Silk:1997xw}, the result would instead be $M\propto \sigma^5$, at the upper end of the measured range.  

\subsubsection{Soltan's argument}
An even more striking connection between QSOs and quiescent MBHs is provided by the Soltan's argument \cite{1982MNRAS.200..115S}, which compares the total mass of quiescent MBHs observed in galaxies today to the total accreted mass as inferred by the overall luminosity emitted by QSOs along the cosmic history. The argument has been later revisited in \cite{Yu:2002sq}.

We have seen in Eq.~\eqref{eq:energyacc} that the total energy radiated during the accretion process can be related to the mass gained by the MBH via $E  = L T = \epsilon/(1-\epsilon) M_{\bullet} c^2$, where $M_{\bullet}$ is the mass accumulated by the MBH via accretion and $T$ is the whole time in which the accretion process has taken place. Let $e(L, t)$ be the energy density (energy per unit volume) produced during the accretion process that we measure by observing the luminosity $L$ at time $t$ and let $\phi(L, t)$ be the luminosity function that counts the number density of MBHs having luminosity $L$ at a given time $t$. Thus, we have
\begin{equation}
e(L, t) dL dt =  \phi(L, t) L~dL dt.
\end{equation}
For observational reasons, it is easier to relate the luminosity function $\phi(L, t)$ to the number density of MBHs with observed flux $S = L/(4\pi r^2)$ at redshift $z$ as
\begin{equation}
n(S, z) dS dz = \phi(L, V) dL dV =  \phi(L, z) dL \frac{dV}{dz} dz,
\end{equation}
where $V(z)$ is the volume measured at redshift $z$. As a consequence,
\begin{equation}
e(L, t) dL dt = 4\pi r^2 n(S, z) ~S ~dS  \left(\frac{dV}{dz}\right)^{-1} dt = \frac{4\pi}{c} (1+z) ~n(S, z) S~ dS dz,
\end{equation}
where in the second step, we have used the expression $r^2 (dV/dz)^{-1} c dt = (1+z)dz$ valid in the Friedman expanding universe. Therefore, the total energy density radiated by accretion is given by
\begin{equation}
e = \frac{4\pi}{c} \int_{0}^{\infty} (1+z)dz \int_{0}^{\infty}n(S, z) S~ dS,
\end{equation}
and the accreted mass density during the optical bright phases is (recall Eq.~\eqref{eq:energyacc})
\begin{equation}
  \rho_{\rm acc} = \frac{1-\epsilon}{\epsilon}\frac{e}{c^2}.
\end{equation}
Notice that, though we used the Friedman cosmological model in the derivation, the final result is independent of it. The energy density, or the accreted mass density, depends only on the efficiency of the accretion process.
The function $e$ can be empirically evaluated by observationally determining $n(S, z)$; this yields
\begin{equation}
\rho_{\rm acc}  \approx 2.2 \times 10^5 \left(\frac{0.1}{\epsilon}\right) \frac{M_{\astrosun}}{Mpc^3}.
\end{equation}
Note that the above estimate does not account for obscured, type II AGN, which account for about half of the AGN population. Including them in the calculation brings the estimate to \cite{2009ApJ...690...20S}
\begin{equation}
\rho_{\rm acc}  \approx 4.5 \times 10^5 \left(\frac{0.1}{\epsilon}\right) \frac{M_{\astrosun}}{Mpc^3}.
\end{equation}
The density of quiescent MBHs in the local universe can be estimated from the local galaxy stellar mass function via the $M-M_\star$ relation. This yields 
\begin{equation}
\rho_{\rm BH}  \approx 3-5 \times 10^5 \frac{M_{\astrosun}}{Mpc^3}.
\end{equation}
By comparing the two expressions, we see that the mass density of MBHs we see today in the universe is totally accounted for by radiative efficient accretion with $\epsilon \approx 0.1$ during the QSO phase, thus strengthening the notion that MBHs lurking in galaxy centres in the local universe are the relics of QSOs that we mostly observe at high redshift (the peak of QSO activity is at $z\approx 2$).

\subsubsection{MBH growth along the cosmic history}
Within the last two decades, a picture emerged in which the MBHs we observe in galaxies today descend from BH {\it seeds} at high redshift to become supermassive as they accrete gas and stars and possibly merge with other compact objects, including other MBHs. The need to start from a seed with $M<10^6$M$_\odot$ is dictated by the fact that no physical mechanism able to monolithically form a billion stellar-mass compact object is known. Moreover the Soltan's argument favours a picture in which most of today's MBH mass has been accreted along the cosmic history. 

We now compute the pace at which MBHs can acquire their mass through accretion. From the definition of Eddington luminosity we can define the Eddington timescale
\begin{equation} \label{eq:tedd}
  t_{\rm Edd}=\frac{Mc^2}{L_{\rm Edd}}=\frac{\sigma_Tc}{4\pi Gm_p}=0.45~{\rm Gyr}.
\end{equation}
The evolution equation for the MBH mass can be then written as \cite{0004-637X-620-1-59}
\begin{equation} \label{eq:mdotbh}
\dot{M} = (1- \epsilon)\dot{M}_{\rm acc} = \frac{1-\epsilon}{\epsilon}f_{\rm Edd}\frac{M}{t_{\rm Edd}},
\end{equation}
where $f_{\rm Edd} = L/L_{\rm Edd}$ is the fraction of the Eddington luminosity being radiated. Eq.~(\ref{eq:mdotbh}) is readily integrated via variable separation to yield
\begin{equation}
M(t)=M_0 e^{\frac{1-\epsilon}{\epsilon}f_{\rm Edd}\frac{t-t_0}{t_{\rm Edd}}},
\end{equation}
where $M_0$ is the initial mass of the MBH at time $t_0$. By setting $f_{\rm Edd}=1$, we can now see what is the fastest rate at which a MBH can grow. This depends on the efficiency parameter $\epsilon$, which is directly related to the spin parameter $0\le a <1$ of the MBH. In fact, by assuming prograde accretion, the innermost stable circular orbit (ISCO) around a maximally rotating MBH 
is located at $GM/c^2$, meaning that the gas loses a much larger fraction of its rest mass energy before being accreted (and thus $\epsilon$ is much larger). By taking this into account one gets:
\begin{equation}
a=0\rightarrow \epsilon \approx 0.06 \rightarrow M=M_0 ~e^{\frac{t}{3\times 10^7{\rm yr}}},
\end{equation}
\begin{equation}
a=0.998\rightarrow \epsilon \approx 0.42 \rightarrow M=M_0 ~e^{\frac{t}{3\times 10^8{\rm yr}}}.
\end{equation}
For comparison, the age of the Universe is $t_{\rm Hubble}=1.4\times 10^{10}$ yr. Therefore, continuous Eddington limited accretion allows 50-to-500 e-folds in mass, depending on the MBH spin. This is more then enough to grow a $10^9 M_\odot$ MBH starting from any reasonable $M_0$.

\subsubsection{Seeding mechanisms}

A detailed description of the proposed MBH seeding mechanisms in protogalaxies at high $z$ is beyond the scope of these lectures. An overview of the subject can be found, \emph{e.g.}, in \cite{2012AdAst2012E..12S}. Here we just mention that three main mechanisms have been proposed in the literature:
\begin{itemize}
\item {\it Population III remnant}. Originally proposed by \cite{2001ApJ...551L..27M}, this scenario relies on the fact that the first generation of essentially metal free stars is expected to have a very top-heavy mass function. As we saw in Section \ref{sec:fragment} the absence of metal disfavours fragmentation and massive stars of $\approx 10^3$M$_\odot$ can form. The natural relics of such massive stars are BHs of several hundred solar masses. The viability of popIII remnant as seeds of the most massive MBHs have recently called into questions for a couple of reasons. First, fragmentation still occurs in gas with primordial metallicity, and popIII stars might not be as massive as originally thought, \emph{e.g.}, \cite{2011ApJ...737...75G}. Second, $\approx 100$M$_\odot$ remnant might not be massive enough to settle in the centre of the protogalaxy and tend to wonder in the outskirts of their parent halos, where the density of gas is too low to trigger significant accretion, \emph{e.g.}, \cite{2018arXiv180406477S}. 
\item {\it Runaway popIII mergers}. If clusters of massive stars are a common occurrence at high redshifts, runaway mergers might still result in the formation of a metal free star with  $M> 10^3$M$_\odot$, leaving behind an $\approx 10^3$M$_\odot$ BH remnant. This scenario was originally proposed in \cite{2009ApJ...694..302D} and has been subsequently revisited in \cite{2011ApJ...740L..42D,2014MNRAS.442.3616L}.
\item {\it Direct collapse}. Under this general label can be included a variety of models, often differing significantly in the key physical processes. The key idea is that in the most massive protogalactic halos at $z\approx 15$, gas accreted from the cosmic web can be supplied to the very centre at a rate of $\approx 1$M$_\odot$ yr$^{-1}$. Such extreme conditions can prompt the formation of a seed BH with mass in the range $10^4-10^5$M$_\odot$ either via collective infall from a marginally stable massive disk \cite{2006MNRAS.371.1813L} or via the formation of a quasi star \cite{2006MNRAS.370..289B} or via direct collapse \cite{2010MNRAS.402.1249S}.   
\end{itemize}

\subsubsection{Challenges to the standard MBH growth model}
One challenge to this simple picture is that MBHs of $10^9$M$_\odot$ are not observed only in the local Universe, but also at high redshift. The current record holder is ULAS J134208.10$+$092838.61, a MBH with mass approximatively of $10^9$M$_\odot$ at $z=7.54$ \cite{2018Natur.553..473B}, and the ultra-luminous QSO J215728.21$-$360215.1 has an estimated mass of $2\times10^{10}$M$_\odot$ already at $z = 4.75$\cite{Wolf:2018ivm}. Note that at $z=7.54$ the Universe was only 0.7 Gyr old. In such a short time, a Schwarzschild MBH can still grow by $(7\times 10^8)/(3\times 10^7)= 23$ e-folds in mass, \emph{i.e.}, by a factor $\approx 10^{10}$. However a maximally spinning MBH can accrete only by  $(7\times 10^8)/(3\times 10^8)= 2.3$ e-folds in mass, \emph{i.e.}, by a factor $\approx 10$. This is a problem because prolonged accretion inevitably results in highly spinning MBHs after about an e-fold in mass growth. It seems therefore unlikely that the highest redshift quasars grew by Eddington limited, prolonged accretion, independently on the seeding model.

The problem is mitigated by two observations. First, the $z>6$ QSOs are the most extreme and rare objects in the Universe, and are probably not be indicative of the standard evolution of MBHs. Conversely, The peak of QSO activity is at $z\approx 2$ \cite{2006AJ....131.2766R} when the Universe is more than 3 Gyr old, allowing much more time for mass growth. Second, the equilibrium spin parameter of an accreting MBH is not necessarily $a=0.998$ (a theoretical value computed in \cite{1974ApJ...191..507T}). For example, \cite{2004ApJ...602..312G} showed that the spin equilibrium of a MBH accreting from a thick disk in fully relativistic magnetohydrodynamic models settles onto $a\approx 0.93$. This result might be particularly relevant since for $L\rightarrow L_{\rm Edd}$, the accretion disk tends to puff-up in the inner regions, thus acquiring a tick geometry. Most importantly $\epsilon$ has a steep dependence on $a$ for $a\rightarrow 1$; in fact, for $a=0.93$ its value is only $\epsilon=0.18$, which implies an e-fold timescale of $\approx 10^8$ yr. Although this is still marginally insufficient to grow the most extreme objects at $z>7$ from a 'direct collapse' seed, it significant alleviates the tension between naive accretion models and QSOs observations.

We close by mentioning other two proposed ways to resolve this rapid growth issue. The first is the concept of 'chaotic accretion', first proposed in \cite{2005MNRAS.363...49K}. The idea is that gas does not have to be accreted in a well defined plane for the whole duration of the accretion process. If gas is accreted in 'pockets' with different orientation, the accretion flow will sometimes align and sometimes counter-align with the MBH spin, keeping its value small (thus keeping $\epsilon$ small and the e-fold time short). The second is the concept of super-critical accretion. In practice, in slim or puffed-up disks \cite{1988ApJ...332..646A}, $\epsilon$ might be much lower than the canonical value expected from thin disk accretion. This is because in the inner rims of the disk, the diffusion time of the photons can be longer than the viscous time, so that photons are 'advected' with the accretion flow. \cite{2014ApJ...784L..38M,2015ApJ...804..148V} showed that these models allow the MBH to grow to $M> 10^9~$M$_\odot$ in much less than a Gyr even starting from popIII seeds.

\subsection{Massive black hole binaries}

Despite growing most (possibly the vast majority) of their mass through accretion along the cosmic history, in the hierarchical structure formation framework \cite{1978MNRAS.183..341W} MBHs will also acquire some mass because of mergers with other MBHs. Galaxies are in fact observed to merge quite regularly, with massive galaxies experiencing at least a major merger (defined as a merger between two galaxies with mass ratio $>1/4$) within their lifetime \cite{2017MNRAS.470.3507M}. 

If each of the progenitor galaxies host a MBH, the outcome of a galaxy merger will be the formation of a MBHBs that eventually coalesces due to gravitational wave emission. The evolution of the system has been first described in \cite{1980Natur.287..307B} and can be divided into three stages:
\begin{figure} [h!]
  \centering
    \includegraphics[width=0.6\textwidth]{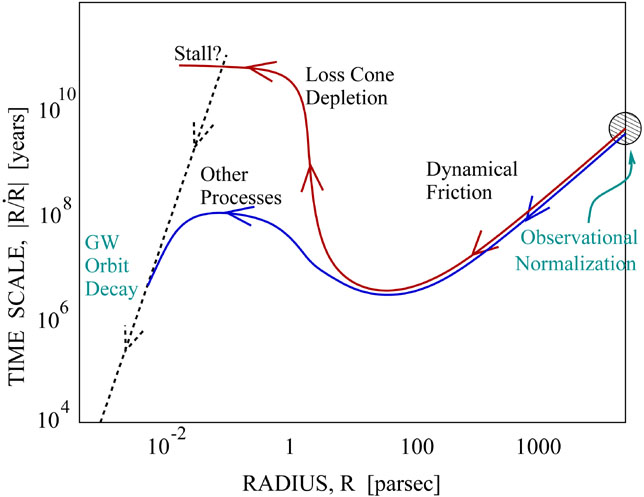}
    \caption{Cartoon showing the different phases of a MBHB evolution following a galaxy merger. Shown on the $y$ axis is the so called 'residence timescale, {\it i.e.} the timescale the binary reside at a given log frequency interval in separation.}
    \label{fig:hardening}
\end{figure}
\begin{enumerate}
\item dynamical friction (DF), 
\item hardening against the stellar and/or gaseous background,
\item gravitational wave inspiral.
\end{enumerate}  
The overall evolution is sketched in figure \ref{fig:hardening}. Initially, DF is efficient in bringing the two MBHs from kpc to pc separations, where they form a bound binary. At this point, DF becomes ineffective and binary hardening must proceed via other mechanisms, including the interaction with stars and gas in the dense nuclear environment. At sub-parsec separation, GW emission takes over, leading to swift coalescence of the MBHB. We now describe the stages of this evolution is some more detail.

\subsubsection{Dynamical friction}\label{sec:dynfric}
A massive object wandering in a sea of light perturber is subject to a collective force that undergoes the name of DF. A detailed first calculation of the force was carried out by Chandrasekhar \cite{1943ApJ....97..255C} and a step by step derivation can be found in \cite{binney2011galactic}, and we do not propose it here. The physical situation is a 'sea' of small particles with mass $m$ and velocity $v$ that interacts with a massive perturbing object of mass $M$ and velocity $V$ with respect to the CoM velocity of the sea of small particles\footnote{For the purpose of the calculation the particles are described by having random motion with collective null bulk velocity, and the massive perturber has a certain velocity $V$ with respect to the particle distribution.}. Since gravity goes with $1/r^2$ and the number of interactions in a uniform density distribution goes with $r^2$, it turns out that the sum of all Newtonian interactions with the light particles exert a collective force on the massive perturber, causing an acceleration given by
\begin{equation}
  \frac{d{\bf V}}{dt}=-16\pi^2{\rm ln}\Lambda\,G^2m(M+m)\frac{\bf V}{V^3}\int_0^Vf(v)v^2dv,
  \label{eq:chandra}
\end{equation}
where, $f(v)$ is the velocity distribution function of the light particles, ${\rm ln}\Lambda$ is the Coulomb logarithm already encountered in Section \ref{subsec:Relaxation_timescale} and bold symbols are used for 3-D vectors.

We can easily check the two limiting cases of equation~\eqref{eq:chandra}.
For $V\rightarrow 0$, $f(v)$ can be approximated with a constant, $f_0$, and the integral in the equation simply gives $f_0V^3/3$ so that
\begin{equation}
\frac{d{\bf V}}{dt}=-\frac{16}{3}\pi^2{\rm ln}\Lambda\,G^2m(M+m)f_0 \bold{V}.
\end{equation}
This has the form $dV/dt\propto -V$ of a Stoke force. So in the limit of small velocities, DF acts like a viscous type of friction.

The distribution function $f(v)$ is characterized by a typical velocity $\bar{v}$. If $V\gg \bar{v}$ then the integral in equation~\eqref{eq:chandra} is performed over the whole distribution function, returning the number density $n$ of the light particles  divided by $4\pi$. Equation~\eqref{eq:chandra} thus becomes  \cite{binney2011galactic}
\begin{equation}
  \frac{d{\bf V}}{dt}=-4\pi{\rm ln}\Lambda\,G^2M\rho\frac{\bf V}{V^3},
\end{equation}
where we defined the mass density $\rho=n\,m$ and considered the limit $M\gg m$. Therefore, at high velocities, $dV/dt\propto -V^{-2}$, and DF quickly becomes ineffective. Note also that $dV/dt\propto M$, so that $F=M(dV/dt)\propto M^2$. This is usually interpreted as the pull of a wake formed by DF {\it behind} the massive object. Gravitational focusing (See section \ref{sec:3body}) causes the formation of a wake with mass proportional to the perturber mass, so that DF can be interpreted as the gravitational pull of a wake of mass $M$ on the massive perturber (also of mass $M$), thus resulting in the characteristic $M^2$ dependence. Given these two limits, it is obvious that the efficiency of DF peaks for some value of $V$. Not surprisingly, it turns out that DF is most efficient when $V\approx \bar{v}$.

To get a sense of the effect of DF, let us consider the ideal situation of an isothermal sphere, that we already encountered in Section \ref{sec:msigma}. This is the equilibrium solution of a self-gravitating system characterized by a Maxwellian velocity distribution. We saw already that in this case $\rho(r)=\sigma^2/(2\pi G r^2)$, and it is straightforward to show that the circular velocity satisfies $v_c(r)=\sqrt{2}\sigma$, \emph{i.e.}, it is independent on $r$ (as a consequence of the fact that $\sigma$ is constant everywhere for an isothermal sphere). 
We can therefore use the distribution function given by Eq.~(\ref{eq:distfunc}) to write \cite{binney2011galactic}
\begin{equation}
  \frac{d{\bf V}}{dt}=-4\pi{\rm ln}\Lambda\,G^2M\rho\frac{\bf V}{V^3}\left[{\rm erf}(x)-\frac{2x}{\sqrt{\pi}}e^{-x^2}\right],
  \label{eq:chandraiso}
\end{equation}
where $x=V/(\sqrt{2}\sigma)$.  Assuming the massive object is in circular orbit, then its velocity is  $V=v_c$.
Thus, the frictional force $F= M\frac{d{V}}{dt}$  is
\begin{equation}
\begin{aligned}
F=&-\frac{4\pi{\rm ln}\Lambda\,G^2M^2\rho}{v_c^2}\left[{\rm erf}(x)-\frac{2x}{\sqrt{\pi}}e^{-x^2}\right]_{V=v_c}
\\
=&-\frac{{\rm ln}\Lambda\,GM^2}{  r^2}\left[{\rm erf}(1)-\frac{2}{\sqrt{\pi}}e^{-1}\right]
\\
\approx &  -0.428\ {\rm ln}\Lambda\frac{GM^2}{r^2}\, ,
  \label{eq:chandraiso2}
\end{aligned}
\end{equation}
where we have substituted the isothermal sphere density $\rho=v_c^2/(4\pi G r^2)$ and evaluated the expression in brackets at $x=v_c/(\sqrt{2}\sigma)=1$.
\\
The angular momentum of the circular orbit of the object $M$ can be written as $L=V\,r$, so that
\begin{equation}
  \frac{dL}{dt}=\frac{dV}{dt}r +V\frac{dr}{t}. 
\end{equation}
Instantaneously, DF does not act on $r$ and only changes the velocity of the massive object so that we can write $dL/dt=(dV/dt)r=(F/M)r$. On the other hand, the massive object is in circular orbit in an isothermal sphere. Under the approximation that the orbit remains circular, $V$ cannot change (it is always $\sqrt{2}\sigma$), so that the actual result of the interaction would be to move the object onto a tighter orbit, thus shrinking $r$. In practice, DF does not change the kinetic energy of $M$ (it cannot), but it eventually extracts its potential energy, so that we can write
\begin{equation}
  \frac{dL}{dt}=\frac{F}{M}r=V\frac{dr}{t}. 
\end{equation}
From Eq. (\ref{eq:chandraiso2}) we therefore have
\begin{equation}
  V\frac{dr}{dt}=-0.428{\rm ln}\Lambda\frac{GM^2}{r},
\end{equation}
that we can integrate by parts to get
\begin{equation}
 t_f=\frac{1.17}{{\rm ln}\Lambda}\frac{r_i^2v_c}{GM}=\frac{19}{{\rm ln}\Lambda}{\rm Gyr}\left(\frac{r_i}{5\,{\rm kpc}}\right)^2\frac{\sigma}{200\,{\rm km\,s}^{-1}}\frac{10^8{\rm M}_\odot}{M}.
\end{equation}
For typical ${{\rm ln}\Lambda}\approx 10-15$, MBHs can inspiral in the centre of the stellar remnant from a 10 kpc initial distance in less than a Gyr.

This is the time it takes to bring a single MBH to the centre of an isothermal distribution of stars. It can be applied to the two MBHs inspiralling in the aftermath of a galaxy merger so long as they evolve independently of each other as individual objects interacting with the stellar distribution. This is no longer true when the two MBHs start to 'see each other', \emph{i.e.}, they feel each other potential. This happens when the mass in stars enclosed in the separation between the two objects is of the order of the binary mass. For an isothermal sphere this amounts to a mutual separation of
\begin{equation}
a\approx \sqrt{GM}{\sigma^2}\approx 30\,{\rm pc}\left(\frac{M}{10^8{\rm M}_\odot}\right)^{1/2},
\end{equation}
where in the last approximation we used the $M-\sigma$ relation in the form $M_6=70\sigma^4_{70}$, where $M_6=M/10^6{\rm M}_\odot$ and $\sigma_{70}=\sigma/70\,{\rm km\,s}^{-1}$. At this point the two MBHs bind in a Keplerian binary that responds to the surrounding environment as a single object. As a consequence, DF acts as a perturbation of the binary CoM without affecting the relative motion of the individual objects, thus being ineffective in extracting angular momentum from the binary. Note that Eq. (\ref{a0}) for an equal mass binary gives
\begin{equation}
a_0\approx0.02\,{\rm pc}\left(\frac{M}{10^8{\rm M}_\odot}\right)^{3/4}\left(\frac{t}{1\,{\rm Gyr}}\right)^{1/4}F(e)^{-1/4},
\end{equation}
implying that GWs are efficient in merging the MBHB in less than an Hubble time only if it can get to sub-pc separations (or if it is extremely eccentric). If no further physical mechanisms were at play, MBHBs would therefore stall at $\sim$pc separation and would not produce significant GW emission.  

\subsubsection{Stellar hardening}

Fortunately, the MBHB has efficient ways to interact with the dense stellar and/or gaseous environment of the galactic nucleus. In particular, further hardening of the system at sub-parsec scales can proceed either via 3-body scattering of background stars or via torques exerted by a circumbinary disk. In the following we have a quick look in both mechanisms.

The physics of MBHB hardening in a stellar background is the same as described in Sections \ref{sec:3body} and \ref{sec:3body_gw}. In there, we derived the hardening equation in the form
\begin{equation}
  \frac{da}{dt}=-\frac{GH\rho}{\sigma}a^2.
  \label{eq:hardreprise}
\end{equation}
The main issue is that, in general MBHBs interact with stars on very different scales, and $\rho$ can be a (rather strong in fact) function of $r$ (and so can be $\sigma$). So to apply this equation to MBHBs we need to understand what $\rho$ and $\sigma$ are appropriate. To get some insight on this issue let us consider the general expression for the relaxation time in Eq. (\ref{trelaxdef}). By identifying $v$ with the typical velocity dispersion of the system $v=\sigma=(GNm/R)^{1/2}$, defining the typical density as $\rho=Nm/R^3$ and the Coulomb logarithm as ${\rm ln}\Lambda={\rm ln}N$, the equation gives\cite{1971ApJ...164..399S}
\begin{equation}
  t_{\rm rlx}\approx \frac{0.34\sigma^3}{G^2m\rho}{\rm ln}\Lambda=10\,{\rm Gyr}\left(\frac{\sigma}{200{\rm km\,s}^{-1}}\right)^{3}\left(\frac{\rho}{10^6{\rm M}_\odot{\rm pc}^{-3}}\right)^{-1}\left(\frac{m}{1{\rm M}_\odot}\right)^{-1}\left(\frac{\Lambda}{15}\right)^{-1}.
  \label{eq:relaxgeneral}
\end{equation}

This is the time a star needs to be deflected by $\delta v\approx v$, \emph{i.e.}, by an angle $\theta \approx \pi/2$. Consider a star orbiting at a distance $r\gg a$ from a MBHB. In this limit, the angle subtended by the binary as seen by the star is $\theta \approx 1/r$. When a star interacts with the MBHB is typically ejected from the system, and we need new stars to be able to interact with it. Those are the stars lying at the edge of the 'loss cone', \emph{i.e.}, those stars orbiting with a closest approach $r_p\gtrsim a$. So we are interested in the time needed to deflect a star by  $\theta \approx 1/r$, so that it can interact with the binary. This is $t_\theta\approx t_{\rm rlx}\theta\approx  t_{\rm rlx}/r$. If we now take an isothermal sphere we can consider two limits. If $r\gg r_{\rm inf}$, then $\sigma$ is constant and by substituting $\rho$ in Eq. (\ref{eq:relaxgeneral}) we get $t_\theta\propto r$. Conversely, if $r \ll r_{\rm inf}$, then $\sigma \propto r^{-1/2}$. The density still goes as $1/r^2$ and we get  $t_\theta\propto r^{-1/2}$. So the main contribution to the stars interacting with the binary comes from $r\approx r_{\rm inf}$. Still, for $a\approx 1$pc and $r_{\rm inf}\approx 30$ pc, we get that $t_\theta>10$ Gyr, unless $\rho_{\rm inf}>10^4$M$_\odot$ pc$^3$. This means that the loss cone is 'empty', and the hardening of the binary can take more than an Hubble time, which is known as {\it last parsec problem}.

Because of triaxiality and other effects, the relaxation time in nuclei of merging galaxies is generally much shorter than $t_{\rm rlx}$, which means that in general the loss cone is full at $r_{\rm inf}$ and binary hardening proceeds roughly according to Eq. (\ref{eq:hardreprise}) where $\rho=\rho(r_{\rm inf})$ and $\sigma=\sigma(r_{\rm inf})$ \cite{2015MNRAS.454L..66S,2015ApJ...810...49V}. If we now consider $\rho=\sigma^2/(2\pi G r^2)$ and $r_{\rm inf}=2GM/\sigma^2$, appropriate for an isothermal sphere, and use $M_6=70\sigma^4_{70}$ we can rewrite Eq. (\ref{atrans}) and {\ref{ttrans}} as
\begin{equation}
    a_{*/{\rm gw}} \approx 0.01\, {\rm pc}\left(\frac{M}{10^8{\rm M}_\odot}\right)^{3/4},
\label{eq:atransgw}
\end{equation}
\begin{equation}
  t(a_{*/{\rm gw}})\approx 10^8{\rm yr}.
\label{eq:ttransgw}  
\end{equation}
The latter being almost independent on the mass scale and other parameters. Below $a_{*/{\rm gw}}$ GW emission takes over leading to final coalescence in a short time, therefore, the timescale for MBHB coalescence (from the end of the DF phase) is set by $t(a_{*/{\rm gw}})$. The derivation is obviously different for different density profiles, but in general it is found that stellar dynamics can lead to final coalescence in less than a Gyr \cite{2015MNRAS.454L..66S,2015ApJ...810...49V}.

\subsubsection{Interaction with gas}

Galaxies are also observed to host large reservoirs of gas. During the merger process, gas infall into the nuclear region is triggered which can result in extensive starbursts and feeding of the newly formed MBHB. The nature of MBHB-gas interaction is a long standing problem, and it has been studied under a number of different configurations and assumptions. Here we just put forward some simple argument, aiming at giving a general idea of how this interaction can efficiently shrink the MBHB. We assume, for simplicity, that the gas infall produces a massive circumbinary disk, rotating in the same plane of the binary. In this scenario, because of Lindblad resonances, the MBHB carves a hollow (usually referred to as gap) into the central region of the disk \cite{1994ApJ...421..651A} of size $r_{\rm gap}\approx 2a$. In terms of torque balance, this implies that the action of the binary counteracts the angular momentum flow within the disk. If we take an unperturbed thin accretion disk, in a steady state situation, then the flow of mass $\dot{m}$ is equal everywhere in the disk. If we sit at radius $r$ then the angular momentum flow across that radius is simply $dL_{\rm gas}/dt=-\dot{m}\sqrt{GMr}$, where we defined $\dot{m}>0$ and the minus sign takes into account for the fact that the gas is spiralling in (so it is losing angular momentum). Torques exerted by the binary act like a dam at $r_{\rm gap}$. If no gas leaks through the dam, then the binary is injecting in the gas an amount of angular momentum $-dL_{\rm gas}/dt$, taken from its own orbital angular momentum. The MBHB angular momentum, thus, evolves according to \cite{2015MNRAS.448.3603D}
\begin{equation}
  \frac{dL}{dt}=-\dot{m}\sqrt{GMr_{\rm gap}}.
  \label{eq:Lcons}
\end{equation}
By writing the angular momentum of the (assumed circular) binary as $L=\mu\sqrt{GMa}$, where $\mu=M_1M_2/M$ is the reduced mass, in the approximation of no accretion onto the MBHB (no leaking through the dam) Eq. (\ref{eq:Lcons}) gives
\begin{equation}
  \frac{da}{a}=-2\sqrt{2}\frac{dM}{\mu},
\end{equation}
so that the binary shrinks by $\approx$ 3 e-folds as a mass of the order $\mu$ is accumulated at the gap rim. Assuming $\dot{m}$ to be Eddington limited and $\epsilon=0.1$ this occurs in $\approx 4\times 10^7$ yr. This simple argument shows that interaction with a massive gaseous disk can in principle shrink the binary to the efficient GW emission stage in less than $10^8$ yr.

This simple argument assumes that mass is piled up at the inner rim of the gap. However, simulations showed that forcing from the quadrupolar potential of the binary triggers the infall of streams from the inner rim of the disk, part of which are accreted by the binary, with the rest being flung back to the disk at super-Keplerian velocities (\emph{e.g.}, \cite{2014MNRAS.439.3476R}). The dynamics of the interaction is quite complex and involves the estimate of various torques (viscous, gravitational, due to accretion, etc). We just make here some simple considerations. Assuming a circular, equal mass binary, the specific angular momentum (\emph{i.e.}, angular momentum per unit mass) of the binary is $l=\sqrt{GMa}/4$, differentiating, we get:
\begin{equation}
  \frac{da}{a}=2\frac{dl}{l}-\frac{dM}{M}.
\end{equation}
Gas is brought from the inner rim to the MBH, forms small mini-disks and is eventually accreted with a specific angular momentum equal to the one of the accreting BH, so that, in first approximation $dl=0$.  Gas is accreted at a rate $\dot{m}$ from the inner rim, so the initial angular momentum of the accreting gas $L_{\rm gas}=m\sqrt{GMr_{\rm gap}}$ has to go somewhere. This is carried back to the disk by the portion of the streams that are flung back, impacting on the disk and heating it up \cite{2014MNRAS.439.3476R}. Under this condition
\begin{equation}
  \frac{da}{a}=-\frac{dM}{M},
\end{equation}
which means that the binary separation still shrinks by an e-folds by accreting of the order of its own mass.

Following galaxy mergers, MBHBs are therefore expected to coalesce in less then 1 Gyr due to the dynamical processes described above. With a rough count of $10^{11}$ galaxies in the Universe, if each of them experienced at least a major merger within its lifetime, then we might expect roughly $10^{11}/t_{\rm Hubble}\approx 10$ yr$^{-1}$ MBHB mergers. The upcoming Laser Interferometer Space Antenna is expected to probe the cosmic evolution of MBHBs by detecting tens-to-hundreds of mergers within its 4-to-10 yr lifetime \cite{2016PhRvD..93b4003K}.


\section*{Acknowledgements}
These lecture notes are based on the course \emph{``Black hole astrophysics"} taught by one of us, Alberto Sesana, in the \emph{``School on Gravitational Waves for Cosmology and Astrophysics''} at the Centro de Ciencias de Benasque - Pedro Pascual in June 2017. The authors are thankful to the organizers: D. Blas, C. Caprini, V. Cardoso, G. Nardini, to the supporting staff of the Centro de Ciencias, to all the lecturers and the students. A.S. is supported by the Royal Society.   M.M.  acknowledges financial support from the MERAC Foundation, from INAF through PRIN-SKA, from MIUR through Progetto Premiale 'FIGARO'  and 'MITiC', and from the Austrian National Science Foundation through FWF stand-alone grant P31154-N27.
\bibliography{references_Roberto}

\end{document}

%% file: binsystem.pdf_tex
\begingroup%
  \makeatletter%
  \providecommand\color[2][]{%
    \errmessage{(Inkscape) Color is used for the text in Inkscape, but the package 'color.sty' is not loaded}%
    \renewcommand\color[2][]{}%
  }%
  \providecommand\transparent[1]{%
    \errmessage{(Inkscape) Transparency is used (non-zero) for the text in Inkscape, but the package 'transparent.sty' is not loaded}%
    \renewcommand\transparent[1]{}%
  }%
  \providecommand\rotatebox[2]{#2}%
  \ifx\svgwidth\undefined%
    \setlength{\unitlength}{218.79704533bp}%
    \ifx\svgscale\undefined%
      \relax%
    \else%
      \setlength{\unitlength}{\unitlength * \real{\svgscale}}%
    \fi%
  \else%
    \setlength{\unitlength}{\svgwidth}%
  \fi%
  \global\let\svgwidth\undefined%
  \global\let\svgscale\undefined%
  \makeatother%
  \begin{picture}(1,0.50209118)%
    \put(0,0){\includegraphics[width=\unitlength,page=1]{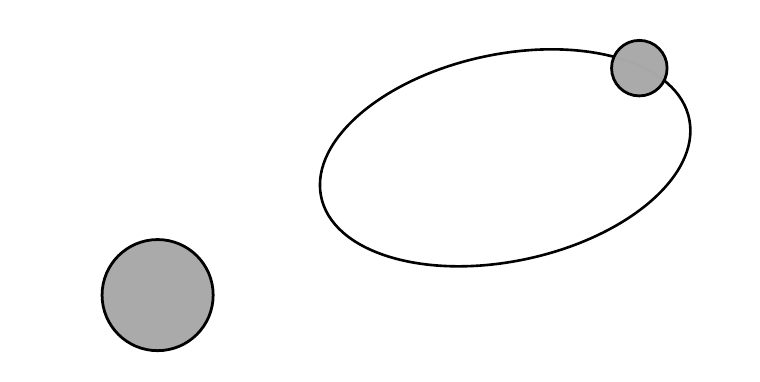}}%
    \put(-0.00379383,0.18947298){\color[rgb]{0,0,0}\makebox(0,0)[lb]{\smash{$M_2$}}}%
    \put(0.85420546,0.46736651){\color[rgb]{0,0,0}\makebox(0,0)[lb]{\smash{$M_1$}}}%
    \put(0.34596237,0.12922389){\color[rgb]{0,0,0}\makebox(0,0)[lb]{\smash{$\bold{r_2}$}}}%
    \put(0.61493828,0.2468713){\color[rgb]{0,0,0}\makebox(0,0)[lb]{\smash{$\bold{r_1}$}}}%
    \put(0.89532921,0.23643571){\color[rgb]{0,0,0}\makebox(0,0)[lb]{\smash{}}}%
    \put(0,0){\includegraphics[width=\unitlength,page=2]{binsystem.pdf}}%
  \end{picture}%
\endgroup%